\documentclass[aps,prx,superscriptaddress,reprint,showpacs,longbibliography,amsmath,amssymb,nofootinbib,nobibnotes]{revtex4-2}
\usepackage[utf8]{inputenc}
\usepackage[T1]{fontenc}
\usepackage{microtype}
\usepackage{graphicx}
\usepackage{mathbbol}
\usepackage{mathrsfs,amssymb,bm}
\usepackage{etoolbox}
\usepackage[usenames,dvipsnames,svgnames,table]{xcolor}
\usepackage[colorlinks,linkcolor=blue,citecolor=blue,urlcolor=blue]{hyperref}

\makeatletter 
    
\renewcommand\onecolumngrid{
\do@columngrid{one}{\@ne}%
\def\set@footnotewidth{\onecolumngrid}
\def\footnoterule{\kern-6pt\hrule width 1.5in\kern6pt}%
}

\renewcommand\twocolumngrid{
        \def\footnoterule{
        \dimen@\skip\footins\divide\dimen@\thr@@
        \kern-\dimen@\hrule width.5in\kern\dimen@}
        \do@columngrid{mlt}{\tw@}
}%

\makeatother    

\newcommand{\nocontentsline}[3]{}
\let\oldaddcontentsline\addcontentsline
\let\addcontentsline\nocontentsline

\renewcommand{\i}{\mathrm{i}}

\def\equationautorefname~#1\null{Eq.~(#1)\null}

\newcommand{\braket}[1]{\ensuremath{\langle{#1}\rangle}}

\newcommand{\ket}[1]{| #1 \rangle}
\newcommand{\abschi}{\lvert\chi\rvert}
\newcommand{\myvec}[1]{\bm{\mathrm{#1}}}
\newcommand{\iu}{\mathrm{i}}

\let\Re\relax
\let\Im\relax
\DeclareMathOperator{\Re}{Re}
\DeclareMathOperator{\Im}{Im}
\DeclareMathOperator{\diag}{diag}

\def\equationautorefname~#1\null{Eq.~(#1)\null}

\begin{abstract}
Nonreciprocity means that the transmission of a signal depends on its direction of propagation.
Despite vastly different platforms and underlying working principles, the realisations of nonreciprocal transport in linear, time-independent systems rely on Aharonov-Bohm interference among several pathways and require breaking time-reversal symmetry.
Here we extend the notion of nonreciprocity to unidirectional bosonic transport in systems with a time-reversal symmetric Hamiltonian by exploiting interference between beamsplitter (excitation preserving) and two-mode-squeezing (excitation non-preserving) interactions. In contrast to standard nonreciprocity, this unidirectional transport manifests when the mode quadratures are resolved with respect to an external reference phase. Hence we dub this phenomenon \emph{quadrature nonreciprocity}.
First, we experimentally demonstrate it in the minimal system of two coupled nanomechanical modes orchestrated by optomechanical interactions.
Next, we develop a theoretical framework to characterise the class of networks exhibiting quadrature nonreciprocity based on features of their particle-hole graphs.
In addition to unidirectionality, these networks can exhibit an even-odd pairing between collective quadratures, which we confirm experimentally in a four-mode system, and an exponential end-to-end gain in the case of arrays of cavities.
Our work opens up new avenues for signal routing and quantum-limited amplification in bosonic systems.
\end{abstract}

\begin{document}
	\title{Quadrature nonreciprocity: unidirectional bosonic transmission without breaking time-reversal symmetry}
	%

\author{Clara C.~Wanjura}
\thanks{C.~C.~W.~and J.~J.~S.~contributed equally to this work.}
\affiliation{Cavendish Laboratory, University of Cambridge, Cambridge CB3 0HE, United Kingdom}

\author{Jesse J.~Slim}
\thanks{C.~C.~W.~and J.~J.~S.~contributed equally to this work.}
\affiliation{Center for Nanophotonics, AMOLF, Science Park 104, 1098 XG Amsterdam, The Netherlands}

\author{Javier del Pino}
\affiliation{Institute for Theoretical Physics, ETH Zürich, 8093 Zürich, Switzerland}
\affiliation{Center for Nanophotonics, AMOLF, Science Park 104, 1098 XG Amsterdam, The Netherlands}

\author{Matteo Brunelli}
\affiliation{Department of Physics, University of Basel, Klingelbergstrasse 82, 4056 Basel, Switzerland}

\author{Ewold Verhagen}
\email{verhagen@amolf.nl}
\affiliation{Center for Nanophotonics, AMOLF, Science Park 104, 1098 XG Amsterdam, The Netherlands}

\author{Andreas Nunnenkamp}
\email{andreas.nunnenkamp@univie.ac.at}
\affiliation{Faculty of Physics, University of Vienna, Boltzmanngasse 5, 1090 Vienna, Austria}
	
	\maketitle
	\DeclareGraphicsExtensions{.pdf,.png,.jpg}
	%
	\begin{figure*}[t]
		\includegraphics[width=\linewidth]{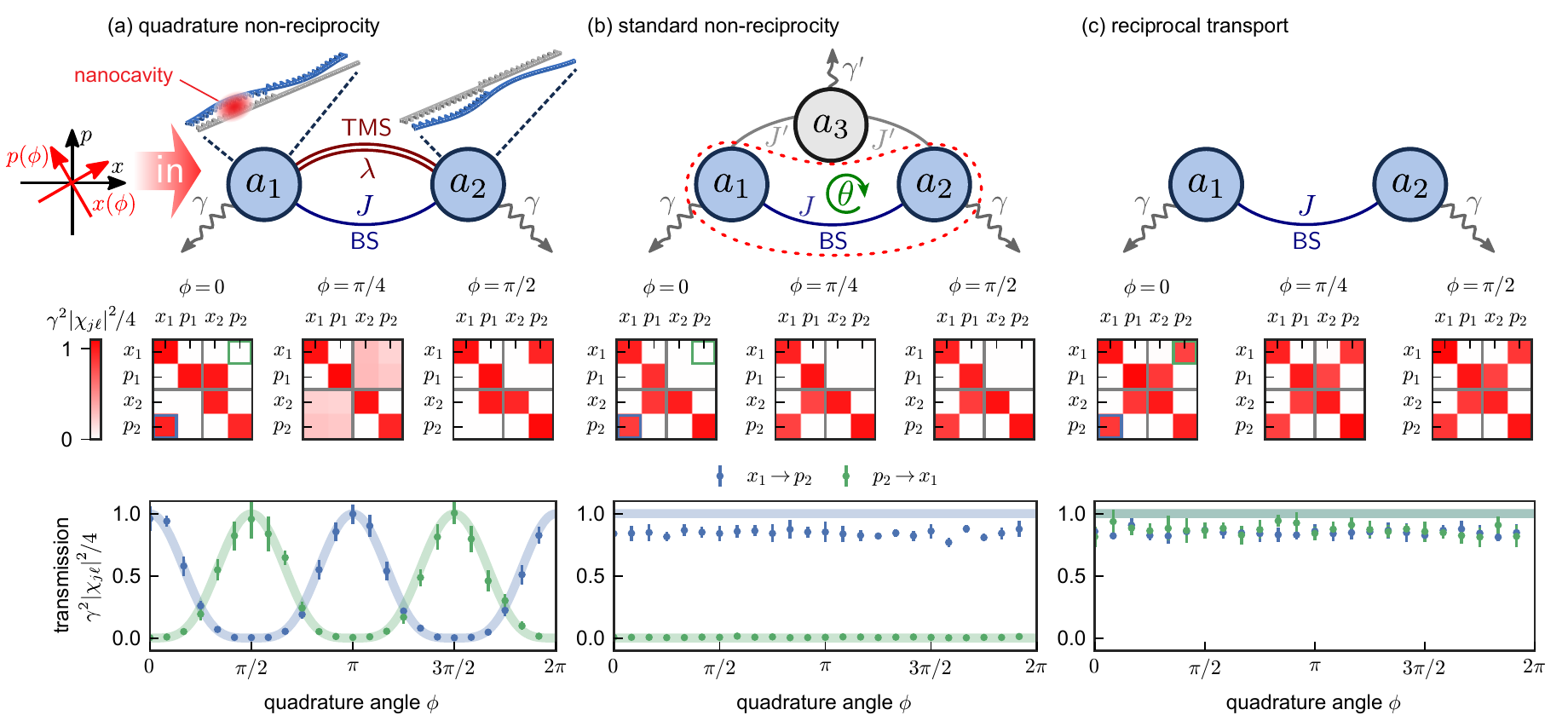}
		\caption{\textbf{Quadrature nonreciprocity (qNR) vs.~standard nonreciprocity (sNR) vs.~reciprocal transport.}
		Coupled-mode diagrams for three systems that feature distinct forms of transport, realized in an experiment with mechanical modes $a_{1,2}$ (equal decay rate $\gamma = 2\pi\times3.7$~kHz) subject to optomechanically-mediated beamsplitter (BS, strength $J$) and two-mode squeezing (TMS, strength $\lambda$) interactions. Panel (a) includes a sketch of the spatial profiles for the mechanical modes coupled to the experimental optical nanocavity. Coherent excitation along quadratures (red) rotated by $\phi$ from the reference gauge (black) is sketched (first row). Corresponding measured susceptibility matrix amplitudes evolve differently with $\phi$ (middle row). Nonreciprocity manifests in an asymmetric modulus of the susceptibility matrix $\lvert\chi\rvert\neq\lvert\chi\rvert^\mathrm{T}$. The behaviour of the elements within the green and blue boxes shows whether reciprocity can be tuned by the incoming quadrature angle (bottom row).
		(a)~Two modes interact through BS and TMS of equal strength ($J = \lambda = \gamma /4$), showing qNR. Nonreciprocity vanishes for $\phi=\pi/4$. Furthermore, setting $\phi=\pi/2$ inverts the transmission direction.
		(b)~An auxiliary mechanical mode $a_3$ (decay rate $\gamma^\prime = 2\pi\times1.3$~kHz) is coupled to $a_{1,2}$ through BS interactions of strength $J^\prime=\sqrt{\gamma\gamma^\prime}/2$, providing a coupling between $a_1$ and $a_2$ via $a_3$ that can interfere with the direct coupling $J=\gamma/2$. SNR requires the breaking of TRS -- realized by picking a non-trivial phase $\theta = \pi/2$ along both paths, akin to a synthetic flux -- so that the susceptibility matrix is nonreciprocal for all~$\phi$. Regardless of the hierarchy of loss rates $\gamma \lessgtr \gamma^\prime$, perfect isolation can be attained (Supplementary Information Sec.~IB).
		(c)~Reciprocal transport between two modes coupled only through BS ($J=\gamma/2$) is characterised by a symmetric susceptibility matrix for any phase $\phi$. Error bars are obtained by repeating the measurement sweep $10$ times and represent the statistical $\pm 2\sigma$ spread around the average value.
		}
			\label{fig:1}
	\end{figure*}%
	
	In a nonreciprocal system, e.g.~an isolator, the transmission varies when interchanging input and output, in an ideal case from unity to zero~\cite{Deak2012,Jalas2013,Caloz2018,Verhagen2017}.
	Nonreciprocity is a resource for many applications, such as sensing~\cite{Lau2018},
	the construction of bosonic networks with general routing capabilities~\cite{Ranzani2015}, and the realisation of topological phases~\cite{Ozawa2019}.
	Complemented with gain, nonreciprocity amplifies weak signals while protecting the source against noise, making it ideal for quantum information processing applications~\cite{Abdo2011}.
	Magnetic-free isolators and directional amplifiers have been proposed based on parametric modulation~\cite{Yu2009,Lira2012,Caloz2018}, interfering parametric processes~\cite{Kamal2011,Malz2018}, and reservoir engineering~\cite{Metelmann2015} and have been experimentally demonstrated in different platforms, e.g. superconducting circuits~\cite{Abdo2013,Sliwa2015,Lecocq2017} and optomechanical systems~\cite{Kim2014,Shen2016,Ruesink2016,Fang2017,Peterson2017,Barzanjeh2017,Bernier2017,Sohn2021}.
	
	In linear systems, achieving a nonreciprocal response relies on breaking time-reversal symmetry (TRS) in the Hamiltonian by employing real or synthetic magnetic fields and dissipation. The directionality in this standard kind of nonreciprocity (sNR) does not depend on the phase of the input signal. 
	In this work we extend the notion of nonreciprocity in linear bosonic systems by identifying a class of systems that show a kind of unidirectional signal transmission, positioned between reciprocal and standard nonreciprocal transmission, whose (uni)directionality depends on the phase of the input signal. We dub the defining property of this class quadrature nonreciprocity (qNR), as it can be revealed when resolving the signal into its quadrature components.
	In contrast to sNR, a qNR Hamiltonian does not break TRS, but achieves unidirectional transport by interfering beamsplitter (excitation-preserving) and two-mode squeezing (excitation non-preserving) interactions.
	It does not require strong Kerr nonlinearity \cite{Fan2012,Caloz2018,Wang2022} or spin-polarized emitters \cite{Sayrin2015,Buddhiraju2020}.
	We report its experimental realisations using an optomechanical network and construct a comprehensive theoretical framework exposing an entire class of qNR systems with exciting properties, including an even-odd pairing between collective quadratures and exponential end-to-end gain in resonator chains.
	Our work introduces a systematic tool to treat unidirectional phase-sensitive transport in bosonic lattice systems, which have recently sparked interest in connection with bosonic analogues of the Kitaev chain~\cite{McDonald2018,Flynn2020,Flynn2021}.
	It further opens the door to studying exotic phenomena in these models, such as multi-mode entanglement~\cite{McDonald2018} and non-Hermitian topology~\cite{Wanjura2020}. From the point of view of applications, the concept of qNR opens up new avenues for signal routing and amplification.

	\subsection*{Defining quadrature nonreciprocity (qNR)}

	We consider a network of $N$ driven-dissipative bosonic modes. Their steady-state response to a coherent probe follows from the Heisenberg-Langevin equations of motion $\dot{\mathbf{q}} = H \mathbf{q} - \sqrt{\gamma} \mathbf{q}_\mathrm{in}$ for the field quadratures $x_j\equiv(a_j+a_j^\dagger)/\sqrt{2}$ and $p_{j}\equiv -\i(a_j-a_j^\dagger)/\sqrt{2}$, where $a_j$ denotes the annihilation operator for the bosonic mode $j\in(1,\cdots,N)$, $H$ the dynamical matrix and $\gamma$ the damping rate.
	We use the vector notation $\mathbf{q}\equiv(x_1,p_1,\cdots,x_N,p_N)^\mathrm{T}$ for the system's quadratures and similarly $\mathbf{q}_\mathrm{in}$ for the quadrature inputs. We assume that each mode dissipates only via coupling to the input-output port, although our analysis extends to general temporal coupled-mode theory~\cite{Fan2003,Ruesink2016}.
	The scattering matrix $S(\omega)$ connects input $\mathbf{q}_\mathrm{in}$ and output quadratures $\mathbf{q}_\mathrm{out}$, oscillating with frequency $\omega$, which satisfy the input-output boundary conditions $\mathbf{q}_{\mathrm{out}} = \mathbf{q}_{\mathrm{in}} + \sqrt{\gamma} \mathbf{q}$ ~\cite{Gardiner1985,Clerk2010}.
	Namely,
	\begin{align}
		\label{eq:SMat}
	   S(\omega) & \equiv \mathbb{1} + \gamma (\i\omega\mathbb{1} + H)^{-1} \equiv \mathbb{1} + \gamma \chi(\omega),
	\end{align}
	with the susceptibility matrix $\chi(\omega)\equiv(\i\omega\mathbb{1} + H)^{-1}$. We will from now on consider driving at resonance ($\omega=0$ in a frame rotating at the mode frequencies) and write $\chi\equiv\chi(0)$, $S\equiv S(0)$ for brevity.
	
	Let us consider a rotation of each pair of quadratures $\{x_j,p_j\}$, with respect to some external phase reference, see Fig.~\ref{fig:1}~(a) inset, namely
	\begin{align}
	   \begin{pmatrix}
	      x_j(\phi_j) \\ p_j(\phi_j)
	   \end{pmatrix}
	   & = \begin{pmatrix}
	      \cos\phi_j & \sin\phi_j \\
	      -\sin\phi_j & \cos\phi_j
	   \end{pmatrix}
	   \begin{pmatrix}
	      x_j(0) \\ p_j(0)
	   \end{pmatrix}
	   \equiv
	   R(\phi_j)
	   \begin{pmatrix}
	      x_j(0) \\ p_j(0)
	   \end{pmatrix}.
	   \label{eq:quadratureRotation}
	\end{align}
    Equation~(\ref{eq:quadratureRotation}) can be understood as a $U(1)$ gauge transformation in the mode basis $\{a_{j},a_{j}^{\dagger}\}$~\cite{Sliwa2015,Fang2017}, see Supplementary Information Sec.~IIA. The susceptibility matrix $\chi$ transforms as
	\begin{align}
	   \chi_\phi \equiv U(\phi_j) \chi U^\mathrm{T}(\phi_j),
	   \label{eq:chiGauge}
	\end{align}
	where we introduced $U(\phi_j)\equiv\oplus_{j=1}^N R(\phi_j)$.
	
	Nonreciprocity is typically defined by asymmetry in the transmission amplitudes~\cite{Deak2012,Caloz2018}.
	We adopt this notion in a generalised sense, with signals split into their $x$, $p$ quadratures, meaning $\lvert S\rvert^\mathrm{T}\neq \lvert S\rvert$ for Eq.~(\ref{eq:SMat}) or, equivalently, $\lvert \chi\rvert^\mathrm{T}\neq \lvert \chi\rvert$. Here $\lvert\cdots\rvert$ denotes taking the element-wise modulus.
	We will say that a system exhibits quadrature nonreciprocity (qNR) if there exist at least \emph{two} different sets of local gauges $\phi_{j}^{(1,2)}$ corresponding to a nonreciprocal and reciprocal susceptibility matrix, respectively. Mathematically, this implies a pair of rotations $U_{1,2}\equiv\oplus_{j=1}^N R(\phi_j^{(1,2)})$ such that
    \begin{align}\label{eq:qNR_cond}
         \lvert U_1\chi U_1^\mathrm{T} \rvert^\mathrm{T} \neq \lvert U_1 \chi U_1^\mathrm{T}\rvert&&\hspace{-2mm} \text{and} \hspace{-2mm}&& \lvert U_2 \chi U_2^\mathrm{T} \rvert^\mathrm{T} = \lvert U_2 \chi U_2^\mathrm{T} \rvert.
     \end{align}

	\subsection*{qNR dimer: the simplest qNR system}

	We now introduce the minimal system displaying qNR: the parametrically-driven dimer shown in Fig.~\ref{fig:1}~(a). It consists of two modes $a_{1,2}$ coupled via beamsplitter (BS) coupling of strength $J$ and two-mode-squeezing (TMS) coupling of strength $\lambda$. In a frame rotating at the mode frequencies, the Hamiltonian reads
	\begin{align}
	   \mathcal{H} & = J a_1^\dagger a_2 + \lambda a_1^\dagger a_2^\dagger + \mathrm{H.c.}, \label{eq:dimerModel}
	\end{align}
	 which corresponds to the most general bi-linear coupling.
	 For $J = \lambda$, we  recover the well-known position-position coupling, which has been extensively studied, e.g.~for implementing quantum non-demolition measurements~\cite{Braginsky1980, Clerk2008} or generating squeezing and entanglement~\cite{Brunelli2019}. Moreover, this Hamiltonian was also introduced as bi-directional phase-sensitive amplifier in~\cite{Metelmann2014Quantum} which was implemented in a superconducting circuit and used for qubit readout~\cite{Chien2020Multiparametric}, and considered as starting point for nonreciprocity with broken time-reversal symmetry in~\cite{Metelmann2015}.
  
	 As we now show, this coupling has important impact in the context of nonreciprocal signal transduction. Eq.~(\ref{eq:dimerModel}) gives the following equations of motion for the quadratures
	\begin{align}
	   \dot x_1 & = -\frac{\gamma}{2} x_1 + (J-\lambda) p_2 - \sqrt{\gamma} x_{1,\mathrm{in}}, \notag\\
	   \dot p_1 & = -\frac{\gamma}{2} p_1 - (J+\lambda) x_2 - \sqrt{\gamma} p_{1,\mathrm{in}}, \notag\\
	   \dot x_2 & = -\frac{\gamma}{2} x_2 + (J-\lambda) p_1 - \sqrt{\gamma} x_{2,\mathrm{in}}, \notag\\
	   \dot p_2 & = -\frac{\gamma}{2} p_2 - (J+\lambda) x_1 - \sqrt{\gamma} p_{2,\mathrm{in}},
	   \label{eq:eomsDimerDirectional}
	\end{align}
    from which we can see that the quadratures can decouple, due to the fact that BS and TMS couplings enter with the opposite sign.
    In particular, setting $J=\lambda$ in Eq.~(\ref{eq:eomsDimerDirectional}) leads to perfect decoupling between $\dot{x}_1$ ($\dot{x}_2$) and $p_2$ ($p_1$), while $\dot{p}_2$ ($\dot{p}_1$) still couples to $x_1$ ($x_2$), in a way which is formally equivalent to a cascaded quantum system~\cite{Carmichael1993,Gardiner1993}.
	This is also reflected in the susceptibility matrix
	\begin{align}
	   \chi & =  \begin{pmatrix}
	      -\frac{2}{\gamma} & 0 & 0 & 0 \\
	      0 & -\frac{2}{\gamma} & \frac{8J}{\gamma^2} & 0 \\
	      0 & 0 & -\frac{2}{\gamma} & 0 \\
	      \frac{8J}{\gamma^2} & 0 & 0 & -\frac{2}{\gamma}
	   \end{pmatrix}.
	   \label{eq:susceptibilityDimer}
	\end{align}
	We interpret this property by saying that a signal encoded in quadrature $x_1$ can propagate from mode 1 to 2, emerging as $p_2$, while the reverse transduction, i.e., $p_2\rightarrow x_1$ does not take place.
	When $J=\lambda>\gamma/4$, we further have $S_{x_2 \to p_1} = \gamma \chi_{x_2 \to p_1} > 1$, which signifies phase-sensitive amplification.
	
	To demonstrate this unidirectional transport between quadratures, we implement Hamiltonian~\eqref{eq:dimerModel} in a sliced photonic crystal nanobeam -- a nano-optomechanical network where multiple non-degenerate, MHz-frequency flexural mechanical modes are coupled to a broad-band, telecom frequency optical mode via radiation pressure. Two mechanical modes with equal linewidth $\gamma$ serve as resonators $a_j$, with effective interactions enabled by temporally modulated radiation pressure of a detuned laser. Specifically, BS coupling (TMS coupling) is stimulated by modulating the intensity of the drive laser at the mechanical frequency difference (sum)~\cite{mathew2020}, allowing control over both strength \emph{and} phase of the interaction through the depth and phase of the modulation tone, respectively. In effect, this protocol can implement arbitrary quadratic bosonic Hamiltonians for mechanical resonators~\cite{delPino2022}.
	
	In addition, the optomechanical interaction allows optical read-out of the nanobeam displacement, as imprinted on the intensity of a spectrally-resolved probe laser reflected from the cavity. The detected displacement signal $h(t) = k_1 z_1(t) + k_2 z_2(t) + \dots$ contains a superposition of the resonator coordinates $z_j(t)$, transduced with strengths $k_j$, which can be resolved in frequency as the resonator frequency separation is much larger than their linewidths.
	
	In our measurements, we define (electronic) local oscillators (LOs) at the resonator frequencies $\omega_j$ that demodulate -- in parallel -- the displacement signal, to obtain the amplitude envelope $\lvert a_j(t)\rvert$ and relative phase $\varphi_j(t)$ of each resonator's harmonic motion $z_j(t) = \lvert a_j(t)\rvert \cos(\omega_j t + \varphi_j(t))$. In effect, the LOs \emph{define} a rotating frame of reference in which the resonator dynamics can be tracked by a complex amplitude $\braket{a_j} = \lvert a_j\rvert e^{i\varphi_j}$, or equivalently by the quadrature amplitudes $\braket{x_j} = \sqrt{2}\Re(\braket{a_j})$ and $\braket{p_j} = \sqrt{2}\Im(\braket{a_j})$. Finally, signals to drive the resonator quadratures coherently through radiation pressure are derived from the same LOs, turning our experiment into a lock-in measurement. 	
	
	To connect the unidirectional response of Eq.~\eqref{eq:susceptibilityDimer} to the definition of qNR~\eqref{eq:qNR_cond}, we need to study how the dimer transforms under a change of gauge. We measure the dimer's quadrature-resolved response, when performing the gauge transformation in Eq.~(\ref{eq:chiGauge}) (see Fig.~\ref{fig:1}~(a) inset). This is experimentally achieved by referring both the interaction tones and LOs to a common time origin and subsequently adding a phase offset $\phi$ to the LOs, rotating the quadratures they define. Note that even though the frequencies of LOs ($\omega_{1,2}$) and interaction tones ($\omega_1\pm\omega_2$) are all distinct, the fact that the latter signals can be derived from the former through mixing leads to a well-defined relation between the LO and interaction phases (Methods). 
	
	For different phases $\phi$, we independently reconstruct the susceptibility matrix~\eqref{eq:chiGauge} for $J=\lambda$ in Fig.~\ref{fig:1}~(a). 
	For the full expression of~\eqref{eq:chiGauge} we refer to the Supplementary Information Sec.~I.A, here we focus on the matrix elements shown in the bottom row of Fig.~\ref{fig:1}, namely
	\begin{align}
	   \chi_{x_1\to p_2} & = \frac{8 J \cos ^2(\phi )}{\gamma ^2}, \label{eq:chiDimerRotated1} \\
	   \chi_{p_2\to x_1} & = -\frac{8 J \sin ^2(\phi )}{\gamma ^2}.
	   \label{eq:chiDimerRotated2}
	\end{align}
	
	It is clear that nonreciprocity only reveals itself in particular rotated quadratures.
	While maximal nonreciprocity is obtained for $\phi=0$,  as in Eq.~\eqref{eq:susceptibilityDimer}, a gauge transformation reduces the `contrast' of the nonreciprocity until, at $\phi=\pi/4$, the transport is completely reciprocal, i.e., $|\chi_{x_1\to p_2}|=|\chi_{p_2\to x_1}|=4 J/\gamma^2$, and in fact $\lvert\chi\rvert^\mathrm{T}=\lvert\chi\rvert$. This confirms that the dimer is indeed a qNR system, as per our definition~\eqref{eq:qNR_cond}. Further increasing $\phi$ swaps the direction of nonreciprocity, $a_1\leftrightarrow a_2$, with complete reversal at $\phi=\frac{\pi}{2}$, when $\chi_{x_1\to p_2}=0$, $|\chi_{p_2\to x_1}|=8 J/\gamma ^2$.
	
	In the qNR dimer, the cancellation at $\phi=\pi/4$ and the reversal of directionality fundamentally stems from TRS.
	We identify a system's TRS from its Hamiltonian, i.e. its dynamics in absence of local dissipation or gain. This is motivated by the fact that dissipation or gain by themselves, while breaking temporal symmetry in a `trivial' way, cannot induce nonreciprocal behaviour.
	In Supplementary Information Sec.~IIB-D we demonstrate that TRS implies the constraints $\lvert \chi_{x_j \to x_\ell}\rvert = \lvert \chi_{p_\ell \to p_j}\rvert $, $\lvert \chi_{x_j \to p_\ell}\rvert = \lvert \chi_{x_\ell \to p_j}\rvert $ for any gauges $\phi_{j}$ and reciprocity for at least one set of gauges. This is strikingly different from a system that breaks TRS, such as the isolator of Ref.~\cite{Metelmann2015}.
	In Fig.~\ref{fig:1}~(b) we show the measured susceptibility matrix for the `sNR isolator', which is implemented in our optomechanical system from two equal-linewidth mechanical modes $a_1$ and $a_2$, coupled directly via a BS interaction of strength $J$ while an auxiliary lower-order mechanical mode $a_3$ is introduced and coupled to both $a_1$ and $a_2$ with BS strength $J'$. Contrary to the qNR dimer, here isolation is enabled by a $U(1)$ gauge-invariant flux $\theta$, the relative phase between the couplings $J$, $J'$ as shown in Fig.~\ref{fig:1}~(b). Since nonreciprocity is controlled by the TRS-breaking flux, it is independent of local rotations in phase space by $\phi$, as reflected by the phase-independent susceptibility matrices in Fig.~\ref{fig:1}~(b) ($\lvert\chi_{x_1\to x_2}\rvert^2=\lvert\chi_{p_1\to p_2}\rvert^2=1$, $\lvert\chi_{x_2\to x_1}\rvert^2=\lvert\chi_{p_2\to p_1}\rvert^2=0$).
	
	For reference, we also contrast both notions of nonreciprocity against a reciprocal system. We display in Fig.~\ref{fig:1}~(c) the susceptibility matrix measured for two beamsplitter-coupled modes, which is completely reciprocal ($\lvert\chi\rvert^\mathrm{T}=\lvert\chi\rvert$) and gauge-invariant ($\chi_\phi=\chi$).

	\subsection*{Time-reversal symmetry and qNR}
	
	\begin{figure}
	   \centering
	   \includegraphics[width=\linewidth]{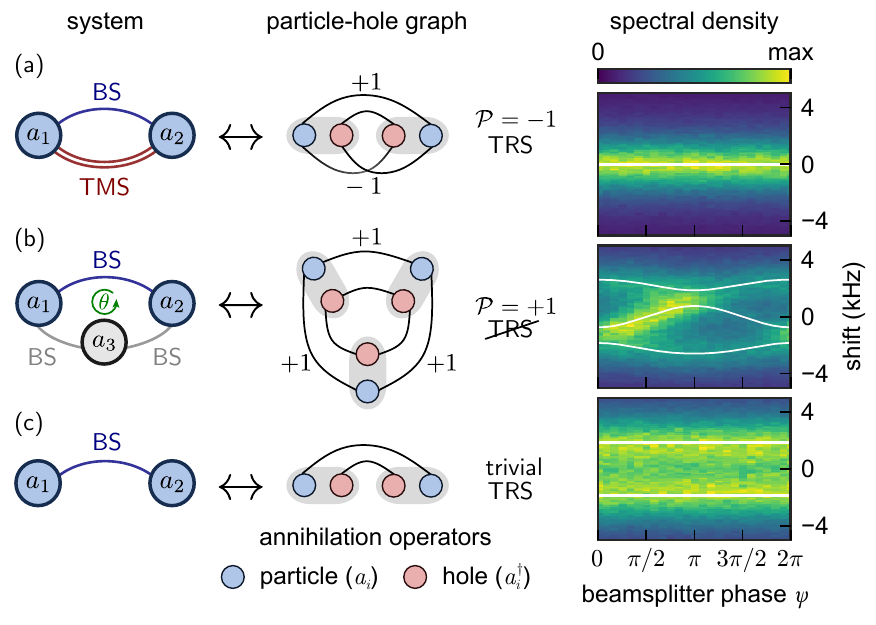}
	   \caption{\textbf{Time-reversal symmetry and qNR.}
	   The coupled-mode diagrams (left column) of the systems studied in Fig.~\ref{fig:1} are expanded using a general graph representation of quadratic Hamiltonians (middle column). Operators $a_j$, $a_\ell$, $a_j^\dagger$,  $a_\ell^\dagger$, annihilating `particles' and `holes' respectively, that are coupled in the equations of motions are connected through a line. (a)~If the graph forms a double loop as is the case for the qNR dimer, TRS is always preserved. (b)~If the resulting graph decomposes into two loops as for the sNR isolator, TRS can be broken by a gauge invariant phase, enabling sNR.
	   (c)~When the graph displays no loops as is the case for the trivial BS dimer, TRS is always preserved. The spectral density of the thermal fluctuations in mode $a_1$ serves as an experimental signature of TRS (right column). In a gauge where all other interactions are real, we vary the phase $\psi$ of the BS coupling between $a_1$ and $a_2$. For (a) and (c), this does not affect the system's eigenfrequencies, given by the real part of the dynamical matrix' eigenvalues (white lines), while for (b) it signals TRS breaking.
	   Experimental parameters are identical to those used in Fig.~\ref{fig:1}.
	   }
	   \label{fig:2}
	\end{figure}%

	\label{sec:qNRFormal}

	The results above point to the fact that preserving TRS in the Hamiltonian is a key element, which sets qNR apart from sNR.
	A time-reversal symmetric Hamiltonian imposes certain symmetries on the dynamical matrix that underlie the unique transformation properties of a qNR susceptibility matrix.
	This motivates us to provide a general characterisation of TRS Hamiltonians in bosonic networks.
	TRS means that there exists a $U(1)$ transformation 
	\begin{align}
	   a_j \to e^{\i\phi_j} a_j,
	   \label{eq:gaugeTransformation}
	\end{align}
	that renders the coefficients in the Hamiltonian real~\cite{Koch2010}.
    In concrete experiments, that realise the Hamiltonian terms through parametric driving, finding the phases $\phi_j$ in Eq.~\eqref{eq:gaugeTransformation} that lead to a real Hamiltonian is equivalent to making the time-dependent parametric drives symmetric in time ($t \to -t$). As we see here, even though there is one gauge in which the drive is symmetric in time, systems with phase-dependent transmission can feature unidirectional transmission. While TRS is often associated with reciprocity~\cite{Caloz2018}, in fact TRS \emph{only} requires $\lvert\chi\rvert$ to be symmetric for one set of phases $\phi_j^{(0)}$ in Eq.~\eqref{eq:gaugeTransformation}, see Supplementary Information Sec.~II.B.
	 Indeed, in qNR systems, reciprocity \emph{only} occurs when $\phi_j=\phi_j^{(0)}$ (in the qNR dimer, $\phi_j^{(0)}=\phi^{(0)}=\pi/4, 3\pi/4$) and nonreciprocity occurs for all $\phi_j\neq\phi^{(0)}$. This feature, which to the best of our knowledge has not been recognised before, positions qNR precisely between reciprocity and gauge-independent sNR.
	
	We develop a criterion to identify TRS for arbitrary quadratic bosonic Hamiltonians, based on the graph representation of the Hamiltonian matrix in the field basis, inspired by Bogoliubov-de-Gennes theory~\cite{delPino2022}, see Supplementary Information~II.B. We associate the Hamiltonian matrix with a graph in which the ladder operators $a_j$---annihilating `particle' excitations---and $a_j^\dagger$---annihilating `hole' excitations---are represented as vertices and the interactions as edges, 
	i.e., connecting $a_j$ to $a_\ell$ (and $a_j^\dagger$ to $a_\ell^\dagger$) for BS between sites $j$, $\ell$ and $a_j$ to $a_\ell^\dagger$ (and $a_j^\dagger$ to $a_\ell$) for TMS. 
	The graph representations of the three systems of Fig.~\ref{fig:1} display manifestly different structures (Fig.~\ref{fig:2}), leading to a general criterion for TRS (Supplementary Information Sec.~III). The graph of the qNR dimer, Fig.~\ref{fig:2}~(a), connects all vertices in a \textit{double} loop visiting each site twice. Such double loops guarantee TRS for arbitrary, complex, coupling constants.
	In contrast, the graph of the sNR isolator, Fig.~\ref{fig:2}~(b), decomposes into two disjoint loops. This structure allows to break TRS through a non-vanishing relative phase between the coupling constants.
	The BS dimer, Fig.~\ref{fig:2}~(c), displays no loops and trivially preserves TRS.
	
	\begin{figure*}[ht!]
	   \centering
	   \includegraphics{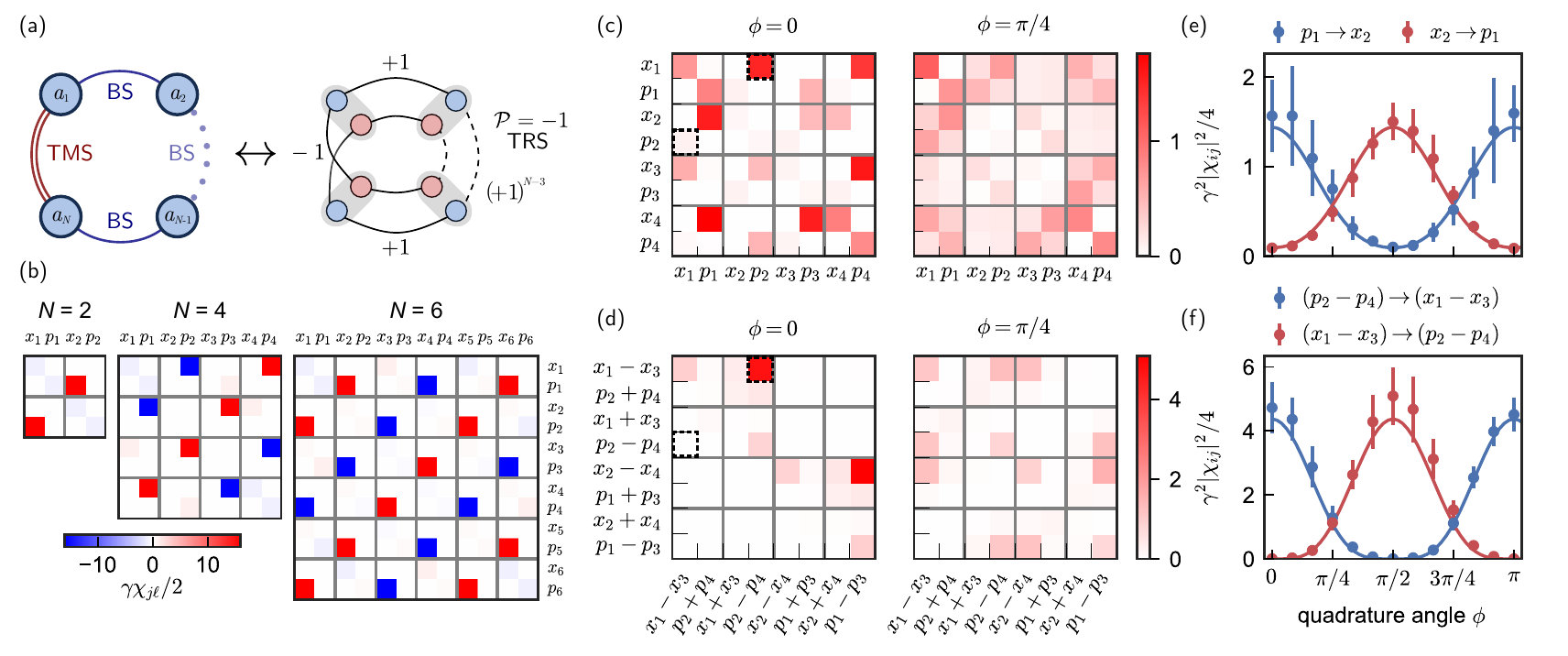}
	   \caption{\textbf{qNR transmission in ring networks.} (a)~Graph representation of a ring with $N$ modes, featuring a single TMS interaction. The ring is closed by $N-1$ BS interactions between the remaining modes. For arbitrary $N$ the loop product $\mathcal{P} = -1$, indicating preservation of TRS. To obtain qNR transmission, $N$ must be even. (b)~A feature of qNR that only becomes apparent for $N>2$ is the pairing between collective quadratures of even and odd sites in the susceptibility matrix $\chi$, shown here in theory for $J=\lambda=2N\cdot\gamma$ and $\phi = 0$. (c)~Susceptibility matrices $\abschi$ for the $N=4$ ring realized in experiment. We equalize all resonator dampings to $\gamma = 2\pi\times10$~kHz using feedback (Methods) and set $J=\lambda=2\pi\times5$~kHz, the largest $\lambda$ that permits a linear response to thermal noise in the experiment. For $\phi=0$, we obtain a nonreciprocal $\chi$ that shows collective pairing, albeit with reduced contrast compared to (b). For $\phi=\pi/4$, $\chi$ is reciprocal, showing the qNR of transmission of the ring. (d)~Susceptibility matrices in the basis of even and odd collective quadratures show their pairing for $\phi=0$, when $\chi$ is block-diagonal. Similarly, reciprocal susceptibility is obtained for $\phi=\pi/4$.
	   (e,f)~The values of selected susceptibility matrix elements (e: resonator basis; dashed boxes in c, f: collective basis; dashed boxes in d) show continuous tuning as function of quadrature angle $\phi$, in theory (solid line) and experiment (circles). Even in theory, the isolation for $\phi = 0, \pi/2, \pi$ in the resonator basis (e)~is not perfect, whereas in the collective basis (f)~it is. Error bars are obtained by repeating the measurement sweep $10$ times and represent the statistical $\pm 2\sigma$ spread around the average value.
	   }
	   \label{fig:3}
	\end{figure*}
	
	The different graph structures, which embody the behaviour under time reversal, are catalogued in general by a $\mathbb{Z}_2$ invariant, which we call \emph{loop product} $\mathcal{P}$. To define such a quantity, we consider the particle-hole graph for a general 
	loop with BS and TMS interactions. Then by multiplying by $(-1)$ for a line crossing (TMS coupling) and $(+1)$ for an uncrossed pair of lines (BS couplings) $(+1)$, the loop product $\mathcal{P}$ distinguishes between an even (disjoint loops) and odd number (double loop) of line crossings
    \begin{align}
      	\mathcal{P} \equiv (-1)^{n_\mathrm{TMS}} (+1)^{n_\mathrm{BS}}
      	=
      	\begin{cases}
      		 -1: & \text{double loop (TRS)} \\
      		 +1: & \text{disjoint loops}
      	\end{cases},
      	\label{eq:frustrationProduct}
    \end{align}
    with $n_\mathrm{BS}$ the number of BS couplings and $n_\mathrm{TMS}$ the number of TMS couplings.
	Eq.~\eqref{eq:frustrationProduct} indicates that the only requirement for TRS in a general loop is an odd number of TMS couplings, i.e. their position in the loop is irrelevant.
	
	Complementing our graph-based theoretical criterion above, the experimental response of a system to incoherent excitation serves as a signature of a TRS Hamiltonian~\cite{delPino2022}, as nontrivial fluxes manifest in the eigenfrequencies (Fig.~\ref{fig:2} (right column)).
	We choose a gauge in which all interactions are real, except for the BS coupling between modes $a_1$ and $a_2$ present in all three systems studied so far, whose phase $\psi$ we vary.
	Thermal fluctuations drive stochastically all mechanical quadratures homogeneously and lead to a power spectrum insensitive to $\psi$ if there is a gauge transformation~\eqref{eq:gaugeTransformation} that removes this BS phase.
	In our experiment this indicates the TRS of the qNR (Fig.~\ref{fig:2}~(a)) and BS Hamiltonians (Fig.~\ref{fig:2}~(c)). Conversely, if the eigenfrequencies tune with $\psi$, such a gauge transformation cannot exist, marking the broken TRS of the sNR isolator (Fig.~\ref{fig:2}~(b)).
	
	Finally, we stress another consequence of TRS, already visible in the dimer (see Eq.~\eqref{eq:chiDimerRotated2}), i.e., that quadratures travel in pairs in opposite directions. In fact, the transduction $x_j \to p_\ell$ is accompanied by $x_\ell \to p_j$ with equal transmission in the opposite direction, e.g., in Eq.~\eqref{eq:susceptibilityDimer}, $\lvert\chi_{x_j \to p_\ell}\rvert^2 = \lvert\chi_{x_\ell \to p_j}\rvert^2$, see the Supplementary Information Sec.~IIC-D for the proof. This requirement of counter-propagating pairs of quadratures is reminiscent of other TRS systems, such as quantum spin Hall systems~\cite{Hasan2010,Ozawa2019}.
	
	\subsection*{Constructing qNR ring networks}
	
	We take now inspiration from the particle-hole graphs and look for qNR in $N$-mode ring networks among TRS systems ($\mathcal{P}=-1$), dubbed `$N$-rings' from now on. To obtain a ring with qNR transmission, a TRS Hamiltonian is only necessary, not sufficient. We also have to guarantee that the BS and TMS couplings can interfere. The particle-hole representation allows a reinterpretation of the quadrature decoupling condition discussed along with Eq.~(\ref{eq:eomsDimerDirectional}) as constructive and destructive interference between particle-conserving and particle non-conserving processes.
	This condition brings us to the necessary and sufficient criterion for qNR. Besides the odd number of TMS couplings (Eq.~\eqref{eq:frustrationProduct}), achieving qNR requires that the ring consists of an \emph{even} number of modes.  To understand why, we take a closer look at the 4-mode ring with a single TMS interaction (Fig.~\ref{fig:3}~(a)).
	The corresponding equations of motion for equal couplings ($J=\lambda$) read
	\begin{align}
	   \dot x_1 & = - J (p_2 - p_4) - \frac{\gamma}{2} x_1 - \sqrt{\gamma} x_{1,\mathrm{in}} \notag \\
	   \dot p_2 - \dot p_4 & = - \frac{\gamma}{2} (p_2 - p_4) - \sqrt{\gamma} (p_{2,\mathrm{in}} - p_{4,\mathrm{in}}).
	   \label{eq:eomsN4}
	\end{align}
	
	The coupling to $x_1$ vanishes in the second equation for the collective quadrature $(p_2-p_4)$ due to the interference of BS and TMS couplings.
	Equations~\eqref{eq:eomsN4} have a similar structure as those for the qNR dimer~\eqref{eq:eomsDimerDirectional}, the main difference being that $x_1$ couples nonreciprocally to the \emph{collective} quadrature $(p_2 - p_4)$ instead of to a local quadrature. Analogously, $x_2$ couples to $(p_1 - p_3)$ nonreciprocally.
	As a consequence, the dominant elements of the susceptibility matrix are coupling $(p_1 - p_3)$ to $-x_2$ and $x_4$, as well as $(p_2-p_4)$ to $-x_1$ and $x_3$. This peculiar non-local pairing between even and odd quadratures is shown in Fig.~\ref{fig:3}~(b) for a 4-ring and carries over to larger $N$-rings with even $N$, independent of system size. 
	The pairing is only possible if $N$ is even, since it requires an equal number of quadratures on even and odd sites, respectively. We rigorously derive this condition using the particle-hole graphs and a reduction technique detailed in the Supplementary Information Sec.~IV. 
	
	We implement the 4-ring shown in Fig.~\ref{fig:3}~(a) experimentally. The measured susceptibility matrix in the resonator quadrature basis, (Fig.~\ref{fig:3}~(c)) illustrates its nonreciprocal response for $\phi=0$, while the nonreciprocity vanishes completely for $\phi=\pi/4$, establishing qNR in this system.
	For maximum nonreciprocity ($\phi=0$), the largest magnitude entries of the susceptibility matrix instantiates the non-local coupling structure between quadratures of even and odd sites. This susceptibility matrix structure is transparent in the basis of collective quadratures (Fig.~\ref{fig:3}~(d)), where $\chi$ is, in fact, block-diagonal for $\phi=0$.
	We note that in this basis the nonreciprocal isolation is perfect in theory (Fig.~\ref{fig:3}f), whereas in the local basis it is not (Fig.~\ref{fig:3}e).
	
	\begin{figure}[ht!]
	    \centering
	    \includegraphics[width=\linewidth]{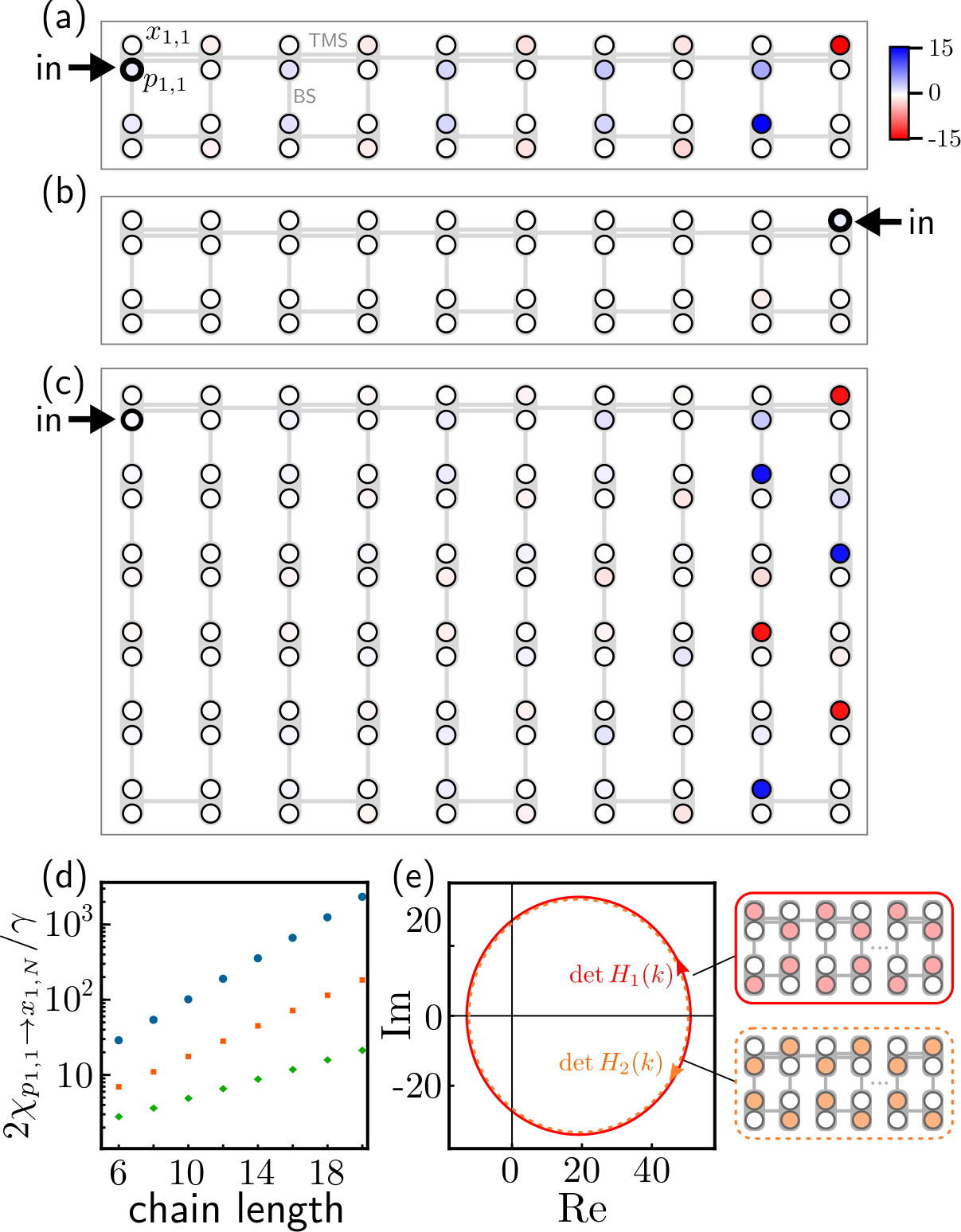}
	    \caption{\textbf{Steady-state response of a qNR chain.}
	    Staggering qNR rings with BS links to form a chain, we obtain a qNR system for which the largest steady-state response is at the opposite end.
	    We show the numerically calculated steady-state field quadratures $\langle x_{j,l}\rangle/(2\sqrt{\gamma}\langle x_{\mathrm{ss};j,l}\rangle)$, $\langle p_{j,l}\rangle/(2\sqrt{\gamma}\langle p_{\mathrm{ss};j,l}\rangle)$ (of the site in row $j$, column $\ell$) for (a)-(b)~a 4-ring qNR chain driven at the (a)~first and (b)~last site; and (c)~a 12-ring qNR chain. The pairing is clearly visible in the steady state response. (d)~End-to-end gain for $4$-ring (blue), $8$-ring (red) and $12$-ring (green) chains and $J=\lambda=4\gamma$.
	    The gain grows exponentially across the system and with system size. (e)~Suggested non-Hermitian topological invariant exploiting the Block-diagonal form of the Fourier-transformed dynamical matrix with blocks $H_{1,2}(k)$ (Methods).
	    }
	    \label{fig:4}
	\end{figure}
	
	\subsection*{Towards qNR lattices}
	
	We now take qNR one step further and use qNR rings as building blocks for constructing qNR lattices. It follows from Supplementary Information Sec.~IIC and IV that if a lattice overall preserves TRS and contains at least one qNR ring, it is itself qNR. 
	
	Of particular interest are translationally invariant chains of qNR $N$-rings, which can combine qNR transmission with non-Hermitian topology~\cite{Wanjura2020}.
	One such example is given by the bosonic Kitaev chain of Ref.~\cite{McDonald2018}, which corresponds to a 2-ring chain with a specific choice of inter-ring coupling phases and is indeed a qNR system. Thanks to our particle-hole graph framework, we can now consider more complex scenarios. As an example, we take a chain made by qNR 4-rings connected through BS interactions. In Figs.~\ref{fig:4}~(a) and (b), we show the simulated steady-state field quadratures for this qNR lattice for $\phi=0$, given a quadrature-resolved input at different sites. Nonreciprocal transmission is clearly visible, while changing the gauge to $\phi=\pi/4$ (not shown), leads to complete reciprocity, i.e., we recover the characteristic feature of qNR, Eq.~\eqref{eq:qNR_cond}. The pairing we characterised in the previous sections also manifests: an input at the $p_{1,1}$ quadrature (in which the indices stand for row and  column in the array), Fig.~\ref{fig:4}~(a), procures the largest steady-state response at two sites at the opposite end of the chain.
	The pairing is evidenced in larger qNR ring sizes, in which the steady-state response of an 8-ring chain is most prominent in every second site of the last plaquette shaping a zig-zag line (Fig.~\ref{fig:4}~(c)).
	
	In this system, qNR is accompanied by amplification with the end-to-end gain growing exponentially with the chain length, see Fig.~\ref{fig:4}~(d). This generalises the phase-dependent directional amplification predicted in a bosonic Kitaev chain~\cite{McDonald2018}. Directional amplification with exponential end-to-end gain has been identified as proxy of non-trivial topology~\cite{Wanjura2020,Porras2019}. Based on previous results from 2-ring chains, we conjecture a connection with a non-trivial non-Hermitian winding number, calculated from the determinant of the non-Hermitian matrix~\cite{Gong2018}, the dynamical matrix in our case (Methods).
	In Fig.~\ref{fig:4}~(e), we plot the determinant of each of the diagonal blocks of the Bloch-matrix (each of the blocks accounts for a set of collective quadratures, see Fig.~\ref{fig:4}~(e)) for the 4-ring qNR chain of Fig.~\ref{fig:4}~(a) under periodic boundary conditions. We find that the determinant of each of these blocks can be assigned a non-trivial winding number which equals in magnitude and differs in sign corresponding to the two directions of directional end-to-end gain. The opposite winding sense of these two matrix blocks is an expression of a TRS Hamiltonian.

	\subsection*{Outlook}

	In conclusion, we introduced the novel phenomenon of quadrature nonreciprocity (qNR).
	In contrast to standard nonreciprocity, qNR does not break time-reversal symmetry in the Hamiltonian and presents a characteristic gauge dependence. We identified the set of bosonic networks that display qNR and reported the first experimental realisations.
	
	Our results point to a close connection of qNR with the existence of QND variables and with back-action evading (BAE) measurements~\cite{Braginsky1980,Clerk2008}. Indeed, our characterisation of qNR simultaneously provides a powerful recipe to design systems with collective QND variables. In this context, exploring the noise properties of qNR networks we introduced will also be interesting. As they rely on interference between coherent interactions, without necessity of dissipation, we can expect quantum-limited performance~\cite{Wang2019}.
	This paves the way towards quantum applications, such as efficient, noiseless sensors, quantum information routers and the generation and measurement of non-classical states, including entangled states~\cite{Gardiner1993,Clerk2010,McDonald2018}.
	Indeed, while we demonstrated the concept in the domain of nanomechanics,
	it could find application in e.g.~classical electrical circuits~\cite{Ningyuan2015}, acoustics~\cite{Zhang2017}, superconducting circuits~\cite{Abdo2013,Sliwa2015,Lecocq2017}, and spin ensembles~\cite{Karg2020}.

    Going forward, we envision the construction of lattices from qNR rings which inherit the qNR properties similar to those of Fig.~\ref{fig:4}.
    Our research opens new avenues for exploring qNR lattices and networks in which new topological phenomena may emerge similar to the quantum spin Hall effect~\cite{Hasan2010,Ozawa2019}.
    

    \subsection*{Acknowledgements}
    We thank A.~A.~Clerk and D.~Malz for useful discussions.
    C.C.W.~acknowledges the funding received from the Winton Programme for the Physics of Sustainability and EPSRC (EP/R513180/1).
    J.d.P.~acknowledges financial support from the ETH Fellowship program (grant no.~20-2 FEL-66). M.B.~acknowledges funding from the Swiss National Science Foundation (PCEFP2$\_$194268).
    This work is part of the research programme of the Netherlands Organisation for Scientific Research (NWO). It is supported by the European Union's Horizon 2020 research and innovation programme under grant agreement No 732894 (FET-Proactive HOT) and the European Research Council (ERC starting grant no.~759644-TOPP).


\renewcommand{\figurename}{EXTENDED DATA FIG.}
\setcounter{figure}{0}
\renewcommand{\theHfigure}{ED.\thefigure}

\section*{Methods}

	    \subsection*{Linear response and interference of beamsplitter and squeezing interactions}
	    A general quadratic, bosonic Hamiltonian can be written in the field basis of creation and annihilation modes, namely $\{a_i,a_i^{\dagger}\}$, as 
		\begin{equation}\label{eq:Ht}
		   	\mathcal{H}=\sum_{i,j}\left\{ a_{i}^{\dagger}\mathcal{A}_{ij}a_{j}+\frac{1}{2}(a_{i}^{\dagger}\mathcal{B}_{ij}a_{j}^{\dagger}+a_{i}\mathcal{B}^*_{ij}a_{j})\right\},
		\end{equation}
		where overall constant shifts have been removed.
		Here, the matrix elements $\mathcal{A}_{ij} = J_{ij} e^{-i\varphi_{ij}}$, $\mathcal{A}_{ji} = \mathcal{A}_{ij}^{*}$ of the Hermitian \emph{hopping matrix} $\mathcal{A}$ encode beamsplitter interactions that conserve the total number of excitations.
		Similarly, we define the symmetric \emph{squeezing matrix} $\mathcal{B}$ that encodes the particle-non-conserving squeezing interactions in its elements $\mathcal{B}_{ij} = \lambda_{ij} e^{i\theta_{ij}}$, $\mathcal{B}_{ji} = \mathcal{B}_{ij}$. 
		
		
		In the quadrature basis $x_j = (a_j + a_j^\dagger)/\sqrt{2}$ and $p_j = \mathrm{i}(a_j^{\dagger} - a_j)/\sqrt{2}$, Eq.~\eqref{eq:Ht} reads
	    \begin{equation}\label{eq:Ht_XY}
		\mathcal{H}=\sum_{i,j}\left\{T_{ij}p_ip_j+V_{ij}x_ix_j+U_{ij}x_ip_j+U_{ij}^Tp_ix_j\right\},
		\end{equation}
	    where we define the effective potential matrices $U=\mathrm{Im}(\mathcal{B}-\mathcal{A}),V=\mathrm{Re}(\mathcal{A}+\mathcal{B})$ and kinetic energy $T=\mathrm{Re}(\mathcal{A}-\mathcal{B})$~\cite{Rossignoli2005}. The corresponding Heisenberg equations of motion read
	     \begin{align}\label{eq:H_evo}
	        \left(\begin{array}{c}
            \dot{x}_{j}\\
            \dot{p}_{j}
            \end{array}\right)=\sum_{k=1}^{N}\left(\begin{array}{c}
            U_{kj}x_{k}+T_{jk}p_{k}\\
            -V_{jk}x_{k}-U_{jk}p_{k}
            \end{array}\right)-\left(\begin{array}{c}
            \gamma_{j}x_{j}\\
            \gamma_{j}p_{j}
            \end{array}\right).
	    \end{align}

	    In our platform, quadratic Hamiltonians of the form~Eq.~\eqref{eq:Ht} are effectively conceived in a rotating frame of reference that oscillates at the natural frequencies of the resonators. As such, free energy terms $\propto a_i^{\dagger}a_i$ are absent, and the matrix elements for $U,V,T$ read
	    \begin{align}
	        V_{ij},T_{ij}=&J_{ij}\cos(\varphi_{ij})\pm\lambda_{ij}\cos(\theta_{ij}),\nonumber\\
	        U_{ij}=&\lambda_{ij}\sin(\theta_{ij})+J_{ij}\sin(\varphi_{ij}).
	    \end{align}
	    These expressions show that matching interaction amplitudes and complex interaction phases can lead to a cancellation of different contributions in~Eq.~\eqref{eq:Ht_XY},
	    decoupling quadratures from different resonators in the dynamics governed by~Eq.~\eqref{eq:H_evo},
	    see Ref.~\cite{McDonald2018,Flynn2020}.
	    This is the case for $\lambda_{ij}=J_{ij}$ and $\theta_{ij}=\varphi_{ij}=n\pi/2$, $n\in\mathbb{Z}$, where $T=V=0$. This can be interpreted as destructive interference in the particle-hole space, see Ref.~\cite{delPino2022} and main text. 
	    
	    From the coefficients in Eq.~\eqref{eq:H_evo}, we define the dynamical matrix $H$ for the quadrature vector $\myvec{q} = (x_1, p_1,\dots,x_N, p_N)^T$, as the matrix that fulfills $\dot{\myvec{q}} = H \myvec{q} - \myvec{f}^{(q)}$. Here $\myvec{f}^{(q)}$ is a column vector grouping driving amplitudes for each quadrature. The quadrature response to excitation is determined by the matrices $U, V$ and $T$ through the susceptibility matrix
	    \begin{equation}
	        \chi(\omega)=(\mathrm{i}\omega\mathbb{1}+H)^{-1}.
	    \end{equation}
	    On resonance (in the rotating frame, this is equivalent to $\omega=0$), this matrix will present asymmetries if and only if $H$ is a nonzero asymmetric matrix. A more exhaustive discussion is provided in Supplementary Information, Sec.~IIB.

	\subsection*{Device, set-up and experimental procedure}
\begin{figure*}[tb]
		\includegraphics[width=\linewidth]{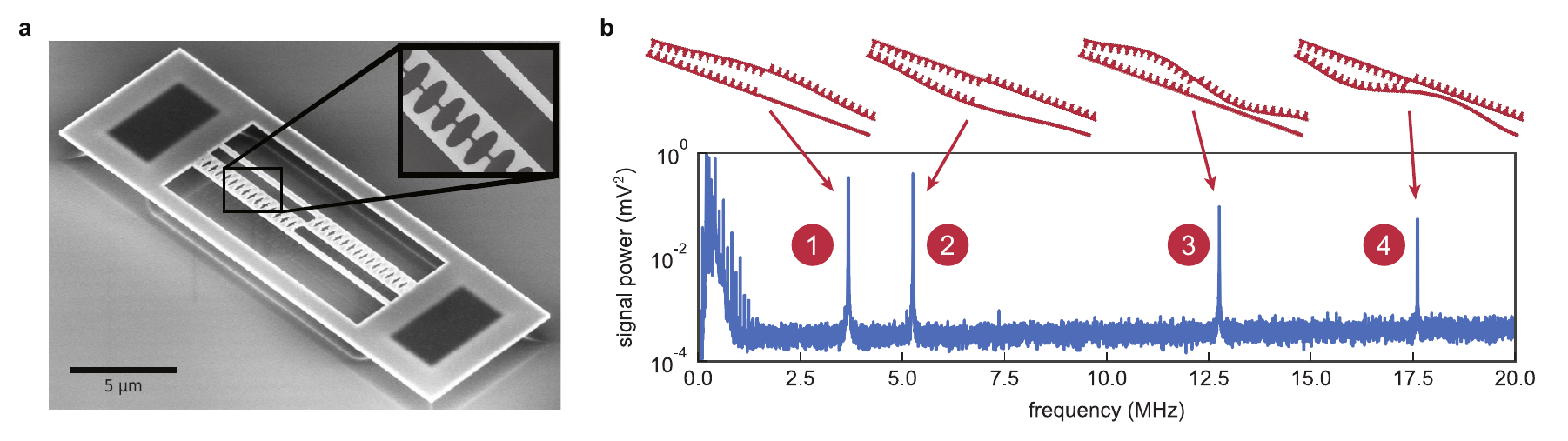}
		\caption{\textbf{Optomechanical device.} (a) Scanning electron micrograph of a device as used in our experiments. Three suspended beams are defined in the top silicon device layer (thickness 220 nm). The flexural motion of each pair of adjacent beams is coupled to an optical resonance, hosted by a point defect in the sliced photonic crystal defined by the teeth. In the presented experiments, we address only one of the cavities from free-space, at normal incidence, to drive and read out the mechanical motion of a single beam pair. The inset shows a top view, revealing the narrow slit separating the beam pair. More details can be found in \cite{delPino2022}. (b) Spectrum of thermal fluctuations imprinted on the intensity of a read-out laser reflected off the optical cavity. In this frequency range, four mechanical resonances can be identified, with the corresponding simulated displacement profiles shown above. These mechanical modes serve as the resonators in the presented experiments.}
		\label{fig:ed-device}
	\end{figure*}

	The device, a suspended sliced photonic crystal nanobeam hosting a telecom-frequency optical mode (free-space wavelength $\lambda_0 \approx 1533.5$ nm, linewidth $\kappa \approx 2\pi\times320$~GHz) coupled simultaneously to multiple non-degenerate flexural mechanical modes (frequencies $\omega_j = 2\pi\times3$--$17$~MHz, dissipation rates $\gamma_j = 2\pi\times1$--$4$~kHz, photon-phonon coupling rates $g_{0,j} = 2\pi\times3$--$6$~MHz), was used in a previous study~\cite{delPino2022}. In the current study, we employ the same set-up and experimental procedures to characterise the system and calibrate the experiments. In particular, we follow the same procedures to attain phase coherence between drives and mechanical signals. For completeness, a micrograph along with the thermomechanical spectrum of the device is shown in Extended Data Fig.~\ref{fig:ed-device}.
	
	For the experiments shown in Figs.\ref{fig:1} and \ref{fig:2}, the mechanical resonances labeled 3 and 4 (frequencies $2\pi\times12.8$ and $2\pi\times17.6$~MHz) were used, with their decay rates matched thermo-optically to $\gamma = 2\pi\times3.7$~kHz by appropriately adjusting the average drive laser intensity. For the three-mode experiments shown in Fig.~\ref{fig:1}b and Fig.~\ref{fig:2}b, the mechanical resonance labeled 2 (frequency $2\pi\times5.3$~MHz, thermo-optically tuned dissipation rate $\gamma^\prime = 2\pi\times1.3$~kHz) is utilized as the auxiliary mode. 
	
	The four-mode experiments shown in Fig.~\ref{fig:3} were performed using the mechanical resonances labelled 1 (frequency $2\pi\times3.7$~MHz) through 4 at maximum drive laser power (thermo-optically tuned dissipation rates $2\pi\times\{1.9, 0.9, 4.2,3.5\}$~kHz, respectively). In this case, simultaneous feedback cooling on all mechanical resonance was employed to increase and equalize their decay rates to $\gamma = 2\pi\times10$~kHz. In addition to homogenizing the system, the feedback cooling reduces the amplitude of thermal fluctuations, thereby ensuring that the cavity photon population responds linearly to mechanical displacement and rendering non-linear transduction~\cite{leijssen2017nonlinear} and non-linear reduction of the optical spring shift unimportant.

	\subsection*{Resonant driving and analysis}
	The central relation used to analyse the results of our experiments is the steady-state response $\myvec{q}_\text{ss}$ of a system of $N$ resonators described by Hamiltionan~\eqref{eq:Ht} to resonant forces driving its quadratures with amplitudes $\myvec{f}^{(q)} = (f^{(q)}_{x_1}, f^{(q)}_{p_1}, \dots, f^{(q)}_{x_N}, f^{(q)}_{p_N})^T$. This response is measured in units of the zero-point fluctuations $x_{\text{zpf},j}$ and given by
	\begin{align}\label{eq:qssAnalysis}
	    \myvec{q}_\text{ss} = \chi_\phi(0) \myvec{f}^{(q)}.
	\end{align}
	
	Harmonic driving forces are generated by modulation of the drive laser intensity $I(t) = I_0 (1 + c_d \cos(\omega_d t + \phi_d))$, where $I_0$ is the average drive laser intensity, $c_d$ the depth of the modulation, $\omega_d$ its frequency, matched to the mechanical resonance of interest, and $\phi_d$ its phase. Due to the short photon lifetime ($\kappa^{-1} \ll \omega_j^{-1}, \omega_d^{-1}$), the cavity population $n_c(t) \propto I(t)$ responds instantaneously to the drive laser intensity on the mechanical timescale, as do the resulting optical forces $F_j(t) = \hbar g_{0,j} n_c(t) / x_{\text{zpf},j}$ on each resonator $j$. Although all mechanical resonators are coupled to the same cavity and thus experience similar optical forces, their large frequency separation ($\lvert\omega_j - \omega_\ell\rvert \ll \gamma_j,\gamma_\ell$) ensures that only a single mode is resonantly driven, allowing to neglect the driving on the other resonators.
	
	The drive forces the evolution $\dot{\myvec{a}} = H^{(a)} \myvec{a} - \myvec{f}^{(a)}(t)$  of the mode vector $\myvec{a} = (a_1, a_1^\dagger, \dots, a_N, a_N^\dagger)^T$ (expressed in the frame rotating along with the resonators) through the driving vector $\myvec{f}^{(a)}(t)$, with dynamical matrix $H^{(a)}$ as defined in Supplementary Information Sec.~IIB. The odd-index elements 
	$f^{(a)}_{2j-1}(t) = \mathrm{i} e^{\mathrm{i}\omega_j t} F_{j}(t) / (2 m \omega_j x_{\text{zpf},j})$
	encode the force $F_{j}(t)$ driving operator $a_j$, whereas the even-index elements $f^{(q)}_{2j}(t) = \left(f^{(q)}_{2j-1}(t)\right)^*$ are related by conjugation and drive the adjoint operator $a_j^\dagger$. For a driving force $F_j(t) = h_j \cos(\omega_j t + \phi_j)$ resonant with resonator $j$, with phase $\phi_j$, the rotating factor $e^{\iu\omega_j t}$ selects the positive frequency sideband of $F_j(t)$. We neglect the non-resonant negative frequency sideband in a similar spirit to the rotating-wave approximation. This results in time-independent driving terms 
	\begin{align}
	    f^{(a)}_{2j-1} = \iu \frac{h_j e^{-\mathrm{i}\phi_j}} {4 m \omega_j x_{\text{zpf},j}}.
	\end{align}
	
	Before each experiment we perform a driven reference measurement for each resonator to quantify how the modulation depth $c_d$ is transduced into the driving term $f^{(a)}_{2j-1}$. We disable all interaction modulations, such that the dynamical matrix is diagonal and only contains damping terms ($H^{(a)}\mapsto -\Gamma/2$). We then apply a modulation of depth between $c_d = 1$--$5 \cdot 10^{-3}$ and sweep its frequency $\omega_d$ across the resonance frequency $\omega_j$ of resonator $j$. We normalize the response to the zero-point fluctuation amplitude, fit it with resonator susceptibility to obtain the driven amplitude $a_\text{max}$ on resonance ($\omega_d = \omega_j$) and determine the force transduction coefficient $K_j = \lvert f^{(a)}_{2j-1}\rvert / c_d = a_\text{max} \gamma_j / (2 c_d)$. 
	
	To reconstruct the on-resonance susceptibility matrix $\chi_\phi(0)$ for a given quadrature angle $\phi$, we first refer all (electronic) interaction and LO tones to a common time origin at which their phases are zero. Then, we shift the LO tone phases by $\phi$ to rotate the quadratures they define. We turn on the modulations that generate the desired interactions, and perform a series of $2N$ experiments where we drive the quadratures one by one, with $\arg(f^{(a)}_{2j-1}) = 0$ to drive $x_j$ and $\arg(f^{(a)}_{2j-1}) = \pi/2$ to drive $p_j$ for each resonator $j$ while the terms driving the other resonators are zero. For each of these driving conditions $\myvec{f}^{(a)}_k$, we record the steady state response amplitudes $\myvec{a}_{\text{ss},k}$ after averaging with a low-pass filter (third-order filter, $3$~dB bandwidth $2\pi\times10$~Hz). 
	
	We collect the driving vectors in the matrix $\mathcal{F}^{(a)} = \left[ \myvec{f}^{(a)}_1 \cdots \myvec{f}^{(a)}_{2N} \right]$ and the steady-state responses in $A^{(a)} = \left[ \myvec{a}_{\text{ss},1} \cdots \myvec{a}_{\text{ss},2N} \right]$. Next we transform to the quadrature basis via a matrix $W$, and obtain the real driving matrix $\mathcal{F}^{(q)} = W\mathcal{F}^{(a)} = \left[ \myvec{f}^{(q)}_{1} \cdots \myvec{f}^{(q)}_{2N} \right]$ that contains the quadrature driving amplitudes for each setting. Similarly, the real matrix $A^{(q)} = WA^{(a)} = \left[ \myvec{q}_{\text{ss},1} \cdots \myvec{q}_{\text{ss},2N} \right]$ contains the steady-state responses in the quadrature basis. Noting that the columns of $\mathcal{F}^{(q)}$ span the entire driving space (and are in fact diagonal), equation \eqref{eq:qssAnalysis} can be used to obtain the susceptibility matrix $\chi_\phi(0) = A^{(q)} (\mathcal{F}^{(q)})^{-1}$.
	
	To estimate experimental errors, the full $\phi$ sweep from $0$ to $2\pi$ was repeated $10$ times. The matrix colour plots in Figs.~\ref{fig:1} and~\ref{fig:3} show average values over all sweeps, whereas the plots of single matrix elements feature error bars that represent the statistical $\pm2 \sigma$ (i.e. a $95\%$ confidence interval) spread around the average value.

	\subsection*{Phase coherent interaction tones}
	Even though the frequencies of LOs ($\omega_{j,\ell}$) and interaction tones ($\omega_j\pm\omega_\ell$) are all distinct, a phase relation between them can nevertheless be established. To do so, we absorb the explicit time dependence of each tone $\cos(\omega_k t + \varphi_k) = \cos\beta_k(t)$ into the instantaneous phase $\beta_k(t) = \omega_k t + \varphi_k$. Phase coherence is now attained between an interaction tone $j\pm\ell$ (at $\omega_j\pm\omega_\ell$) and the \emph{combination} of both LOs $j$,$\ell$ ($\omega_{j,\ell}$), since the phase difference
	\begin{align}
	    \Delta\varphi_{j\pm\ell} &= \beta_{j\pm\ell}(t) - \big(\beta_j(t) \pm \beta_\ell(t)\big) \nonumber \\
	    &= \varphi_{j\pm\ell} - \left( \varphi_j \pm \varphi_\ell \right)
	\end{align}
	is stable (time-independent). Physically, this phase difference may be evaluated by generating a tone with the combined instantaneous phase $\beta_j(t) \pm \beta_\ell(t)$ through mixing of the LOs and subsequent high-pass (low-pass) filtering, and comparing that to the interaction tone.

\subsection*{Non-Hermitian topological invariant}
To analyse the topological properties of the chain shown in the main text Fig.~\ref{fig:4}, we write the dynamical equations for the quadratures under periodic boundary conditions, $J=\lambda$, and obtain the dynamical matrix in the plane-wave basis.
We label the sites of each unit cell $A_j$ for odd columns with $j\in\{1,2\}$ the row index, and $B_j$ for even columns.
The resulting dynamical equations decouple into two respectively closed sets of equations for the sets of quadratures $\{x_j^{A_1}, x_j^{B_2}, p_j^{B_1}, p_j^{A_2}\}$, namely,
\begin{align}
   \dot x_j^{A_1} =& -\frac{\gamma}{2} x_j^{A_1} + J p_j^{A_2} - J p_j^{B_1} + J p_{j-1}^{B_1} - \sqrt{\gamma} x_{j,\mathrm{in}}^{A_1},\notag \\[2ex]
   \dot x_j^{B_2} =& -\frac{\gamma}{2} x_j^{B_2} + J p_j^{A_2} - J p_j^{B_1} - \sqrt{\gamma} x_{j,\mathrm{in}}^{B_2},\notag \\[2ex]
   \dot p_j^{B_1} =& -\frac{\gamma}{2} p_j^{B_1} - J x_j^{A_1} - J x_j^{B_2} - J x_{j+1}^{A_1} - \sqrt{\gamma} p_{j,\mathrm{in}}^{B_1},\notag \\[2ex]
   \dot p_j^{A_2} =& -\frac{\gamma}{2} p_j^{A_2} - J x_j^{B_2} - J x_j^{A_1} - \sqrt{\gamma} p_{j,\mathrm{in}}^{A_2},
   \intertext{and $\{x_j^{A_2}, x_j^{B_1}, x_j^{A_1}, x_j^{B_2}\}$, reading}
   \dot x_j^{A_2} =& -\frac{\gamma}{2} x_j^{A_2} + J p_j^{A_1} + J p_j^{B_2} - \sqrt{\gamma} x_{j,\mathrm{in}}^{A_2},\notag \\[2ex]
   \dot x_j^{B_1} =& -\frac{\gamma}{2} x_j^{B_1} - J p_j^{A_1} + J p_j^{B_2} + J p_{j+1}^{A_1} - \sqrt{\gamma} x_{j,\mathrm{in}}^{B_1},\notag \\[2ex]
   \dot p_j^{A_1} =& -\frac{\gamma}{2} p_j^{A_1} - J x_j^{A_2} - J x_j^{B_1} - J x_{j-1}^{B_1} - \sqrt{\gamma} p_{j,\mathrm{in}}^{A_1},\notag \\[2ex]
   \dot p_j^{B_2} =& -\frac{\gamma}{2} p_j^{B_2} - J x_j^{A_2} - J x_j^{B_1} - \sqrt{\gamma} p_{j,\mathrm{in}}^{B_2}.
\end{align}
The fact that these two sets decouple is a consequence of the pairing we found for qNR rings.
We now switch to reciprocal space, $\ket{j}\equiv \frac{1}{\sqrt{N}} \sum_k e^{-\mathrm{i} ka} \ket{k}$, in which we used Dirac notation to denote the basis vectors $\{\ket{j}\}$ and $\{\ket{k}\}$ and $a$ mimics a lattice spacing (we set $a=1$). We obtain two dynamical matrices for each of these sets
\begin{align}
   &
   H_1(k)  = \begin{pmatrix}
      -\frac{\gamma}{2} & 0 & J (-1 + e^{-\mathrm{i} k}) & J \\
      0 & -\frac{\gamma}{2} & J & J \\
      -J (1+e^{\mathrm{i}k}) & -J & -\frac{\gamma}{2} & 0 \\
      -J & -J & 0 & -\frac{\gamma}{2}
   \end{pmatrix}, \notag \\
   &
   H_2(k)  = \begin{pmatrix}
      -\frac{\gamma}{2} & 0 & J & J \\
      0 & -\frac{\gamma}{2} & J (-1 + e^{\mathrm{i} k}) & J \\
      -J & -J (1+e^{-\mathrm{i}k}) & -\frac{\gamma}{2} & 0 \\
      -J & -J & 0 & -\frac{\gamma}{2}
   \end{pmatrix}.
\end{align}
We calculate a topological invariant from each matrix, namely the winding number of their determinant
\begin{align}
   \nu_{1,2} & = \frac{1}{2\pi\mathrm{i}}\int_0^{2\pi} \mathrm{d} k\, \det H_{1,2}(k).
\end{align}
We find
\begin{align}
   \det H_{1,2} (k) = & -2 J^4 \cos (k)+\frac{1}{16} \left(\gamma ^4+16 J^4+12 \gamma ^2 J^2\right) \notag \\
   & \pm i \left(2 J^4+\frac{\gamma ^2 J^2}{2}\right) \sin (k),
\end{align}
in which we choose the $(-)$ sign for $\det H_1(k)$ and the $(+)$ sign for $\det H_2(k)$.
These two curves wind in opposite directions in the complex plane, as $k$ evolves from $0$ to $2\pi$, inducing the opposite sign for $\nu_1$ and $\nu_2$ (see Fig.~\ref{fig:4}~(e) in the main text).

Above, we used the plane-wave basis to identify the non-trivial topology of each block $H_1(k)$, $H_2(k)$.
The eigenvalues of $H_1(k)$ are also the eigenvalues of $H_2(k)$, albeit for a different $k$. Together they form a degenerate sub-space. To obtain the physical (and real) eigenvectors, i.e. pairs of canonically conjugated, real valued quadratures, we need to superpose eigenvectors from these sub-spaces.

\clearpage
\pagebreak
\bibliography{refs_extracted_v2}

\begin{thebibliography}{54}%
\makeatletter
\providecommand \@ifxundefined [1]{%
 \@ifx{#1\undefined}
}%
\providecommand \@ifnum [1]{%
 \ifnum #1\expandafter \@firstoftwo
 \else \expandafter \@secondoftwo
 \fi
}%
\providecommand \@ifx [1]{%
 \ifx #1\expandafter \@firstoftwo
 \else \expandafter \@secondoftwo
 \fi
}%
\providecommand \natexlab [1]{#1}%
\providecommand \enquote  [1]{``#1''}%
\providecommand \bibnamefont  [1]{#1}%
\providecommand \bibfnamefont [1]{#1}%
\providecommand \citenamefont [1]{#1}%
\providecommand \href@noop [0]{\@secondoftwo}%
\providecommand \href [0]{\begingroup \@sanitize@url \@href}%
\providecommand \@href[1]{\@@startlink{#1}\@@href}%
\providecommand \@@href[1]{\endgroup#1\@@endlink}%
\providecommand \@sanitize@url [0]{\catcode `\\12\catcode `\$12\catcode
  `\&12\catcode `\#12\catcode `\^12\catcode `\_12\catcode `\%12\relax}%
\providecommand \@@startlink[1]{}%
\providecommand \@@endlink[0]{}%
\providecommand \url  [0]{\begingroup\@sanitize@url \@url }%
\providecommand \@url [1]{\endgroup\@href {#1}{\urlprefix }}%
\providecommand \urlprefix  [0]{URL }%
\providecommand \Eprint [0]{\href }%
\providecommand \doibase [0]{https://doi.org/}%
\providecommand \selectlanguage [0]{\@gobble}%
\providecommand \bibinfo  [0]{\@secondoftwo}%
\providecommand \bibfield  [0]{\@secondoftwo}%
\providecommand \translation [1]{[#1]}%
\providecommand \BibitemOpen [0]{}%
\providecommand \bibitemStop [0]{}%
\providecommand \bibitemNoStop [0]{.\EOS\space}%
\providecommand \EOS [0]{\spacefactor3000\relax}%
\providecommand \BibitemShut  [1]{\csname bibitem#1\endcsname}%
\let\auto@bib@innerbib\@empty
\bibitem [{\citenamefont {De\'ak}\ and\ \citenamefont
  {F{\"u}l{\"o}p}(2012)}]{Deak2012}%
  \BibitemOpen
  \bibfield  {author} {\bibinfo {author} {\bibfnamefont {L.}~\bibnamefont
  {De\'ak}}\ and\ \bibinfo {author} {\bibfnamefont {T.}~\bibnamefont
  {F{\"u}l{\"o}p}},\ }\bibfield  {title} {\bibinfo {title} {Reciprocity in
  quantum, electromagnetic and other wave scattering},\ }\href
  {https://doi.org/10.1016/j.aop.2011.10.013} {\bibfield  {journal} {\bibinfo
  {journal} {Annals of Physics}\ }\textbf {\bibinfo {volume} {327}},\ \bibinfo
  {pages} {1050} (\bibinfo {year} {2012})}\BibitemShut {NoStop}%
\bibitem [{\citenamefont {Jalas}\ \emph {et~al.}(2013)\citenamefont {Jalas},
  \citenamefont {Petrov}, \citenamefont {Eich}, \citenamefont {Freude},
  \citenamefont {Fan}, \citenamefont {Yu}, \citenamefont {Baets}, \citenamefont
  {Popovic}, \citenamefont {Melloni}, \citenamefont {Joannopoulos},
  \citenamefont {Vanwolleghem}, \citenamefont {Doerr},\ and\ \citenamefont
  {Renner}}]{Jalas2013}%
  \BibitemOpen
  \bibfield  {author} {\bibinfo {author} {\bibfnamefont {D.}~\bibnamefont
  {Jalas}}, \bibinfo {author} {\bibfnamefont {A.}~\bibnamefont {Petrov}},
  \bibinfo {author} {\bibfnamefont {M.}~\bibnamefont {Eich}}, \bibinfo {author}
  {\bibfnamefont {W.}~\bibnamefont {Freude}}, \bibinfo {author} {\bibfnamefont
  {S.}~\bibnamefont {Fan}}, \bibinfo {author} {\bibfnamefont {Z.}~\bibnamefont
  {Yu}}, \bibinfo {author} {\bibfnamefont {R.}~\bibnamefont {Baets}}, \bibinfo
  {author} {\bibfnamefont {M.}~\bibnamefont {Popovic}}, \bibinfo {author}
  {\bibfnamefont {A.}~\bibnamefont {Melloni}}, \bibinfo {author} {\bibfnamefont
  {J.}~\bibnamefont {Joannopoulos}}, \bibinfo {author} {\bibfnamefont
  {M.}~\bibnamefont {Vanwolleghem}}, \bibinfo {author} {\bibfnamefont
  {C.}~\bibnamefont {Doerr}},\ and\ \bibinfo {author} {\bibfnamefont
  {H.}~\bibnamefont {Renner}},\ }\bibfield  {title} {\bibinfo {title} {What is
  -- and what is not -- an optical isolator},\ }\href
  {https://doi.org/10.1038/nphoton.2013.185} {\bibfield  {journal} {\bibinfo
  {journal} {Nature Photonics}\ }\textbf {\bibinfo {volume} {7}},\ \bibinfo
  {pages} {579–582} (\bibinfo {year} {2013})}\BibitemShut {NoStop}%
\bibitem [{\citenamefont {Caloz}\ \emph {et~al.}(2018)\citenamefont {Caloz},
  \citenamefont {Al\`u}, \citenamefont {Tretyakov}, \citenamefont {Sounas},
  \citenamefont {Achouri},\ and\ \citenamefont {Deck-L\'eger}}]{Caloz2018}%
  \BibitemOpen
  \bibfield  {author} {\bibinfo {author} {\bibfnamefont {C.}~\bibnamefont
  {Caloz}}, \bibinfo {author} {\bibfnamefont {A.}~\bibnamefont {Al\`u}},
  \bibinfo {author} {\bibfnamefont {S.}~\bibnamefont {Tretyakov}}, \bibinfo
  {author} {\bibfnamefont {D.}~\bibnamefont {Sounas}}, \bibinfo {author}
  {\bibfnamefont {K.}~\bibnamefont {Achouri}},\ and\ \bibinfo {author}
  {\bibfnamefont {Z.}~\bibnamefont {Deck-L\'eger}},\ }\bibfield  {title}
  {\bibinfo {title} {Electromagnetic nonreciprocity},\ }\href
  {https://doi.org/10.1103/PhysRevApplied.10.047001} {\bibfield  {journal}
  {\bibinfo  {journal} {Phys. Rev. Applied}\ }\textbf {\bibinfo {volume}
  {10}},\ \bibinfo {pages} {047001} (\bibinfo {year} {2018})}\BibitemShut
  {NoStop}%
\bibitem [{\citenamefont {Verhagen}\ and\ \citenamefont
  {Al{\`u}}(2017)}]{Verhagen2017}%
  \BibitemOpen
  \bibfield  {author} {\bibinfo {author} {\bibfnamefont {E.}~\bibnamefont
  {Verhagen}}\ and\ \bibinfo {author} {\bibfnamefont {A.}~\bibnamefont
  {Al{\`u}}},\ }\bibfield  {title} {\bibinfo {title} {Optomechanical
  nonreciprocity},\ }\href {https://doi.org/10.1038/nphys4283} {\bibfield
  {journal} {\bibinfo  {journal} {Nature Physics}\ }\textbf {\bibinfo {volume}
  {13}},\ \bibinfo {pages} {922} (\bibinfo {year} {2017})}\BibitemShut
  {NoStop}%
\bibitem [{\citenamefont {Lau}\ and\ \citenamefont {Clerk}(2018)}]{Lau2018}%
  \BibitemOpen
  \bibfield  {author} {\bibinfo {author} {\bibfnamefont {H.-K.}\ \bibnamefont
  {Lau}}\ and\ \bibinfo {author} {\bibfnamefont {A.~A.}\ \bibnamefont
  {Clerk}},\ }\bibfield  {title} {\bibinfo {title} {Fundamental limits and
  non-reciprocal approaches in non-{H}ermitian quantum sensing},\ }\href
  {https://doi.org/10.1038/s41467-018-06477-7} {\bibfield  {journal} {\bibinfo
  {journal} {Nature Communications}\ }\textbf {\bibinfo {volume} {9}},\
  \bibinfo {pages} {4320} (\bibinfo {year} {2018})}\BibitemShut {NoStop}%
\bibitem [{\citenamefont {Ranzani}\ and\ \citenamefont
  {Aumentado}(2015)}]{Ranzani2015}%
  \BibitemOpen
  \bibfield  {author} {\bibinfo {author} {\bibfnamefont {L.}~\bibnamefont
  {Ranzani}}\ and\ \bibinfo {author} {\bibfnamefont {J.}~\bibnamefont
  {Aumentado}},\ }\bibfield  {title} {\bibinfo {title} {Graph-based analysis of
  nonreciprocity in coupled-mode systems},\ }\href
  {https://doi.org/10.1088/1367-2630/17/2/023024} {\bibfield  {journal}
  {\bibinfo  {journal} {New Journal of Physics}\ }\textbf {\bibinfo {volume}
  {17}},\ \bibinfo {pages} {023024} (\bibinfo {year} {2015})}\BibitemShut
  {NoStop}%
\bibitem [{\citenamefont {Ozawa}\ \emph {et~al.}(2019)\citenamefont {Ozawa},
  \citenamefont {Price}, \citenamefont {Amo}, \citenamefont {Goldman},
  \citenamefont {Hafezi}, \citenamefont {Lu}, \citenamefont {Rechtsman},
  \citenamefont {Schuster}, \citenamefont {Simon}, \citenamefont {Zilberberg},\
  and\ \citenamefont {Carusotto}}]{Ozawa2019}%
  \BibitemOpen
  \bibfield  {author} {\bibinfo {author} {\bibfnamefont {T.}~\bibnamefont
  {Ozawa}}, \bibinfo {author} {\bibfnamefont {H.~M.}\ \bibnamefont {Price}},
  \bibinfo {author} {\bibfnamefont {A.}~\bibnamefont {Amo}}, \bibinfo {author}
  {\bibfnamefont {N.}~\bibnamefont {Goldman}}, \bibinfo {author} {\bibfnamefont
  {M.}~\bibnamefont {Hafezi}}, \bibinfo {author} {\bibfnamefont
  {L.}~\bibnamefont {Lu}}, \bibinfo {author} {\bibfnamefont {M.~C.}\
  \bibnamefont {Rechtsman}}, \bibinfo {author} {\bibfnamefont {D.}~\bibnamefont
  {Schuster}}, \bibinfo {author} {\bibfnamefont {J.}~\bibnamefont {Simon}},
  \bibinfo {author} {\bibfnamefont {O.}~\bibnamefont {Zilberberg}},\ and\
  \bibinfo {author} {\bibfnamefont {I.}~\bibnamefont {Carusotto}},\ }\bibfield
  {title} {\bibinfo {title} {Topological photonics},\ }\href
  {https://doi.org/10.1103/RevModPhys.91.015006} {\bibfield  {journal}
  {\bibinfo  {journal} {Rev. Mod. Phys.}\ }\textbf {\bibinfo {volume} {91}},\
  \bibinfo {pages} {015006} (\bibinfo {year} {2019})}\BibitemShut {NoStop}%
\bibitem [{\citenamefont {Abdo}\ \emph {et~al.}(2011)\citenamefont {Abdo},
  \citenamefont {Schackert}, \citenamefont {Hatridge}, \citenamefont
  {Rigetti},\ and\ \citenamefont {Devoret}}]{Abdo2011}%
  \BibitemOpen
  \bibfield  {author} {\bibinfo {author} {\bibfnamefont {B.}~\bibnamefont
  {Abdo}}, \bibinfo {author} {\bibfnamefont {F.}~\bibnamefont {Schackert}},
  \bibinfo {author} {\bibfnamefont {M.}~\bibnamefont {Hatridge}}, \bibinfo
  {author} {\bibfnamefont {C.}~\bibnamefont {Rigetti}},\ and\ \bibinfo {author}
  {\bibfnamefont {M.}~\bibnamefont {Devoret}},\ }\bibfield  {title} {\bibinfo
  {title} {{J}osephson amplifier for qubit readout},\ }\href
  {https://doi.org/10.1063/1.3653473} {\bibfield  {journal} {\bibinfo
  {journal} {Applied Physics Letters}\ }\textbf {\bibinfo {volume} {99}},\
  \bibinfo {pages} {162506} (\bibinfo {year} {2011})}\BibitemShut {NoStop}%
\bibitem [{\citenamefont {Yu}\ and\ \citenamefont {Fan}(2009)}]{Yu2009}%
  \BibitemOpen
  \bibfield  {author} {\bibinfo {author} {\bibfnamefont {Z.}~\bibnamefont
  {Yu}}\ and\ \bibinfo {author} {\bibfnamefont {S.}~\bibnamefont {Fan}},\
  }\bibfield  {title} {\bibinfo {title} {Complete optical isolation created by
  indirect interband photonic transitions},\ }\href
  {https://doi.org/10.1038/nphoton.2008.273} {\bibfield  {journal} {\bibinfo
  {journal} {Nature Photonics}\ }\textbf {\bibinfo {volume} {3}},\ \bibinfo
  {pages} {91–94} (\bibinfo {year} {2009})}\BibitemShut {NoStop}%
\bibitem [{\citenamefont {Lira}\ \emph {et~al.}(2012)\citenamefont {Lira},
  \citenamefont {Yu}, \citenamefont {Fan},\ and\ \citenamefont
  {Lipson}}]{Lira2012}%
  \BibitemOpen
  \bibfield  {author} {\bibinfo {author} {\bibfnamefont {H.}~\bibnamefont
  {Lira}}, \bibinfo {author} {\bibfnamefont {Z.}~\bibnamefont {Yu}}, \bibinfo
  {author} {\bibfnamefont {S.}~\bibnamefont {Fan}},\ and\ \bibinfo {author}
  {\bibfnamefont {M.}~\bibnamefont {Lipson}},\ }\bibfield  {title} {\bibinfo
  {title} {Electrically driven nonreciprocity induced by interband photonic
  transition on a silicon chip},\ }\href
  {https://doi.org/10.1103/PhysRevLett.109.033901} {\bibfield  {journal}
  {\bibinfo  {journal} {Phys. Rev. Lett.}\ }\textbf {\bibinfo {volume} {109}},\
  \bibinfo {pages} {033901} (\bibinfo {year} {2012})}\BibitemShut {NoStop}%
\bibitem [{\citenamefont {Kamal}\ \emph {et~al.}(2011)\citenamefont {Kamal},
  \citenamefont {Clarke},\ and\ \citenamefont {Devoret}}]{Kamal2011}%
  \BibitemOpen
  \bibfield  {author} {\bibinfo {author} {\bibfnamefont {A.}~\bibnamefont
  {Kamal}}, \bibinfo {author} {\bibfnamefont {J.}~\bibnamefont {Clarke}},\ and\
  \bibinfo {author} {\bibfnamefont {M.}~\bibnamefont {Devoret}},\ }\bibfield
  {title} {\bibinfo {title} {Noiseless non-reciprocity in a parametric active
  device},\ }\href {https://doi.org/10.1038/nphys1893} {\bibfield  {journal}
  {\bibinfo  {journal} {Nature Physics}\ }\textbf {\bibinfo {volume} {7}},\
  \bibinfo {pages} {311–315} (\bibinfo {year} {2011})}\BibitemShut {NoStop}%
\bibitem [{\citenamefont {Malz}\ \emph {et~al.}(2018)\citenamefont {Malz},
  \citenamefont {T\'oth}, \citenamefont {Bernier}, \citenamefont {Feofanov},
  \citenamefont {Kippenberg},\ and\ \citenamefont {Nunnenkamp}}]{Malz2018}%
  \BibitemOpen
  \bibfield  {author} {\bibinfo {author} {\bibfnamefont {D.}~\bibnamefont
  {Malz}}, \bibinfo {author} {\bibfnamefont {L.~D.}\ \bibnamefont {T\'oth}},
  \bibinfo {author} {\bibfnamefont {N.~R.}\ \bibnamefont {Bernier}}, \bibinfo
  {author} {\bibfnamefont {A.~K.}\ \bibnamefont {Feofanov}}, \bibinfo {author}
  {\bibfnamefont {T.~J.}\ \bibnamefont {Kippenberg}},\ and\ \bibinfo {author}
  {\bibfnamefont {A.}~\bibnamefont {Nunnenkamp}},\ }\bibfield  {title}
  {\bibinfo {title} {Quantum-limited directional amplifiers with
  optomechanics},\ }\href {https://doi.org/10.1103/PhysRevLett.120.023601}
  {\bibfield  {journal} {\bibinfo  {journal} {Phys. Rev. Lett.}\ }\textbf
  {\bibinfo {volume} {120}},\ \bibinfo {pages} {023601} (\bibinfo {year}
  {2018})}\BibitemShut {NoStop}%
\bibitem [{\citenamefont {Metelmann}\ and\ \citenamefont
  {Clerk}(2015)}]{Metelmann2015}%
  \BibitemOpen
  \bibfield  {author} {\bibinfo {author} {\bibfnamefont {A.}~\bibnamefont
  {Metelmann}}\ and\ \bibinfo {author} {\bibfnamefont {A.~A.}\ \bibnamefont
  {Clerk}},\ }\bibfield  {title} {\bibinfo {title} {Nonreciprocal photon
  transmission and amplification via reservoir engineering},\ }\href
  {https://doi.org/10.1103/PhysRevX.5.021025} {\bibfield  {journal} {\bibinfo
  {journal} {Phys. Rev. X}\ }\textbf {\bibinfo {volume} {5}},\ \bibinfo {pages}
  {021025} (\bibinfo {year} {2015})}\BibitemShut {NoStop}%
\bibitem [{\citenamefont {Abdo}\ \emph {et~al.}(2013)\citenamefont {Abdo},
  \citenamefont {Kamal},\ and\ \citenamefont {Devoret}}]{Abdo2013}%
  \BibitemOpen
  \bibfield  {author} {\bibinfo {author} {\bibfnamefont {B.}~\bibnamefont
  {Abdo}}, \bibinfo {author} {\bibfnamefont {A.}~\bibnamefont {Kamal}},\ and\
  \bibinfo {author} {\bibfnamefont {M.}~\bibnamefont {Devoret}},\ }\bibfield
  {title} {\bibinfo {title} {Nondegenerate three-wave mixing with the
  {J}osephson ring modulator},\ }\href
  {https://doi.org/10.1103/PhysRevB.87.014508} {\bibfield  {journal} {\bibinfo
  {journal} {Phys. Rev. B}\ }\textbf {\bibinfo {volume} {87}},\ \bibinfo
  {pages} {014508} (\bibinfo {year} {2013})}\BibitemShut {NoStop}%
\bibitem [{\citenamefont {Sliwa}\ \emph {et~al.}(2015)\citenamefont {Sliwa},
  \citenamefont {Hatridge}, \citenamefont {Narla}, \citenamefont {Shankar},
  \citenamefont {Frunzio}, \citenamefont {Schoelkopf},\ and\ \citenamefont
  {Devoret}}]{Sliwa2015}%
  \BibitemOpen
  \bibfield  {author} {\bibinfo {author} {\bibfnamefont {K.~M.}\ \bibnamefont
  {Sliwa}}, \bibinfo {author} {\bibfnamefont {M.}~\bibnamefont {Hatridge}},
  \bibinfo {author} {\bibfnamefont {A.}~\bibnamefont {Narla}}, \bibinfo
  {author} {\bibfnamefont {S.}~\bibnamefont {Shankar}}, \bibinfo {author}
  {\bibfnamefont {L.}~\bibnamefont {Frunzio}}, \bibinfo {author} {\bibfnamefont
  {R.~J.}\ \bibnamefont {Schoelkopf}},\ and\ \bibinfo {author} {\bibfnamefont
  {M.~H.}\ \bibnamefont {Devoret}},\ }\bibfield  {title} {\bibinfo {title}
  {Reconfigurable josephson circulator/directional amplifier},\ }\href
  {https://doi.org/10.1103/PhysRevX.5.041020} {\bibfield  {journal} {\bibinfo
  {journal} {Phys. Rev. X}\ }\textbf {\bibinfo {volume} {5}},\ \bibinfo {pages}
  {041020} (\bibinfo {year} {2015})}\BibitemShut {NoStop}%
\bibitem [{\citenamefont {Lecocq}\ \emph {et~al.}(2017)\citenamefont {Lecocq},
  \citenamefont {Ranzani}, \citenamefont {Peterson}, \citenamefont {Cicak},
  \citenamefont {Simmonds}, \citenamefont {Teufel},\ and\ \citenamefont
  {Aumentado}}]{Lecocq2017}%
  \BibitemOpen
  \bibfield  {author} {\bibinfo {author} {\bibfnamefont {F.}~\bibnamefont
  {Lecocq}}, \bibinfo {author} {\bibfnamefont {L.}~\bibnamefont {Ranzani}},
  \bibinfo {author} {\bibfnamefont {G.~A.}\ \bibnamefont {Peterson}}, \bibinfo
  {author} {\bibfnamefont {K.}~\bibnamefont {Cicak}}, \bibinfo {author}
  {\bibfnamefont {R.~W.}\ \bibnamefont {Simmonds}}, \bibinfo {author}
  {\bibfnamefont {J.~D.}\ \bibnamefont {Teufel}},\ and\ \bibinfo {author}
  {\bibfnamefont {J.}~\bibnamefont {Aumentado}},\ }\bibfield  {title} {\bibinfo
  {title} {Nonreciprocal microwave signal processing with a field-programmable
  josephson amplifier},\ }\href
  {https://doi.org/10.1103/PhysRevApplied.7.024028} {\bibfield  {journal}
  {\bibinfo  {journal} {Phys. Rev. Applied}\ }\textbf {\bibinfo {volume} {7}},\
  \bibinfo {pages} {024028} (\bibinfo {year} {2017})}\BibitemShut {NoStop}%
\bibitem [{\citenamefont {Kim}\ \emph {et~al.}(2015)\citenamefont {Kim},
  \citenamefont {Kuzyk}, \citenamefont {Han}, \citenamefont {Wang},\ and\
  \citenamefont {Bahl}}]{Kim2014}%
  \BibitemOpen
  \bibfield  {author} {\bibinfo {author} {\bibfnamefont {J.}~\bibnamefont
  {Kim}}, \bibinfo {author} {\bibfnamefont {M.~C.}\ \bibnamefont {Kuzyk}},
  \bibinfo {author} {\bibfnamefont {K.}~\bibnamefont {Han}}, \bibinfo {author}
  {\bibfnamefont {H.}~\bibnamefont {Wang}},\ and\ \bibinfo {author}
  {\bibfnamefont {G.}~\bibnamefont {Bahl}},\ }\bibfield  {title} {\bibinfo
  {title} {Non-reciprocal {B}rillouin scattering induced transparency},\ }\href
  {https://doi.org/10.1038/nphys3236} {\bibfield  {journal} {\bibinfo
  {journal} {Nature Physics}\ }\textbf {\bibinfo {volume} {11}},\ \bibinfo
  {pages} {275} (\bibinfo {year} {2015})}\BibitemShut {NoStop}%
\bibitem [{\citenamefont {Shen}\ \emph {et~al.}(2016)\citenamefont {Shen},
  \citenamefont {Zhang}, \citenamefont {Chen}, \citenamefont {Zou},
  \citenamefont {Xiao}, \citenamefont {Zou}, \citenamefont {Sun}, \citenamefont
  {Guo},\ and\ \citenamefont {Dong}}]{Shen2016}%
  \BibitemOpen
  \bibfield  {author} {\bibinfo {author} {\bibfnamefont {Z.}~\bibnamefont
  {Shen}}, \bibinfo {author} {\bibfnamefont {Y.-L.}\ \bibnamefont {Zhang}},
  \bibinfo {author} {\bibfnamefont {Y.}~\bibnamefont {Chen}}, \bibinfo {author}
  {\bibfnamefont {C.-L.}\ \bibnamefont {Zou}}, \bibinfo {author} {\bibfnamefont
  {Y.-F.}\ \bibnamefont {Xiao}}, \bibinfo {author} {\bibfnamefont {X.-B.}\
  \bibnamefont {Zou}}, \bibinfo {author} {\bibfnamefont {F.-W.}\ \bibnamefont
  {Sun}}, \bibinfo {author} {\bibfnamefont {G.-C.}\ \bibnamefont {Guo}},\ and\
  \bibinfo {author} {\bibfnamefont {C.-H.}\ \bibnamefont {Dong}},\ }\bibfield
  {title} {\bibinfo {title} {Experimental realization of optomechanically
  induced non-reciprocity},\ }\href {https://doi.org/10.1038/nphoton.2016.161}
  {\bibfield  {journal} {\bibinfo  {journal} {Nat. Photonics}\ }\textbf
  {\bibinfo {volume} {10}},\ \bibinfo {pages} {657} (\bibinfo {year}
  {2016})}\BibitemShut {NoStop}%
\bibitem [{\citenamefont {Ruesink}\ \emph {et~al.}(2016)\citenamefont
  {Ruesink}, \citenamefont {Miri}, \citenamefont {Al{\`u}},\ and\ \citenamefont
  {Verhagen}}]{Ruesink2016}%
  \BibitemOpen
  \bibfield  {author} {\bibinfo {author} {\bibfnamefont {F.}~\bibnamefont
  {Ruesink}}, \bibinfo {author} {\bibfnamefont {M.-A.}\ \bibnamefont {Miri}},
  \bibinfo {author} {\bibfnamefont {A.}~\bibnamefont {Al{\`u}}},\ and\ \bibinfo
  {author} {\bibfnamefont {E.}~\bibnamefont {Verhagen}},\ }\bibfield  {title}
  {\bibinfo {title} {Nonreciprocity and magnetic-free isolation based on
  optomechanical interactions},\ }\href {https://doi.org/10.1038/ncomms13662}
  {\bibfield  {journal} {\bibinfo  {journal} {Nature Communications}\ }\textbf
  {\bibinfo {volume} {7}},\ \bibinfo {pages} {13662} (\bibinfo {year}
  {2016})}\BibitemShut {NoStop}%
\bibitem [{\citenamefont {Fang}\ \emph {et~al.}(2017)\citenamefont {Fang},
  \citenamefont {Luo}, \citenamefont {Metelmann}, \citenamefont {Matheny},
  \citenamefont {Marquardt}, \citenamefont {Clerk},\ and\ \citenamefont
  {Painter}}]{Fang2017}%
  \BibitemOpen
  \bibfield  {author} {\bibinfo {author} {\bibfnamefont {K.}~\bibnamefont
  {Fang}}, \bibinfo {author} {\bibfnamefont {J.}~\bibnamefont {Luo}}, \bibinfo
  {author} {\bibfnamefont {A.}~\bibnamefont {Metelmann}}, \bibinfo {author}
  {\bibfnamefont {M.~H.}\ \bibnamefont {Matheny}}, \bibinfo {author}
  {\bibfnamefont {F.}~\bibnamefont {Marquardt}}, \bibinfo {author}
  {\bibfnamefont {A.~A.}\ \bibnamefont {Clerk}},\ and\ \bibinfo {author}
  {\bibfnamefont {O.}~\bibnamefont {Painter}},\ }\bibfield  {title} {\bibinfo
  {title} {Generalized non-reciprocity in an optomechanical circuit via
  synthetic magnetism and reservoir engineering},\ }\href
  {https://doi.org/10.1038/nphys4009} {\bibfield  {journal} {\bibinfo
  {journal} {Nature Physics}\ }\textbf {\bibinfo {volume} {13}},\ \bibinfo
  {pages} {465 EP } (\bibinfo {year} {2017})}\BibitemShut {NoStop}%
\bibitem [{\citenamefont {Peterson}\ \emph {et~al.}(2017)\citenamefont
  {Peterson}, \citenamefont {Lecocq}, \citenamefont {Cicak}, \citenamefont
  {Simmonds}, \citenamefont {Aumentado},\ and\ \citenamefont
  {Teufel}}]{Peterson2017}%
  \BibitemOpen
  \bibfield  {author} {\bibinfo {author} {\bibfnamefont {G.~A.}\ \bibnamefont
  {Peterson}}, \bibinfo {author} {\bibfnamefont {F.}~\bibnamefont {Lecocq}},
  \bibinfo {author} {\bibfnamefont {K.}~\bibnamefont {Cicak}}, \bibinfo
  {author} {\bibfnamefont {R.~W.}\ \bibnamefont {Simmonds}}, \bibinfo {author}
  {\bibfnamefont {J.}~\bibnamefont {Aumentado}},\ and\ \bibinfo {author}
  {\bibfnamefont {J.~D.}\ \bibnamefont {Teufel}},\ }\bibfield  {title}
  {\bibinfo {title} {Demonstration of efficient nonreciprocity in a microwave
  optomechanical circuit},\ }\href {https://doi.org/10.1103/PhysRevX.7.031001}
  {\bibfield  {journal} {\bibinfo  {journal} {Phys. Rev. X}\ }\textbf {\bibinfo
  {volume} {7}},\ \bibinfo {pages} {031001} (\bibinfo {year}
  {2017})}\BibitemShut {NoStop}%
\bibitem [{\citenamefont {Barzanjeh}\ \emph {et~al.}(2017)\citenamefont
  {Barzanjeh}, \citenamefont {Wulf}, \citenamefont {Peruzzo}, \citenamefont
  {Kalaee}, \citenamefont {Dieterle}, \citenamefont {Painter},\ and\
  \citenamefont {Fink}}]{Barzanjeh2017}%
  \BibitemOpen
  \bibfield  {author} {\bibinfo {author} {\bibfnamefont {S.}~\bibnamefont
  {Barzanjeh}}, \bibinfo {author} {\bibfnamefont {M.}~\bibnamefont {Wulf}},
  \bibinfo {author} {\bibfnamefont {M.}~\bibnamefont {Peruzzo}}, \bibinfo
  {author} {\bibfnamefont {M.}~\bibnamefont {Kalaee}}, \bibinfo {author}
  {\bibfnamefont {P.~B.}\ \bibnamefont {Dieterle}}, \bibinfo {author}
  {\bibfnamefont {O.}~\bibnamefont {Painter}},\ and\ \bibinfo {author}
  {\bibfnamefont {J.~M.}\ \bibnamefont {Fink}},\ }\bibfield  {title} {\bibinfo
  {title} {Mechanical on-chip microwave circulator},\ }\href
  {https://doi.org/10.1038/s41467-017-01304-x} {\bibfield  {journal} {\bibinfo
  {journal} {Nature Communications}\ }\textbf {\bibinfo {volume} {8}},\
  \bibinfo {pages} {953} (\bibinfo {year} {2017})}\BibitemShut {NoStop}%
\bibitem [{\citenamefont {Bernier}\ \emph {et~al.}(2017)\citenamefont
  {Bernier}, \citenamefont {T{\'o}th}, \citenamefont {Koottandavida},
  \citenamefont {Ioannou}, \citenamefont {Malz}, \citenamefont {Nunnenkamp},
  \citenamefont {Feofanov},\ and\ \citenamefont {Kippenberg}}]{Bernier2017}%
  \BibitemOpen
  \bibfield  {author} {\bibinfo {author} {\bibfnamefont {N.~R.}\ \bibnamefont
  {Bernier}}, \bibinfo {author} {\bibfnamefont {L.~D.}\ \bibnamefont
  {T{\'o}th}}, \bibinfo {author} {\bibfnamefont {A.}~\bibnamefont
  {Koottandavida}}, \bibinfo {author} {\bibfnamefont {M.~A.}\ \bibnamefont
  {Ioannou}}, \bibinfo {author} {\bibfnamefont {D.}~\bibnamefont {Malz}},
  \bibinfo {author} {\bibfnamefont {A.}~\bibnamefont {Nunnenkamp}}, \bibinfo
  {author} {\bibfnamefont {A.~K.}\ \bibnamefont {Feofanov}},\ and\ \bibinfo
  {author} {\bibfnamefont {T.~J.}\ \bibnamefont {Kippenberg}},\ }\bibfield
  {title} {\bibinfo {title} {Nonreciprocal reconfigurable microwave
  optomechanical circuit},\ }\href {https://doi.org/10.1038/s41467-017-00447-1}
  {\bibfield  {journal} {\bibinfo  {journal} {Nature Communications}\ }\textbf
  {\bibinfo {volume} {8}},\ \bibinfo {pages} {604} (\bibinfo {year}
  {2017})}\BibitemShut {NoStop}%
\bibitem [{\citenamefont {Sohn}\ \emph {et~al.}(2021)\citenamefont {Sohn},
  \citenamefont {{\"O}rsel},\ and\ \citenamefont {Bahl}}]{Sohn2021}%
  \BibitemOpen
  \bibfield  {author} {\bibinfo {author} {\bibfnamefont {D.~B.}\ \bibnamefont
  {Sohn}}, \bibinfo {author} {\bibfnamefont {O.~E.}\ \bibnamefont
  {{\"O}rsel}},\ and\ \bibinfo {author} {\bibfnamefont {G.}~\bibnamefont
  {Bahl}},\ }\bibfield  {title} {\bibinfo {title} {Electrically driven optical
  isolation through phonon-mediated photonic {A}utler--{T}ownes splitting},\
  }\href {https://doi.org/10.1038/s41566-021-00884-x} {\bibfield  {journal}
  {\bibinfo  {journal} {Nature Photonics}\ }\textbf {\bibinfo {volume} {15}},\
  \bibinfo {pages} {822} (\bibinfo {year} {2021})}\BibitemShut {NoStop}%
\bibitem [{\citenamefont {Fan}\ \emph {et~al.}(2012)\citenamefont {Fan},
  \citenamefont {Wang}, \citenamefont {Varghese}, \citenamefont {Shen},
  \citenamefont {Niu}, \citenamefont {Xuan}, \citenamefont {Weiner},\ and\
  \citenamefont {Qi}}]{Fan2012}%
  \BibitemOpen
  \bibfield  {author} {\bibinfo {author} {\bibfnamefont {L.}~\bibnamefont
  {Fan}}, \bibinfo {author} {\bibfnamefont {J.}~\bibnamefont {Wang}}, \bibinfo
  {author} {\bibfnamefont {L.~T.}\ \bibnamefont {Varghese}}, \bibinfo {author}
  {\bibfnamefont {H.}~\bibnamefont {Shen}}, \bibinfo {author} {\bibfnamefont
  {B.}~\bibnamefont {Niu}}, \bibinfo {author} {\bibfnamefont {Y.}~\bibnamefont
  {Xuan}}, \bibinfo {author} {\bibfnamefont {A.~M.}\ \bibnamefont {Weiner}},\
  and\ \bibinfo {author} {\bibfnamefont {M.}~\bibnamefont {Qi}},\ }\bibfield
  {title} {\bibinfo {title} {An all-silicon passive optical diode},\ }\href
  {https://doi.org/10.1126/science.1214383} {\bibfield  {journal} {\bibinfo
  {journal} {Science}\ }\textbf {\bibinfo {volume} {335}},\ \bibinfo {pages}
  {447} (\bibinfo {year} {2012})}\BibitemShut {NoStop}%
\bibitem [{\citenamefont {Wang}\ \emph {et~al.}(2022)\citenamefont {Wang},
  \citenamefont {Wang},\ and\ \citenamefont {Clerk}}]{Wang2022}%
  \BibitemOpen
  \bibfield  {author} {\bibinfo {author} {\bibfnamefont {Y.-X.}\ \bibnamefont
  {Wang}}, \bibinfo {author} {\bibfnamefont {C.}~\bibnamefont {Wang}},\ and\
  \bibinfo {author} {\bibfnamefont {A.~A.}\ \bibnamefont {Clerk}},\ }\href
  {https://doi.org/10.48550/ARXIV.2203.09392} {\bibinfo {title} {Quantum
  nonreciprocal interactions via dissipative gauge symmetry}} (\bibinfo {year}
  {2022})\BibitemShut {NoStop}%
\bibitem [{\citenamefont {Sayrin}\ \emph {et~al.}(2015)\citenamefont {Sayrin},
  \citenamefont {Junge}, \citenamefont {Mitsch}, \citenamefont {Albrecht},
  \citenamefont {O'Shea}, \citenamefont {Schneeweiss}, \citenamefont {Volz},\
  and\ \citenamefont {Rauschenbeutel}}]{Sayrin2015}%
  \BibitemOpen
  \bibfield  {author} {\bibinfo {author} {\bibfnamefont {C.}~\bibnamefont
  {Sayrin}}, \bibinfo {author} {\bibfnamefont {C.}~\bibnamefont {Junge}},
  \bibinfo {author} {\bibfnamefont {R.}~\bibnamefont {Mitsch}}, \bibinfo
  {author} {\bibfnamefont {B.}~\bibnamefont {Albrecht}}, \bibinfo {author}
  {\bibfnamefont {D.}~\bibnamefont {O'Shea}}, \bibinfo {author} {\bibfnamefont
  {P.}~\bibnamefont {Schneeweiss}}, \bibinfo {author} {\bibfnamefont
  {J.}~\bibnamefont {Volz}},\ and\ \bibinfo {author} {\bibfnamefont
  {A.}~\bibnamefont {Rauschenbeutel}},\ }\bibfield  {title} {\bibinfo {title}
  {Nanophotonic optical isolator controlled by the internal state of cold
  atoms},\ }\href {https://doi.org/10.1103/PhysRevX.5.041036} {\bibfield
  {journal} {\bibinfo  {journal} {Phys. Rev. X}\ }\textbf {\bibinfo {volume}
  {5}},\ \bibinfo {pages} {041036} (\bibinfo {year} {2015})}\BibitemShut
  {NoStop}%
\bibitem [{\citenamefont {Buddhiraju}\ \emph {et~al.}(2020)\citenamefont
  {Buddhiraju}, \citenamefont {Song}, \citenamefont {Papadakis},\ and\
  \citenamefont {Fan}}]{Buddhiraju2020}%
  \BibitemOpen
  \bibfield  {author} {\bibinfo {author} {\bibfnamefont {S.}~\bibnamefont
  {Buddhiraju}}, \bibinfo {author} {\bibfnamefont {A.}~\bibnamefont {Song}},
  \bibinfo {author} {\bibfnamefont {G.~T.}\ \bibnamefont {Papadakis}},\ and\
  \bibinfo {author} {\bibfnamefont {S.}~\bibnamefont {Fan}},\ }\bibfield
  {title} {\bibinfo {title} {Nonreciprocal metamaterial obeying time-reversal
  symmetry},\ }\href {https://doi.org/10.1103/PhysRevLett.124.257403}
  {\bibfield  {journal} {\bibinfo  {journal} {Phys. Rev. Lett.}\ }\textbf
  {\bibinfo {volume} {124}},\ \bibinfo {pages} {257403} (\bibinfo {year}
  {2020})}\BibitemShut {NoStop}%
\bibitem [{\citenamefont {McDonald}\ \emph {et~al.}(2018)\citenamefont
  {McDonald}, \citenamefont {Pereg-Barnea},\ and\ \citenamefont
  {Clerk}}]{McDonald2018}%
  \BibitemOpen
  \bibfield  {author} {\bibinfo {author} {\bibfnamefont {A.}~\bibnamefont
  {McDonald}}, \bibinfo {author} {\bibfnamefont {T.}~\bibnamefont
  {Pereg-Barnea}},\ and\ \bibinfo {author} {\bibfnamefont {A.~A.}\ \bibnamefont
  {Clerk}},\ }\bibfield  {title} {\bibinfo {title} {Phase-dependent chiral
  transport and effective non-{H}ermitian dynamics in a bosonic
  {K}itaev-{M}ajorana chain},\ }\href
  {https://doi.org/10.1103/PhysRevX.8.041031} {\bibfield  {journal} {\bibinfo
  {journal} {Phys. Rev. X}\ }\textbf {\bibinfo {volume} {8}},\ \bibinfo {pages}
  {041031} (\bibinfo {year} {2018})}\BibitemShut {NoStop}%
\bibitem [{\citenamefont {Flynn}\ \emph {et~al.}(2020)\citenamefont {Flynn},
  \citenamefont {Cobanera},\ and\ \citenamefont {Viola}}]{Flynn2020}%
  \BibitemOpen
  \bibfield  {author} {\bibinfo {author} {\bibfnamefont {V.~P.}\ \bibnamefont
  {Flynn}}, \bibinfo {author} {\bibfnamefont {E.}~\bibnamefont {Cobanera}},\
  and\ \bibinfo {author} {\bibfnamefont {L.}~\bibnamefont {Viola}},\ }\bibfield
   {title} {\bibinfo {title} {Deconstructing effective non-{H}ermitian dynamics
  in quadratic bosonic {H}amiltonians},\ }\href
  {https://doi.org/10.1088/1367-2630/ab9e87} {\bibfield  {journal} {\bibinfo
  {journal} {New J. Phys.}\ } (\bibinfo {year} {2020})}\BibitemShut {NoStop}%
\bibitem [{\citenamefont {Flynn}\ \emph {et~al.}(2021)\citenamefont {Flynn},
  \citenamefont {Cobanera},\ and\ \citenamefont {Viola}}]{Flynn2021}%
  \BibitemOpen
  \bibfield  {author} {\bibinfo {author} {\bibfnamefont {V.~P.}\ \bibnamefont
  {Flynn}}, \bibinfo {author} {\bibfnamefont {E.}~\bibnamefont {Cobanera}},\
  and\ \bibinfo {author} {\bibfnamefont {L.}~\bibnamefont {Viola}},\ }\bibfield
   {title} {\bibinfo {title} {Topology by dissipation: {M}ajorana bosons in
  metastable quadratic {M}arkovian dynamics},\ }\href
  {https://doi.org/10.1103/PhysRevLett.127.245701} {\bibfield  {journal}
  {\bibinfo  {journal} {Phys. Rev. Lett.}\ }\textbf {\bibinfo {volume} {127}},\
  \bibinfo {pages} {245701} (\bibinfo {year} {2021})}\BibitemShut {NoStop}%
\bibitem [{\citenamefont {Wanjura}\ \emph {et~al.}(2020)\citenamefont
  {Wanjura}, \citenamefont {Brunelli},\ and\ \citenamefont
  {Nunnenkamp}}]{Wanjura2020}%
  \BibitemOpen
  \bibfield  {author} {\bibinfo {author} {\bibfnamefont {C.~C.}\ \bibnamefont
  {Wanjura}}, \bibinfo {author} {\bibfnamefont {M.}~\bibnamefont {Brunelli}},\
  and\ \bibinfo {author} {\bibfnamefont {A.}~\bibnamefont {Nunnenkamp}},\
  }\bibfield  {title} {\bibinfo {title} {Topological framework for directional
  amplification in driven-dissipative cavity arrays},\ }\href
  {https://doi.org/10.1038/s41467-020-16863-9} {\bibfield  {journal} {\bibinfo
  {journal} {Nature Communications}\ }\textbf {\bibinfo {volume} {11}},\
  \bibinfo {pages} {3149} (\bibinfo {year} {2020})}\BibitemShut {NoStop}%
\bibitem [{\citenamefont {Fan}\ \emph {et~al.}(2003)\citenamefont {Fan},
  \citenamefont {Suh},\ and\ \citenamefont {Joannopoulos}}]{Fan2003}%
  \BibitemOpen
  \bibfield  {author} {\bibinfo {author} {\bibfnamefont {S.}~\bibnamefont
  {Fan}}, \bibinfo {author} {\bibfnamefont {W.}~\bibnamefont {Suh}},\ and\
  \bibinfo {author} {\bibfnamefont {J.~D.}\ \bibnamefont {Joannopoulos}},\
  }\bibfield  {title} {\bibinfo {title} {Temporal coupled-mode theory for the
  fano resonance in optical resonators},\ }\href
  {https://doi.org/10.1364/JOSAA.20.000569} {\bibfield  {journal} {\bibinfo
  {journal} {J. Opt. Soc. Am. A}\ }\textbf {\bibinfo {volume} {20}},\ \bibinfo
  {pages} {569} (\bibinfo {year} {2003})}\BibitemShut {NoStop}%
\bibitem [{\citenamefont {Gardiner}\ and\ \citenamefont
  {Collett}(1985)}]{Gardiner1985}%
  \BibitemOpen
  \bibfield  {author} {\bibinfo {author} {\bibfnamefont {C.~W.}\ \bibnamefont
  {Gardiner}}\ and\ \bibinfo {author} {\bibfnamefont {M.~J.}\ \bibnamefont
  {Collett}},\ }\bibfield  {title} {\bibinfo {title} {Input and output in
  damped quantum systems: {Q}uantum stochastic differential equations and the
  master equation},\ }\href {https://doi.org/10.1103/PhysRevA.31.3761}
  {\bibfield  {journal} {\bibinfo  {journal} {Phys. Rev. A}\ }\textbf {\bibinfo
  {volume} {31}},\ \bibinfo {pages} {3761} (\bibinfo {year}
  {1985})}\BibitemShut {NoStop}%
\bibitem [{\citenamefont {Clerk}\ \emph {et~al.}(2010)\citenamefont {Clerk},
  \citenamefont {Devoret}, \citenamefont {Girvin}, \citenamefont {Marquardt},\
  and\ \citenamefont {Schoelkopf}}]{Clerk2010}%
  \BibitemOpen
  \bibfield  {author} {\bibinfo {author} {\bibfnamefont {A.~A.}\ \bibnamefont
  {Clerk}}, \bibinfo {author} {\bibfnamefont {M.~H.}\ \bibnamefont {Devoret}},
  \bibinfo {author} {\bibfnamefont {S.~M.}\ \bibnamefont {Girvin}}, \bibinfo
  {author} {\bibfnamefont {F.}~\bibnamefont {Marquardt}},\ and\ \bibinfo
  {author} {\bibfnamefont {R.~J.}\ \bibnamefont {Schoelkopf}},\ }\bibfield
  {title} {\bibinfo {title} {Introduction to quantum noise, measurement, and
  amplification},\ }\href {https://doi.org/10.1103/RevModPhys.82.1155}
  {\bibfield  {journal} {\bibinfo  {journal} {Rev. Mod. Phys.}\ }\textbf
  {\bibinfo {volume} {82}},\ \bibinfo {pages} {1155} (\bibinfo {year}
  {2010})}\BibitemShut {NoStop}%
\bibitem [{\citenamefont {Braginsky}\ \emph {et~al.}(1980)\citenamefont
  {Braginsky}, \citenamefont {Vorontsov},\ and\ \citenamefont
  {Thorne}}]{Braginsky1980}%
  \BibitemOpen
  \bibfield  {author} {\bibinfo {author} {\bibfnamefont {V.~B.}\ \bibnamefont
  {Braginsky}}, \bibinfo {author} {\bibfnamefont {Y.~I.}\ \bibnamefont
  {Vorontsov}},\ and\ \bibinfo {author} {\bibfnamefont {K.~S.}\ \bibnamefont
  {Thorne}},\ }\bibfield  {title} {\bibinfo {title} {Quantum nondemolition
  measurements},\ }\href
  {https://www.science.org/doi/10.1126/science.209.4456.547} {\bibfield
  {journal} {\bibinfo  {journal} {Science}\ }\textbf {\bibinfo {volume}
  {209}},\ \bibinfo {pages} {547} (\bibinfo {year} {1980})}\BibitemShut
  {NoStop}%
\bibitem [{\citenamefont {Clerk}\ \emph {et~al.}(2008)\citenamefont {Clerk},
  \citenamefont {Marquardt},\ and\ \citenamefont {Jacobs}}]{Clerk2008}%
  \BibitemOpen
  \bibfield  {author} {\bibinfo {author} {\bibfnamefont {A.~A.}\ \bibnamefont
  {Clerk}}, \bibinfo {author} {\bibfnamefont {F.}~\bibnamefont {Marquardt}},\
  and\ \bibinfo {author} {\bibfnamefont {K.}~\bibnamefont {Jacobs}},\
  }\bibfield  {title} {\bibinfo {title} {Back-action evasion and squeezing of a
  mechanical resonator using a cavity detector},\ }\href
  {https://doi.org/10.1088/1367-2630/10/9/095010} {\bibfield  {journal}
  {\bibinfo  {journal} {New Journal of Physics}\ }\textbf {\bibinfo {volume}
  {10}},\ \bibinfo {pages} {095010} (\bibinfo {year} {2008})}\BibitemShut
  {NoStop}%
\bibitem [{\citenamefont {Brunelli}\ \emph {et~al.}(2019)\citenamefont
  {Brunelli}, \citenamefont {Malz},\ and\ \citenamefont
  {Nunnenkamp}}]{Brunelli2019}%
  \BibitemOpen
  \bibfield  {author} {\bibinfo {author} {\bibfnamefont {M.}~\bibnamefont
  {Brunelli}}, \bibinfo {author} {\bibfnamefont {D.}~\bibnamefont {Malz}},\
  and\ \bibinfo {author} {\bibfnamefont {A.}~\bibnamefont {Nunnenkamp}},\
  }\bibfield  {title} {\bibinfo {title} {Conditional dynamics of optomechanical
  two-tone backaction-evading measurements},\ }\href
  {https://doi.org/10.1103/PhysRevLett.123.093602} {\bibfield  {journal}
  {\bibinfo  {journal} {Phys. Rev. Lett.}\ }\textbf {\bibinfo {volume} {123}},\
  \bibinfo {pages} {093602} (\bibinfo {year} {2019})}\BibitemShut {NoStop}%
\bibitem [{\citenamefont {Metelmann}\ and\ \citenamefont
  {Clerk}(2014)}]{Metelmann2014Quantum}%
  \BibitemOpen
  \bibfield  {author} {\bibinfo {author} {\bibfnamefont {A.}~\bibnamefont
  {Metelmann}}\ and\ \bibinfo {author} {\bibfnamefont {A.~A.}\ \bibnamefont
  {Clerk}},\ }\bibfield  {title} {\bibinfo {title} {Quantum-limited
  amplification via reservoir engineering},\ }\href
  {https://doi.org/10.1103/PhysRevLett.112.133904} {\bibfield  {journal}
  {\bibinfo  {journal} {Phys. Rev. Lett.}\ }\textbf {\bibinfo {volume} {112}},\
  \bibinfo {pages} {133904} (\bibinfo {year} {2014})}\BibitemShut {NoStop}%
\bibitem [{\citenamefont {Chien}\ \emph {et~al.}(2020)\citenamefont {Chien},
  \citenamefont {Lanes}, \citenamefont {Liu}, \citenamefont {Cao},
  \citenamefont {Lu}, \citenamefont {Motz}, \citenamefont {Liu}, \citenamefont
  {Pekker},\ and\ \citenamefont {Hatridge}}]{Chien2020Multiparametric}%
  \BibitemOpen
  \bibfield  {author} {\bibinfo {author} {\bibfnamefont {T.-C.}\ \bibnamefont
  {Chien}}, \bibinfo {author} {\bibfnamefont {O.}~\bibnamefont {Lanes}},
  \bibinfo {author} {\bibfnamefont {C.}~\bibnamefont {Liu}}, \bibinfo {author}
  {\bibfnamefont {X.}~\bibnamefont {Cao}}, \bibinfo {author} {\bibfnamefont
  {P.}~\bibnamefont {Lu}}, \bibinfo {author} {\bibfnamefont {S.}~\bibnamefont
  {Motz}}, \bibinfo {author} {\bibfnamefont {G.}~\bibnamefont {Liu}}, \bibinfo
  {author} {\bibfnamefont {D.}~\bibnamefont {Pekker}},\ and\ \bibinfo {author}
  {\bibfnamefont {M.}~\bibnamefont {Hatridge}},\ }\bibfield  {title} {\bibinfo
  {title} {Multiparametric amplification and qubit measurement with a kerr-free
  josephson ring modulator},\ }\href
  {https://doi.org/10.1103/PhysRevA.101.042336} {\bibfield  {journal} {\bibinfo
   {journal} {Phys. Rev. A}\ }\textbf {\bibinfo {volume} {101}},\ \bibinfo
  {pages} {042336} (\bibinfo {year} {2020})}\BibitemShut {NoStop}%
\bibitem [{\citenamefont {Carmichael}(1993)}]{Carmichael1993}%
  \BibitemOpen
  \bibfield  {author} {\bibinfo {author} {\bibfnamefont {H.~J.}\ \bibnamefont
  {Carmichael}},\ }\bibfield  {title} {\bibinfo {title} {Quantum trajectory
  theory for cascaded open systems},\ }\href
  {https://doi.org/10.1103/PhysRevLett.70.2273} {\bibfield  {journal} {\bibinfo
   {journal} {Phys. Rev. Lett.}\ }\textbf {\bibinfo {volume} {70}},\ \bibinfo
  {pages} {2273} (\bibinfo {year} {1993})}\BibitemShut {NoStop}%
\bibitem [{\citenamefont {Gardiner}(1993)}]{Gardiner1993}%
  \BibitemOpen
  \bibfield  {author} {\bibinfo {author} {\bibfnamefont {C.~W.}\ \bibnamefont
  {Gardiner}},\ }\bibfield  {title} {\bibinfo {title} {Driving a quantum system
  with the output field from another driven quantum system},\ }\href
  {https://doi.org/10.1103/PhysRevLett.70.2269} {\bibfield  {journal} {\bibinfo
   {journal} {Phys. Rev. Lett.}\ }\textbf {\bibinfo {volume} {70}},\ \bibinfo
  {pages} {2269} (\bibinfo {year} {1993})}\BibitemShut {NoStop}%
\bibitem [{\citenamefont {Mathew}\ \emph {et~al.}(2020)\citenamefont {Mathew},
  \citenamefont {Pino},\ and\ \citenamefont {Verhagen}}]{mathew2020}%
  \BibitemOpen
  \bibfield  {author} {\bibinfo {author} {\bibfnamefont {J.~P.}\ \bibnamefont
  {Mathew}}, \bibinfo {author} {\bibfnamefont {J.~d.}\ \bibnamefont {Pino}},\
  and\ \bibinfo {author} {\bibfnamefont {E.}~\bibnamefont {Verhagen}},\
  }\bibfield  {title} {\bibinfo {title} {Synthetic gauge fields for phonon
  transport in a nano-optomechanical system},\ }\href
  {https://doi.org/10.1038/s41565-019-0630-8} {\bibfield  {journal} {\bibinfo
  {journal} {Nature Nanotechnology}\ }\textbf {\bibinfo {volume} {15}},\
  \bibinfo {pages} {198} (\bibinfo {year} {2020})}\BibitemShut {NoStop}%
\bibitem [{\citenamefont {del Pino}\ \emph {et~al.}(2022)\citenamefont {del
  Pino}, \citenamefont {Slim},\ and\ \citenamefont {Verhagen}}]{delPino2022}%
  \BibitemOpen
  \bibfield  {author} {\bibinfo {author} {\bibfnamefont {J.}~\bibnamefont {del
  Pino}}, \bibinfo {author} {\bibfnamefont {J.~J.}\ \bibnamefont {Slim}},\ and\
  \bibinfo {author} {\bibfnamefont {E.}~\bibnamefont {Verhagen}},\ }\bibfield
  {title} {\bibinfo {title} {Non-{H}ermitian chiral phononics through
  optomechanically induced squeezing},\ }\href
  {https://doi.org/10.1038/s41586-022-04609-0} {\bibfield  {journal} {\bibinfo
  {journal} {Nature}\ }\textbf {\bibinfo {volume} {606}},\ \bibinfo {pages}
  {82} (\bibinfo {year} {2022})}\BibitemShut {NoStop}%
\bibitem [{\citenamefont {Koch}\ \emph {et~al.}(2010)\citenamefont {Koch},
  \citenamefont {Houck}, \citenamefont {Hur},\ and\ \citenamefont
  {Girvin}}]{Koch2010}%
  \BibitemOpen
  \bibfield  {author} {\bibinfo {author} {\bibfnamefont {J.}~\bibnamefont
  {Koch}}, \bibinfo {author} {\bibfnamefont {A.~A.}\ \bibnamefont {Houck}},
  \bibinfo {author} {\bibfnamefont {K.~L.}\ \bibnamefont {Hur}},\ and\ \bibinfo
  {author} {\bibfnamefont {S.~M.}\ \bibnamefont {Girvin}},\ }\bibfield  {title}
  {\bibinfo {title} {Time-reversal-symmetry breaking in circuit-{QED}-based
  photon lattices},\ }\href {https://doi.org/10.1103/PhysRevA.82.043811}
  {\bibfield  {journal} {\bibinfo  {journal} {Phys. Rev. A}\ }\textbf {\bibinfo
  {volume} {82}},\ \bibinfo {pages} {043811} (\bibinfo {year}
  {2010})}\BibitemShut {NoStop}%
\bibitem [{\citenamefont {Hasan}\ and\ \citenamefont {Kane}(2010)}]{Hasan2010}%
  \BibitemOpen
  \bibfield  {author} {\bibinfo {author} {\bibfnamefont {M.~Z.}\ \bibnamefont
  {Hasan}}\ and\ \bibinfo {author} {\bibfnamefont {C.~L.}\ \bibnamefont
  {Kane}},\ }\bibfield  {title} {\bibinfo {title} {Colloquium: {T}opological
  insulators},\ }\href {https://doi.org/10.1103/RevModPhys.82.3045} {\bibfield
  {journal} {\bibinfo  {journal} {Rev. Mod. Phys.}\ }\textbf {\bibinfo {volume}
  {82}},\ \bibinfo {pages} {3045} (\bibinfo {year} {2010})}\BibitemShut
  {NoStop}%
\bibitem [{\citenamefont {Porras}\ and\ \citenamefont
  {Fern\'andez-Lorenzo}(2019)}]{Porras2019}%
  \BibitemOpen
  \bibfield  {author} {\bibinfo {author} {\bibfnamefont {D.}~\bibnamefont
  {Porras}}\ and\ \bibinfo {author} {\bibfnamefont {S.}~\bibnamefont
  {Fern\'andez-Lorenzo}},\ }\bibfield  {title} {\bibinfo {title} {Topological
  amplification in photonic lattices},\ }\href
  {https://doi.org/10.1103/PhysRevLett.122.143901} {\bibfield  {journal}
  {\bibinfo  {journal} {Phys. Rev. Lett.}\ }\textbf {\bibinfo {volume} {122}},\
  \bibinfo {pages} {143901} (\bibinfo {year} {2019})}\BibitemShut {NoStop}%
\bibitem [{\citenamefont {Gong}\ \emph {et~al.}(2018)\citenamefont {Gong},
  \citenamefont {Ashida}, \citenamefont {Kawabata}, \citenamefont {Takasan},
  \citenamefont {Higashikawa},\ and\ \citenamefont {Ueda}}]{Gong2018}%
  \BibitemOpen
  \bibfield  {author} {\bibinfo {author} {\bibfnamefont {Z.}~\bibnamefont
  {Gong}}, \bibinfo {author} {\bibfnamefont {Y.}~\bibnamefont {Ashida}},
  \bibinfo {author} {\bibfnamefont {K.}~\bibnamefont {Kawabata}}, \bibinfo
  {author} {\bibfnamefont {K.}~\bibnamefont {Takasan}}, \bibinfo {author}
  {\bibfnamefont {S.}~\bibnamefont {Higashikawa}},\ and\ \bibinfo {author}
  {\bibfnamefont {M.}~\bibnamefont {Ueda}},\ }\bibfield  {title} {\bibinfo
  {title} {Topological phases of non-{H}ermitian systems},\ }\href
  {https://doi.org/10.1103/PhysRevX.8.031079} {\bibfield  {journal} {\bibinfo
  {journal} {Phys. Rev. X}\ }\textbf {\bibinfo {volume} {8}},\ \bibinfo {pages}
  {031079} (\bibinfo {year} {2018})}\BibitemShut {NoStop}%
\bibitem [{\citenamefont {Wang}\ and\ \citenamefont {Clerk}(2019)}]{Wang2019}%
  \BibitemOpen
  \bibfield  {author} {\bibinfo {author} {\bibfnamefont {Y.-X.}\ \bibnamefont
  {Wang}}\ and\ \bibinfo {author} {\bibfnamefont {A.~A.}\ \bibnamefont
  {Clerk}},\ }\bibfield  {title} {\bibinfo {title} {Non-{H}ermitian dynamics
  without dissipation in quantum systems},\ }\href
  {https://doi.org/10.1103/PhysRevA.99.063834} {\bibfield  {journal} {\bibinfo
  {journal} {Phys. Rev. A}\ }\textbf {\bibinfo {volume} {99}},\ \bibinfo
  {pages} {063834} (\bibinfo {year} {2019})}\BibitemShut {NoStop}%
\bibitem [{\citenamefont {Ningyuan}\ \emph {et~al.}(2015)\citenamefont
  {Ningyuan}, \citenamefont {Owens}, \citenamefont {Sommer}, \citenamefont
  {Schuster},\ and\ \citenamefont {Simon}}]{Ningyuan2015}%
  \BibitemOpen
  \bibfield  {author} {\bibinfo {author} {\bibfnamefont {J.}~\bibnamefont
  {Ningyuan}}, \bibinfo {author} {\bibfnamefont {C.}~\bibnamefont {Owens}},
  \bibinfo {author} {\bibfnamefont {A.}~\bibnamefont {Sommer}}, \bibinfo
  {author} {\bibfnamefont {D.}~\bibnamefont {Schuster}},\ and\ \bibinfo
  {author} {\bibfnamefont {J.}~\bibnamefont {Simon}},\ }\bibfield  {title}
  {\bibinfo {title} {Time- and site-resolved dynamics in a topological
  circuit},\ }\href {https://doi.org/10.1103/PhysRevX.5.021031} {\bibfield
  {journal} {\bibinfo  {journal} {Phys. Rev. X}\ }\textbf {\bibinfo {volume}
  {5}},\ \bibinfo {pages} {021031} (\bibinfo {year} {2015})}\BibitemShut
  {NoStop}%
\bibitem [{\citenamefont {Zhang}\ \emph {et~al.}(2017)\citenamefont {Zhang},
  \citenamefont {Wei}, \citenamefont {Cheng}, \citenamefont {Zhang},
  \citenamefont {Wu},\ and\ \citenamefont {Liu}}]{Zhang2017}%
  \BibitemOpen
  \bibfield  {author} {\bibinfo {author} {\bibfnamefont {Z.}~\bibnamefont
  {Zhang}}, \bibinfo {author} {\bibfnamefont {Q.}~\bibnamefont {Wei}}, \bibinfo
  {author} {\bibfnamefont {Y.}~\bibnamefont {Cheng}}, \bibinfo {author}
  {\bibfnamefont {T.}~\bibnamefont {Zhang}}, \bibinfo {author} {\bibfnamefont
  {D.}~\bibnamefont {Wu}},\ and\ \bibinfo {author} {\bibfnamefont
  {X.}~\bibnamefont {Liu}},\ }\bibfield  {title} {\bibinfo {title} {Topological
  creation of acoustic pseudospin multipoles in a flow-free symmetry-broken
  metamaterial lattice},\ }\href
  {https://doi.org/10.1103/PhysRevLett.118.084303} {\bibfield  {journal}
  {\bibinfo  {journal} {Phys. Rev. Lett.}\ }\textbf {\bibinfo {volume} {118}},\
  \bibinfo {pages} {084303} (\bibinfo {year} {2017})}\BibitemShut {NoStop}%
\bibitem [{\citenamefont {Karg}\ \emph {et~al.}(2020)\citenamefont {Karg},
  \citenamefont {Gouraud}, \citenamefont {Ngai}, \citenamefont {Schmid},
  \citenamefont {Hammerer},\ and\ \citenamefont {Treutlein}}]{Karg2020}%
  \BibitemOpen
  \bibfield  {author} {\bibinfo {author} {\bibfnamefont {T.~M.}\ \bibnamefont
  {Karg}}, \bibinfo {author} {\bibfnamefont {B.}~\bibnamefont {Gouraud}},
  \bibinfo {author} {\bibfnamefont {C.~T.}\ \bibnamefont {Ngai}}, \bibinfo
  {author} {\bibfnamefont {G.-L.}\ \bibnamefont {Schmid}}, \bibinfo {author}
  {\bibfnamefont {K.}~\bibnamefont {Hammerer}},\ and\ \bibinfo {author}
  {\bibfnamefont {P.}~\bibnamefont {Treutlein}},\ }\bibfield  {title} {\bibinfo
  {title} {Light-mediated strong coupling between a mechanical oscillator and
  atomic spins 1 meter apart},\ }\href
  {https://doi.org/10.1126/science.abb0328} {\bibfield  {journal} {\bibinfo
  {journal} {Science}\ }\textbf {\bibinfo {volume} {369}},\ \bibinfo {pages}
  {174} (\bibinfo {year} {2020})}\BibitemShut {NoStop}%
\bibitem [{\citenamefont {Rossignoli}\ and\ \citenamefont
  {Kowalski}(2005)}]{Rossignoli2005}%
  \BibitemOpen
  \bibfield  {author} {\bibinfo {author} {\bibfnamefont {R.}~\bibnamefont
  {Rossignoli}}\ and\ \bibinfo {author} {\bibfnamefont {A.~M.}\ \bibnamefont
  {Kowalski}},\ }\bibfield  {title} {\bibinfo {title} {Complex modes in
  unstable quadratic bosonic forms},\ }\href
  {https://doi.org/10.1103/PhysRevA.72.032101} {\bibfield  {journal} {\bibinfo
  {journal} {Phys. Rev. A}\ }\textbf {\bibinfo {volume} {72}},\ \bibinfo
  {pages} {032101} (\bibinfo {year} {2005})}\BibitemShut {NoStop}%
\bibitem [{\citenamefont {Leijssen}\ \emph {et~al.}(2017)\citenamefont
  {Leijssen}, \citenamefont {La~Gala}, \citenamefont {Freisem}, \citenamefont
  {Muhonen},\ and\ \citenamefont {Verhagen}}]{leijssen2017nonlinear}%
  \BibitemOpen
  \bibfield  {author} {\bibinfo {author} {\bibfnamefont {R.}~\bibnamefont
  {Leijssen}}, \bibinfo {author} {\bibfnamefont {G.~R.}\ \bibnamefont
  {La~Gala}}, \bibinfo {author} {\bibfnamefont {L.}~\bibnamefont {Freisem}},
  \bibinfo {author} {\bibfnamefont {J.~T.}\ \bibnamefont {Muhonen}},\ and\
  \bibinfo {author} {\bibfnamefont {E.}~\bibnamefont {Verhagen}},\ }\bibfield
  {title} {\bibinfo {title} {Nonlinear cavity optomechanics with nanomechanical
  thermal fluctuations},\ }\href {https://doi.org/10.1038/ncomms16024}
  {\bibfield  {journal} {\bibinfo  {journal} {Nature Communications}\ }\textbf
  {\bibinfo {volume} {8}},\ \bibinfo {pages} {16024} (\bibinfo {year}
  {2017})}\BibitemShut {NoStop}%
\end{thebibliography}%


\begin{thebibliography}{3}%
\makeatletter
\providecommand \@ifxundefined [1]{%
 \@ifx{#1\undefined}
}%
\providecommand \@ifnum [1]{%
 \ifnum #1\expandafter \@firstoftwo
 \else \expandafter \@secondoftwo
 \fi
}%
\providecommand \@ifx [1]{%
 \ifx #1\expandafter \@firstoftwo
 \else \expandafter \@secondoftwo
 \fi
}%
\providecommand \natexlab [1]{#1}%
\providecommand \enquote  [1]{``#1''}%
\providecommand \bibnamefont  [1]{#1}%
\providecommand \bibfnamefont [1]{#1}%
\providecommand \citenamefont [1]{#1}%
\providecommand \href@noop [0]{\@secondoftwo}%
\providecommand \href [0]{\begingroup \@sanitize@url \@href}%
\providecommand \@href[1]{\@@startlink{#1}\@@href}%
\providecommand \@@href[1]{\endgroup#1\@@endlink}%
\providecommand \@sanitize@url [0]{\catcode `\\12\catcode `\$12\catcode
  `\&12\catcode `\#12\catcode `\^12\catcode `\_12\catcode `\%12\relax}%
\providecommand \@@startlink[1]{}%
\providecommand \@@endlink[0]{}%
\providecommand \url  [0]{\begingroup\@sanitize@url \@url }%
\providecommand \@url [1]{\endgroup\@href {#1}{\urlprefix }}%
\providecommand \urlprefix  [0]{URL }%
\providecommand \Eprint [0]{\href }%
\providecommand \doibase [0]{https://doi.org/}%
\providecommand \selectlanguage [0]{\@gobble}%
\providecommand \bibinfo  [0]{\@secondoftwo}%
\providecommand \bibfield  [0]{\@secondoftwo}%
\providecommand \translation [1]{[#1]}%
\providecommand \BibitemOpen [0]{}%
\providecommand \bibitemStop [0]{}%
\providecommand \bibitemNoStop [0]{.\EOS\space}%
\providecommand \EOS [0]{\spacefactor3000\relax}%
\providecommand \BibitemShut  [1]{\csname bibitem#1\endcsname}%
\let\auto@bib@innerbib\@empty
\bibitem [{\citenamefont {Koch}\ \emph {et~al.}(2010)\citenamefont {Koch},
  \citenamefont {Houck}, \citenamefont {Hur},\ and\ \citenamefont
  {Girvin}}]{Koch2010_SI}%
  \BibitemOpen
  \bibfield  {author} {\bibinfo {author} {\bibfnamefont {J.}~\bibnamefont
  {Koch}}, \bibinfo {author} {\bibfnamefont {A.~A.}\ \bibnamefont {Houck}},
  \bibinfo {author} {\bibfnamefont {K.~L.}\ \bibnamefont {Hur}},\ and\ \bibinfo
  {author} {\bibfnamefont {S.~M.}\ \bibnamefont {Girvin}},\ }\bibfield  {title}
  {\bibinfo {title} {Time-reversal-symmetry breaking in circuit-QED-based
  photon lattices},\ }\href {https://doi.org/10.1103/PhysRevA.82.043811}
  {\bibfield  {journal} {\bibinfo  {journal} {Phys. Rev. A}\ }\textbf {\bibinfo
  {volume} {82}},\ \bibinfo {pages} {043811} (\bibinfo {year}
  {2010})}\BibitemShut {NoStop}%
\bibitem [{\citenamefont {Ranzani}\ and\ \citenamefont
  {Aumentado}(2015)}]{Ranzani2015_SI}%
  \BibitemOpen
  \bibfield  {author} {\bibinfo {author} {\bibfnamefont {L.}~\bibnamefont
  {Ranzani}}\ and\ \bibinfo {author} {\bibfnamefont {J.}~\bibnamefont
  {Aumentado}},\ }\bibfield  {title} {\bibinfo {title} {Graph-based analysis of
  nonreciprocity in coupled-mode systems},\ }\href
  {https://doi.org/10.1088/1367-2630/17/2/023024} {\bibfield  {journal}
  {\bibinfo  {journal} {New Journal of Physics}\ }\textbf {\bibinfo {volume}
  {17}},\ \bibinfo {pages} {023024} (\bibinfo {year} {2015})}\BibitemShut
  {NoStop}%
\bibitem [{\citenamefont {Naaman}\ and\ \citenamefont
  {Aumentado}(2022)}]{Naaman2022_SI}%
  \BibitemOpen
  \bibfield  {author} {\bibinfo {author} {\bibfnamefont {O.}~\bibnamefont
  {Naaman}}\ and\ \bibinfo {author} {\bibfnamefont {J.}~\bibnamefont
  {Aumentado}},\ }\bibfield  {title} {\bibinfo {title} {Synthesis of
  parametrically coupled networks},\ }\href
  {https://doi.org/10.1103/PRXQuantum.3.020201} {\bibfield  {journal} {\bibinfo
   {journal} {PRX Quantum}\ }\textbf {\bibinfo {volume} {3}},\ \bibinfo {pages}
  {020201} (\bibinfo {year} {2022})}\BibitemShut {NoStop}%
\end{thebibliography}


\clearpage
\pagebreak
\renewcommand{\theequation}{S\arabic{equation}}
\renewcommand{\thefigure}{S\arabic{figure}}
\setcounter{equation}{0}
\setcounter{table}{0}
\renewcommand{\figurename}{FIG.}
\setcounter{figure}{0}
\renewcommand{\theHfigure}{S.\thefigure}

\patchcmd{\section}{\centering}{\raggedright}{}{}
\patchcmd{\subsection}{\centering}{\raggedright}{}{}
\patchcmd{\subsubsection}{\centering}{\raggedright}{}{}


\newpage
\onecolumngrid
\part*{\large\centering Supplementary Information for ``Quadrature nonreciprocity: unidirectional bosonic transmission without breaking time-reversal symmetry''}
\tableofcontents
\let\addcontentsline\oldaddcontentsline
\setcounter{secnumdepth}{2} 
    \section{Explicit results}
	\subsection{Susceptibility in the qNR dimer}
	\label{sec:detailsDimer}
	The full susceptibility matrix of the qNR dimer rotated by a phase $\phi$ is given by
	\begin{align}
	   \chi_\phi & =
        \begin{pmatrix}
         -\frac{2}{\gamma } & 0 & -\frac{4 J \sin (2 \phi )}{\gamma ^2} & -\frac{8 J \sin ^2(\phi )}{\gamma ^2} \\
         0 & -\frac{2}{\gamma } & \frac{8 J \cos ^2(\phi )}{\gamma ^2} & \frac{4 J \sin (2 \phi )}{\gamma ^2} \\
         -\frac{4 J \sin (2 \phi )}{\gamma ^2} & -\frac{8 J \sin ^2(\phi )}{\gamma ^2} & -\frac{2}{\gamma } & 0 \\
         \frac{8 J \cos ^2(\phi )}{\gamma ^2} & \frac{4 J \sin (2 \phi )}{\gamma ^2} & 0 & -\frac{2}{\gamma } \\
        \end{pmatrix}.
	\end{align}
	We see that reciprocity occurs for $\phi=\frac{\pi}{4},\frac{3\pi}{4},\dots$ while the transport is non-reciprocal otherwise. Furthermore, the directionality occurs for pairs of quadratures in opposite directions. For instance, for $\phi=0$, $\lvert\chi_{x_1\to p_2}\rvert=\lvert\chi_{x_2\to p_1}\rvert=\frac{16 J^2}{\gamma^4}$ and $\lvert\chi_{p_2\to x_1}\rvert=\lvert\chi_{p_1\to x_2}\rvert=0$.
	
	\subsection{Standard non-reciprocity in the TRS-breaking isolator}
The TRS-breaking isolator shown in Figure~1~(b) does not feature squeezing ($\mathcal{B} = 0$) and can therefore be described by a dynamical matrix in the reduced basis $\{a_1,a_2,a_3\}$
	    \begin{align}
	        H^{(a^\prime)} = -\i\mathcal{A} - \frac{\Gamma^\prime}{2} = -\i\begin{pmatrix}
	        -i\frac{\gamma_1}{2} & e^{i\Phi} J_{12} & J_{31} \\
	        e^{-i\Phi} J_{12} & -i\frac{\gamma_2}{2} & J_{23} \\
	        J_{31} & J_{23} & -i\frac{\gamma_3}{2}
	        \end{pmatrix},
	    \end{align}
	    for the reduced operator vector $\myvec{a}^\prime = (a_1, a_2, a_3)^\mathrm{T}$ through the equation of motion $\dot{\myvec{a}}^\prime = H^{(a^\prime)} \myvec{a}^\prime - \myvec{f}^{(a^\prime)}(t)$, with input fields grouped in the vector $\myvec{f}^{(a^\prime)}(t)$ ($f_j^{(a^\prime)}=\sqrt{\gamma_j}a_{j,\text{in}}$) and dissipation $\Gamma^\prime = \diag(\gamma_1, \gamma_2, \gamma_3)$. We assume that the flux $\Phi$ is carried only by the coupling between resonators $1$ and $2$, denoting the real coupling amplitudes by $J_{ij}$.
	    
	    The on-resonance steady-state susceptibility of the isolator to the drive $\myvec{f}^{(a^\prime)}$ is given by $\chi^{(a^\prime)} = (H^{(a^\prime)})^{-1}$. For the ideal flux $\Phi = \pi/2$, the relevant susceptibilities are given by
	    \begin{align}
	        \chi^{(a^\prime)}_{12} &= \left(J_{23}J_{31} - J_{12}\gamma_3/2\right) \det(\chi^{(a^\prime)}),&& 	        \chi^{(a^\prime)}_{21} = \left(J_{23}J_{31} + J_{12}\gamma_3/2\right)\det(\chi^{(a^\prime)}),
	    \end{align}
	    with $\det(\chi^{(a^\prime)}) = \left(J_{23}^2\gamma_1/2 + J_{31}^2\gamma_2/2 +  J_{12}^2\gamma_3/2 + \gamma_1\gamma_2\gamma_3/8\right)^{-1}$. Perfect isolation ($\chi^{(a^\prime)}_{12} = 0$, $\chi^{(a^\prime)}_{21} \neq 0$) is obtained when $J_{12}\gamma_3/2 = J_{23}J_{31}$, which can be satisfied for any hierarchy of the loss rates $\gamma_i$. Furthermore, perfect \textit{circulation} between the resonators $a_1$, $a_2$ and the auxiliary resonator $a_3$ -- i.e. $\chi^{(a^\prime)}_{12} = \chi^{(a^\prime)}_{23} = \chi^{(a^\prime)}_{31} = 0$ while $\chi^{(a^\prime)}_{21}, \chi^{(a^\prime)}_{32}, \chi^{(a^\prime)}_{13} \neq 0$ -- can be obtained by setting $J_{ij} = \sqrt{\gamma_i\gamma_j}/2$. In this situation, the mechanical cooperativity $\mathcal{C}_{ij}=(4J_{ij}^2)/(\gamma_i\gamma_j)$ for each link is unity. 
	    
	    Finally, for unity cooperativity, the non-zero elements of the reduced susceptibility matrix
	    \begin{align}
	        \overline{\chi}^{(a^\prime)} = \begin{pmatrix}
	            -\gamma_1^{-1} & 0 \\
	            \sqrt{\gamma_1\gamma_2}^{-1} & -\gamma_2^{-1}
	        \end{pmatrix},
	    \end{align}
	    which describes the transfer between resonators $1$ and $2$ while disregarding the response of the auxiliary resonator $3$, have equal magnitude if $\gamma_1 = \gamma_2$, as is the case in our experiments.

	\section{Gauge transformations, rotations in phase space and TRS}
	Here we derive the implications a TRS preserving Hamiltonian has on the dynamical matrix of a system with added local decay. While (local) gain and loss explicitly break TRS, they only modify the diagonal of the dynamical matrix such that the underlying TRS preserving Hamiltonian still imposes certain symmetries on the dynamical matrix and susceptibility matrix. This results in the unique transformation properties of qNR.
	
	In the main text, we defined TRS based on gauge transformations that make the coefficients in the Hamiltonian real. We first show that a $U(1)$ gauge transformation $e^{\mathrm{i}\phi a_j^{\dagger}a_j}$ in phase space manifests as rotation by $\phi$ and then show that TRS implies that the susceptibility matrix, or scattering matrix, for at least one $\phi$, should be reciprocal on resonance.
	
	\subsection{Gauge transformations and phase rotations}
	For a dynamical matrix $H^{(a)}$ expressed in the basis of the fields $\{a_j, a_j^\dagger\}$, a gauge transformation performed on a multi-mode system with $N$ sites can be expressed via the unitary $V \equiv \bigoplus_{j=1}^N V_j$, with $V_j= \mathrm{diag}\,(e^{\mathrm{i}\phi_j},e^{-\mathrm{i}\phi_j})$ acting on an individual site. The dynamical matrix transforms according to $(H^{(a)})'(\{\phi_j\}) = V H^{(a)} V^\dagger$.
	Moving into the quadrature basis, $x_j=(a_j+a_j^\dagger)/\sqrt{2}$, $p_j=-\mathrm{i}(a_j-a_j^\dagger)/\sqrt{2}$, we transform $(H^{(a)})'$ via the unitary
	\begin{align}
	   W \equiv
	   \bigoplus_{j=1}^N\frac{1}{\sqrt{2}}
	   \begin{pmatrix}
	   1 & 1 \\
	   -\mathrm{i} & \mathrm{i}
	   \end{pmatrix},
	\end{align}
	such that the transformed dynamical matrix in the quadrature basis $H(\{\phi_j\})$ is given by
	\begin{align}
	   H(\{\phi_j\}) & = W V H^{(a)} V^\dagger W^\dagger = W V W^\dagger (W H^{(a)} W^\dagger) W V^\dagger W^\dagger = W V W^\dagger H(0) W V^\dagger W^\dagger \equiv U H(0) U^\dagger.
	\end{align}
	Therefore, in the quadrature basis, the dynamical matrix transforms under a gauge transformation according to
	\begin{align}
	   U \equiv W V W^\dagger
	   & =
	   \frac{1}{2}
	   \bigoplus_{j=1}^N
	   \begin{pmatrix}
	   1 & 1 \\
	   -\mathrm{i} & \mathrm{i}
	   \end{pmatrix}
	   \begin{pmatrix}
	   e^{\mathrm{i}\phi_j} & 0 \\
	   0 & e^{-\mathrm{i}\phi_j}
	   \end{pmatrix}
	   \begin{pmatrix}
	   1 & \mathrm{i} \\
	   \mathrm{i} & -\mathrm{i}
	   \end{pmatrix}
	   = \bigoplus_{j=1}^N
	   \begin{pmatrix}
	   \cos(\phi_j) & -\sin(\phi_j) \\
	   \sin(\phi_j) & \cos(\phi_j)
	   \end{pmatrix}
	   = \bigoplus_{j=1}^N R(\phi_j).
	\end{align}
	This is a rotation in phase space by $\phi_j$ for each mode $j$. Setting $\phi=\phi_j$ we obtain the transformation $U(\phi)= \oplus_{j=1}^N R(\phi)$ given in the main text.
	
	\subsection{The scattering matrix of a time-reversal symmetric system}
	\label{sec:TRSSMat}
	Here we show that a time-reversal symmetric Hamiltonian imposes certain properties on the scattering matrix of the system with added local gain and loss, namely there exists at least one set of gauges for which the system's transport is reciprocal on resonance.
	
	If the system preserves TRS, there exists a set of phases $\{\phi_j\}$ that make the Hamiltonian real~\cite{Koch2010_SI}. Formally, we can express this in terms of the Hamiltonian matrix $\mathscr{H}$
	\begin{align}
	   \mathcal{H} & = \sum_j \sum_\ell \left(
	      a_j^\dagger \mathscr{H}_{a_j^\dagger,a_\ell} a_\ell
	      +
	      a_j^\dagger \mathscr{H}_{a_j^\dagger,a_\ell^\dagger} a_\ell^\dagger
	      + \mathrm{H.c.}
	      \right).
	\end{align}
	TRS is preserved if we can find a gauge transformation V that maps $\mathscr{H}$ to its conjugate, i.e.~makes $\mathscr{H}$ real
	\begin{align}
	   (V \mathscr{H} V^\dagger)^* \stackrel{!}= V \mathscr{H} V^\dagger.
	   \label{eq:TRSHamiltonianMatrix}
	\end{align}
	Note that since $V$ is diagonal, we have $V=V^\mathrm{T}$ and $V^\dagger=V^*$.
	For the systems we are interested in here (coherent couplings with only local decay), the dynamical matrix is closely related to the Hamiltonian matrix
    \begin{align}
        H^{(a)} & = -\i \left(\oplus_{j=1}^N \sigma_z\right) \, \mathscr{H} - \frac{\Gamma}{2},
    \end{align}
    with the Pauli matrix $\sigma_z\equiv\begin{pmatrix}1&0\\0&-1\end{pmatrix}$ and dissipation matrix $\Gamma=\mathrm{diag}\,(\gamma_1,\gamma_1,\dots,\gamma_N,\gamma_N)$.
    Analogously,
    \begin{align}
        \mathscr{H} & = \i \left(\oplus_{j=1}^N \sigma_z\right) \, \left[H^{(a)} + \frac{\Gamma}{2} \right].
        \label{eq:connectionHamiltonianDynMat}
    \end{align}
    Applying the TRS condition~\eqref{eq:TRSHamiltonianMatrix}, we find an analogous relation for the dynamical matrix
    \begin{align}
        \left(V \left[H^{(a)}+\frac{\Gamma}{2}\right] V^*\right)^* = - V \left[H^{(a)}+\frac{\Gamma}{2}\right] V^*.
    \end{align}
    Converting this to an expression for the dynamical matrix in the quadrature basis
    \begin{align}
       V \left[H^{(a)}+\frac{\Gamma}{2}\right] V^* & = V W^\dagger W \left[H^{(a)}+\frac{\Gamma}{2}\right] W^\dagger W V^* = W^\dagger U \left[H+\frac{\Gamma}{2}\right] U^\mathrm{T} W \notag\\
       & \stackrel{!}= - \left(V\left[H^{(a)}+\frac{\Gamma}{2}\right]V^*\right)^* = - (W^\dagger \underbrace{U \left[H+\frac{\Gamma}{2}\right] U^\mathrm{T}}_{\in\mathbb{R}^{N\times N}} W)^* = - W^\mathrm{T} U \left[H+\frac{\Gamma}{2}\right] U^\mathrm{T} W^*.
    \end{align}
    Note that we used that $U$ is real such that $U^\dagger = U^\mathrm{T}$.   Using the identities
    \begin{align}
       WW^\mathrm{T} & = \bigoplus_{j=1}^N \frac{1}{2}
       \begin{pmatrix}
          1 & 1 \\ -\mathrm{i} & \mathrm{i}
       \end{pmatrix}
       \begin{pmatrix}
          1 & -\mathrm{i} \\ 1 & \mathrm{i}
       \end{pmatrix}
       = \bigoplus_{j=1}^N
       \sigma_z = Z, &&
       W^*W^\dagger  = \bigoplus_{j=1}^N \frac{1}{2}
       \begin{pmatrix}
          1 & 1 \\ \mathrm{i} & -\mathrm{i}
       \end{pmatrix}
       \begin{pmatrix}
          1 & \mathrm{i} \\ 1 & -\mathrm{i}
       \end{pmatrix}
       = \bigoplus_{j=1}^N
       \sigma_z = Z,
    \end{align}
    we obtain a TRS condition for the dynamical matrix
    \begin{align}
       \boxed{Z U \left[H+\frac{\Gamma}{2}\right]
       U^\mathrm{T} Z = - U \left[H+\frac{\Gamma}{2}\right]
       U^\mathrm{T}.}
       \label{eq:conditionHdynTRS}
    \end{align}
    
    In addition to TRS, we have a second requirement on $H$ which follows since $\mathscr{H}$ is Hermitian, $\mathscr{H}=\mathscr{H}^\dagger$. With Eq.~\eqref{eq:connectionHamiltonianDynMat}, we find
    \begin{align}
       \left[H^{(a)}+\frac{\Gamma}{2}\right]^\dagger Z = - Z \left[H^{(a)}+\frac{\Gamma}{2}\right]
       \quad\quad\Leftrightarrow\quad\quad
       Z \left[H^{(a)}+\frac{\Gamma}{2}\right] Z = - \left[H^{(a)}+\frac{\Gamma}{2}\right]^\dagger
    \end{align}
    with $Z\equiv\oplus_{j=1}^N \sigma_z$.
    It follows for the dynamical matrix of the field quadratures
    \begin{align}
       \left[H+\frac{\Gamma}{2}\right] & = - W Z W^\dagger \left[H+\frac{\Gamma}{2}\right]^\mathrm{T} W Z W^\dagger.
    \end{align}
    With the definition of $W$ and $Z$, we obtain
    \begin{align}
       W Z W^\dagger & = \bigoplus_{j=1}^N \frac{1}{2}
       \begin{pmatrix}
          1 & 1 \\
          -\mathrm{i} & \mathrm{i}
       \end{pmatrix}
       \begin{pmatrix}
          1 & 0 \\
          0 & -1
       \end{pmatrix}
       \begin{pmatrix}
          1 & \mathrm{i} \\
          1 & -\mathrm{i}
       \end{pmatrix}
       = \bigoplus_{j=1}^N
       \begin{pmatrix}
          0 & \mathrm{i} \\
          -\mathrm{i} & 0
       \end{pmatrix}
       \equiv \bigoplus_{j=1}^N \sigma_y^* \equiv Y^*,
    \end{align}
    with the Pauli matrix $\sigma_y$.
    Therefore, as a consequence of only coherent couplings and local dissipation, the dynamical matrix satisfies in the quadrature basis
    \begin{align}
       \boxed{\left[H+\frac{\Gamma}{2}\right] = - Y^* \left[H+\frac{\Gamma}{2}\right]^\mathrm{T} Y^*.}
       \label{eq:ConditionHdynHermitian}
    \end{align}
    
    Combining this result~\eqref{eq:ConditionHdynHermitian} with condition~\eqref{eq:conditionHdynTRS}, which we found as a consequence of TRS, the term $\Gamma/2$ drops out and we obtain
    \begin{align}
       Z U H U^\mathrm{T} Z = U Y^* H^\mathrm{T} Y^* U^\mathrm{T}.
       \label{eq:TRSCondition1}
    \end{align}
    Note that any diagonal matrix that we add to $H$ drops out in this step, so diagonal modifications, such as local decay, do not change the transformation properties of the dynamical matrix and hence of the susceptibility matrix.
    Now, examining $UY^*$ and $Y^*U^\mathrm{T}$, we find
    \begin{align}
       UY^* & = \bigoplus_{j=1}^N
       \begin{pmatrix}
          \cos\phi_j & -\sin\phi_j \\
          \sin\phi_j & \cos\phi_j
       \end{pmatrix}
       \begin{pmatrix}
          0 & \mathrm{i} \\
          -\mathrm{i} & 0
       \end{pmatrix}
       =  \mathrm{i} \, \bigoplus_{j=1}^N R(\phi_j-\pi/2)
       \equiv \mathrm{i} \, U(\{\phi_j-\pi/2\}), \\
       Y^*U^\mathrm{T} & = \bigoplus_{j=1}^N
       \mathrm{i}\begin{pmatrix}
          \cos(-[\phi_j-\pi/2]) & -\sin(-[\phi_j-\pi/2]) \\
          \sin(-[\phi_j-\pi/2]) & \cos(-[\phi_j-\pi/2])
       \end{pmatrix}
       = -\mathrm{i}\, \bigoplus_{j=1}^N R(-[\phi_j-\pi/2])
       \equiv -\mathrm{i}\, U(\{-[\phi_j-\pi/2]\}).
    \end{align}
    $U(\{\phi_j-\pi/2\})$ is a direct sum of rotation matrices that change the local phase of each mode $j$ by $\phi_j-\pi/2$.
    With this, condition~\eqref{eq:TRSCondition1} becomes
    \begin{align}
       Z U(\{\phi_j\}) H U(\{-\phi_j\}) Z & = U(\{\phi_j-\pi/2\}) H^\mathrm{T} U(\{-[\phi_j-\pi/2]\}).
    \end{align}
    Splitting the rotation matrices $U(\{\phi_j-\pi/2\})=U(\{-\pi/4\})U(\{\phi_j-\pi/4\})$
    and
    $U(\{-[\phi_j-\pi/2]\})=U(\{-[\phi_j-\pi/4]\})U(\{\pi/4\})$,
    and further realising that
    \begin{align}
       ZU(\{-\pi/4\}) & = \bigoplus_{j=!}^N
       \begin{pmatrix}
          1 & 0 \\
          0 & -1
       \end{pmatrix}
       \begin{pmatrix}
          \cos(\pi/4) & \sin(\pi/4) \\
          -\sin(\pi/4) & \cos(\pi/4)
       \end{pmatrix}
       =
       \begin{pmatrix}
          \cos(\pi/4) & \sin(\pi/4) \\
          \sin(\pi/4) & -\cos(\pi/4)
       \end{pmatrix}
       = U(\{\pi/4\}) Z, \\
       U(\{\pi/4\})Z & = \bigoplus_{j=!}^N
       \begin{pmatrix}
          \cos(\pi/4) & -\sin(\pi/4) \\
          \sin(\pi/4) & \cos(\pi/4)
       \end{pmatrix}
       \begin{pmatrix}
          1 & 0 \\
          0 & -1
       \end{pmatrix}
       =
       \begin{pmatrix}
          \cos(\pi/4) & \sin(\pi/4) \\
          \sin(\pi/4) & -\cos(\pi/4)
       \end{pmatrix}
       = ZU(\{-\pi/4\}),
    \end{align}
    we find
    \begin{align}
       Z U(\{\phi_j-\pi/4\}) H U(\{-[\phi_j-\pi/4]\}) Z & = U(\{\phi_j-\pi/4\}) H^\mathrm{T} U(\{-[\phi_j-\pi/4]\}).
    \end{align}
    We redefine
    $\tilde U \equiv U(\{\phi_j-\pi/4\})$
    and
    $\tilde U^{-1} = U^\mathrm{T} \equiv U(\{-[\phi_j-\pi/4]\})$
    which are rotation matrices $U$ in which each phase $\phi_j$ has been shifted by $-\pi/4$. $\tilde U$ therefore also constitutes a gauge transformation. This leads to a condition for the dynamical matrix of a TRS preserving system which only consists of coherent couplings and local dissipation
    \begin{align}
       \boxed{Z\tilde U H \tilde U^\mathrm{T} Z = (\tilde U H^\mathrm{T} \tilde U^\mathrm{T})^\mathrm{T}.}
    \end{align}
    In the final step, we consider how the susceptibility matrix transforms under a gauge transformation
    \begin{align}
       \tilde U \chi(\omega) \tilde U^\mathrm{T}
       & = (\mathrm{i}\omega\mathbb{1} + H)^{-1}
       = (\tilde U \mathrm{i}\omega\mathbb{1}\tilde U^\mathrm{T} + \tilde U H \tilde U^\mathrm{T})^{-1}
       = (\mathrm{i}\omega\mathbb{1} + Z \tilde U H^\mathrm{T} \tilde U^\mathrm{T}Z)^{-1}
       = Z \tilde U (\mathrm{i}\omega\mathbb{1} + H^\mathrm{T})^{-1} \tilde U^\mathrm{T}Z.
    \end{align}
    This yields the condition
    \begin{align}
        \boxed{\tilde U \chi(\omega) \tilde U^\mathrm{T} = Z (\tilde U \chi(-\omega)\tilde U^\mathrm{T})^\mathrm{T} Z.}
        \label{eq:chiMatConditionTRSCoherent}
    \end{align}
    Taking the absolute value for each element in the site basis and noting that $Z$ can only change the sign of each element but neither swaps entries nor changes their modulus, we obtain a condition for the susceptibility matrix of TRS preserving system
    \begin{align}
       \boxed{\left\lvert \tilde U \chi(\omega) \tilde U^\mathrm{T}\right\rvert
       = \left\lvert \tilde U \chi(-\omega) \tilde U^\mathrm{T}\right\rvert^\mathrm{T}.}
       \label{eq:TRSConditionSusceptibilityMatrix}
    \end{align}
    And on resonance,
    $\lvert \tilde U \chi(0) \tilde U^\mathrm{T}\rvert = \lvert \tilde U \chi(0) \tilde U^\mathrm{T}\rvert^\mathrm{T}$.
    This implies that for a TRS preserving there exists at least one gauge in which, on resonance, the system's transport is reciprocal, while off resonance, $\omega\neq 0$, $\omega\to-\omega$.
	
	\subsection{TRS results in equal transmission for pairs of quadratures}
	A general susceptibility matrix $\chi$ in the quadrature basis is of the form
	\begin{align}
	    \chi & = \begin{pmatrix}
	       \chi_{1,1} & \dots & \chi_{1,j} & \dots & \chi_{1,N} \\
	       \vdots & & & & \vdots \\
	       & & \ddots & & \\
	       \vdots & & & & \vdots \\
	       \chi_{N,1} & \dots & \chi_{N,j} & \dots & \chi_{N,N}
	    \end{pmatrix},
	\end{align}
	with blocks
	\begin{align}
	    \chi_{j,\ell} \equiv \begin{pmatrix}
	       \chi_{x_j, x_\ell} & \chi_{x_j, p_\ell} \\
	       \chi_{p_j, x_\ell} & \chi_{p_j, p_\ell}.
	    \end{pmatrix},
	\end{align}
	Here $\chi_{q_j, q_\ell}$ with $q\in\{x,p\}$ relates an input quadrature of $q_\ell$ to the steady state of $q_j$.
	
	Using the TRS conditions on the dynamical matrix $H$ that follow from combining only coherent couplings, Eqs.~\eqref{eq:ConditionHdynHermitian}, we obtain a condition on the susceptibility matrix
	\begin{align}
	    \chi = - Y^* \chi^\mathrm{T}(-\omega) Y^*,
	    \label{eq:chiOnlyCoherentCouplings}
	\end{align}
	or, specifically, for each block $\chi_{j,\ell}$, $\chi_{\ell,j} = -\sigma_y^* \chi_{j,\ell}^\mathrm{T} \sigma_y^*$. It follows for the elements of each block that
	\begin{align}
	    \chi_{\ell,j} =
	    \begin{pmatrix}
	       \chi_{x_\ell, x_j} & \chi_{x_\ell, p_j} \\
	       \chi_{p_\ell, x_j} & \chi_{p_\ell, p_j}
	    \end{pmatrix}
	    =
	    \begin{pmatrix}
	       0 & 1 \\
	       -1 & 0
	    \end{pmatrix}
	    \begin{pmatrix}
	       \chi_{x_j, x_\ell} & \chi_{p_j, x_\ell} \\
	       \chi_{x_j, p_\ell} & \chi_{p_j, p_\ell}
	    \end{pmatrix}
	    \begin{pmatrix}
	       0 & 1 \\
	       -1 & 0
	    \end{pmatrix}
	    =
	    \begin{pmatrix}
	       - \chi_{p_j, p_\ell} & \chi_{x_j, p_\ell} \\
	       \chi_{p_j, x_\ell} & - \chi_{x_j, x_\ell}
	    \end{pmatrix}.
	\end{align}
	On the other hand, assuming that TRS is preserved imposes another condition on $\chi$, namely, there exists a gauge in which $\chi$ satisfies condition~\eqref{eq:chiMatConditionTRSCoherent} such that the individual blocks $\chi_{j,\ell}$ are constrained by
	\begin{align}
	    \chi_{\ell,j}^{\mathrm{r}} & = \sigma_z (\chi_{j,\ell}^\mathrm{r})^\mathrm{T} \sigma_z
	    =
	    \begin{pmatrix}
	       \chi_{x_j, x_\ell}^\mathrm{r} & -\chi_{p_j, x_\ell}^\mathrm{r} \\
	       -\chi_{x_j, p_\ell}^\mathrm{r} & \chi_{p_j, p_\ell}^\mathrm{r}
	    \end{pmatrix}.
	\end{align}
	Here, the superscript $\mathrm{r}$ refers to the set of local gauges $ \{\phi_1,\dots,\phi_N\}$ in which $\chi$ is reciprocal, see Eq.~\eqref{eq:TRSConditionSusceptibilityMatrix}.
	
	Equating these two constraints for the elements of $\chi_{j,\ell}$ (in the gauge in which $\chi$ is reciprocal), we obtain the following conditions
	\begin{align}
	    \chi_{x_j, x_\ell}^\mathrm{r} = -\chi_{p_j, p_\ell}^\mathrm{r} = t_1^\mathrm{r}
	    \quad\quad & \text{and}\quad\quad
	    \chi_{x_j, p_\ell}^\mathrm{r} = -\chi_{p_j, x_\ell}^\mathrm{r} = t_2^\mathrm{r}, 
	    \intertext{with $t_{1,2}^\mathrm{r}$ the susceptibility matrix element at the reciprocal gauge. Similarly,}
	    \chi_{j,\ell} =
	    \begin{pmatrix}
	       t_1^\mathrm{r} & t_2^\mathrm{r} \\
	       -t_2^\mathrm{r} & -t_1^\mathrm{r}
	    \end{pmatrix},
	    \quad\quad & \text{and}\quad\quad
	    \chi_{\ell,j} =
	    \begin{pmatrix}
	       t_1^\mathrm{r} & t_2^\mathrm{r} \\
	       -t_2^\mathrm{r} & -t_1^\mathrm{r}
	    \end{pmatrix}.
	    \label{eq:TRSConstrainedChi}
	\end{align}
	with $t_{1,2}^\mathrm{r}$ the susceptibility matrix element in the reciprocal gauge.
	We now consider how the matrix block $\chi_{j,\ell}$ changes under a gauge transformation $U=\oplus_{j=1}^N R(\phi_j)$ which acts on each individual block according to
	\begin{align}
	   \chi_{j,\ell}(\phi_j, \phi_\ell)  = R(\phi_j) \chi_{j,\ell}^\mathrm{r} R(-\phi_\ell),
	   \chi_{\ell,j}(\phi_j, \phi_\ell)  = R(\phi_\ell) \chi_{\ell,j}^\mathrm{r} R(-\phi_j).
	\end{align}
	Applying these rotations to the blocks of the susceptibility matrix~\eqref{eq:TRSConstrainedChi} and the resulting matrix entries, we obtain constraints on the susceptibility matrix in any gauge $\chi(\phi_1,\dots,\phi_N)$
	\begin{alignat}{2}
	    \chi_{x_j,x_\ell}(\phi_1,\dots,\phi_N) & = - \chi_{p_\ell,p_j}(\phi_1,\dots,\phi_N), \quad\quad\quad
	    \chi_{p_j,p_\ell}(\phi_1,\dots,\phi_N) && = - \chi_{x_\ell,x_j}(\phi_1,\dots,\phi_N), \notag \\
	    \chi_{x_j,p_\ell}(\phi_1,\dots,\phi_N) & = \chi_{x_\ell,p_j}(\phi_1,\dots,\phi_N), \quad\quad\quad
	    \chi_{p_j,x_\ell}(\phi_1,\dots,\phi_N) && = \chi_{p_\ell,x_j}(\phi_1,\dots,\phi_N).
	\end{alignat}
	This does not automatically imply reciprocity, $\lvert\chi\rvert=\lvert\chi\rvert^\mathrm{T}$, and gives room for qNR, while the transmission between certain pairs of quadratures is the same in opposite directions.

	\subsection{Equal transmission for pairs of quadratures implies TRS}
	We now prove that the previous statement actually applies in both directions, i.e.~there exists an equivalence between TRS and the equal transmission of certain pairs of quadratures in opposite directions.
	We start from a system whose susceptibility matrix satisfies for any set of gauges $\{\phi_j\}$
	\begin{alignat}{2}
	    \lvert\chi_{x_j,x_\ell}(\phi_1,\dots,\phi_N)\rvert & = \lvert\chi_{p_\ell,p_j}(\phi_1,\dots,\phi_N)\rvert, \quad\quad\quad
	    \lvert\chi_{p_j,p_\ell}(\phi_1,\dots,\phi_N)\rvert && = \lvert\chi_{x_\ell,x_j}(\phi_1,\dots,\phi_N)\rvert, \notag \\
	    \lvert\chi_{x_j,p_\ell}(\phi_1,\dots,\phi_N)\rvert & = \lvert\chi_{x_\ell,p_j}(\phi_1,\dots,\phi_N)\rvert, \quad\quad\quad
	    \lvert\chi_{p_j,x_\ell}(\phi_1,\dots,\phi_N)\rvert && = \lvert\chi_{p_\ell,x_j}(\phi_1,\dots,\phi_N)\rvert,
	    \label{eq:transmissionConditionsModulus}
	\end{alignat}
	and we ask if this always implies TRS. Indeed, we can answer this question in the affirmative provided that all the couplings in the system are coherent, i.e.~Eq.~\eqref{eq:chiOnlyCoherentCouplings} is valid\footnote{If we examine a system with dissipative couplings, our arguments can still be applied if we expand the dissipative coupling into coherent coupling via an auxiliary lossy bath mode which, if eliminated, would yield the desired non-local dissipator.}.
	Conditions~\eqref{eq:transmissionConditionsModulus} only fix the modulus of the elements of matrix blocks $\chi_{j,\ell}$ and $\chi_{\ell,j}$ such that they can, in full generality, be written as
	\begin{align}
	    \chi_{j,\ell} & = 
	    \begin{pmatrix}
	       \chi_{x_j, x_\ell} & \chi_{p_j, x_\ell} \\
	       \chi_{x_j, p_\ell} & \chi_{p_j, p_\ell}
	    \end{pmatrix},
	    \quad\quad\quad
	    \chi_{\ell,j} = 
	    \begin{pmatrix}
	       \chi_{p_j, p_\ell} e^{\mathrm{i}\xi^{(1)}_{j,\ell}} & \chi_{x_j, p_\ell} e^{\mathrm{i}\xi^{(2)}_{j,\ell}} \\
	       \chi_{p_j, x_\ell} e^{\mathrm{i}\xi^{(3)}_{j,\ell}} & \chi_{x_j, x_\ell} e^{\mathrm{i}\xi^{(4)}_{j,\ell}}
	    \end{pmatrix}
	\end{align}
	with some phases
	$\xi^{(1)}_{j,\ell} = - \xi^{(4)}_{\ell,j}$, $\xi^{(2)}_{j,\ell} = - \xi^{(3)}_{\ell,j}$.
	Since conditions~\eqref{eq:transmissionConditionsModulus} should hold for any gauges $\{\phi_j\}$, we check how these matrices transform and under which conditions for $\xi^{(1,2)}_{j,\ell}$ the susceptibility matrix satisfies conditions~\eqref{eq:transmissionConditionsModulus} in any gauge. We find that we need to require
	\begin{align}
	    \xi^{(1)}_{j,\ell} = \delta_{j,\ell},
	    \quad
	    \xi^{(2)}_{j,\ell} = \pi + \delta_{j,\ell},
	    \quad
	    \xi^{(3)}_{j,\ell} = \pi + \delta_{j,\ell},
	    \quad
	    \xi^{(4)}_{j,\ell} = \delta_{j,\ell}
	\end{align}
	with $\delta_{j,\ell}=\pi$ corresponding to the TRS preserving case.
	
	Furthermore, condition~\eqref{eq:chiOnlyCoherentCouplings} has to be fulfilled, i.e.~$\chi_{\ell,j} = -\sigma_y^* \chi_{j,\ell}^\mathrm{T} \sigma_y^*$,
	\begin{align}
	    \sigma_y^* \chi_{\ell,j} \sigma_y^*
	    = \sigma_y^*
	    \begin{pmatrix}
	       \chi_{p_j, p_\ell} e^{\mathrm{i}\delta_{j,\ell}} & -\chi_{x_j, p_\ell} e^{\mathrm{i}\delta_{j,\ell}} \\
	       -\chi_{p_j, x_\ell} e^{\mathrm{i}\delta_{j,\ell}} & \chi_{x_j, x_\ell} e^{\mathrm{i}\delta_{j,\ell}}
	    \end{pmatrix}
	    \sigma_y^*
	    =
	    \begin{pmatrix}
	       -\chi_{x_j, x_\ell} e^{\mathrm{i}\delta_{j,\ell}} & -\chi_{x_j, p_\ell} e^{\mathrm{i}\delta_{j,\ell}} \\
	       -\chi_{p_j, x_\ell} e^{\mathrm{i}\delta_{j,\ell}} & -\chi_{p_j, p_\ell} e^{\mathrm{i}\delta_{j,\ell}}
	    \end{pmatrix}
	    \stackrel{!}=
	    \begin{pmatrix}
	       -\chi_{x_j, x_\ell} & -\chi_{x_j, p_\ell} \\
	       -\chi_{p_j, x_\ell} & -\chi_{p_j, p_\ell}
	    \end{pmatrix}
	    = - \chi_{j,\ell}.
	\end{align}
	$\delta_{j,\ell}=\pi$ is the only solution which automatically corresponds to a TRS preserving system. Therefore, conditions~\eqref{eq:transmissionConditionsModulus} can only be satisfied if TRS is preserved.
	
	\section{TRS criterion---full loops vs. disjoint loops}
	Here, we prove the statement of the main text that TRS can be identified from the graph representing the system's dynamical matrix in the field basis $H^{(a)}$.
	For this analysis, we break the system down into its minimal simple rings, i.e.~such that there is only one coupling per mode pair (with the exception of the two-mode ring).
	
	The graph representing the block of the dynamical matrix of this ring then forms one full loop, or consists of disjoint loops.
	We now show that full loops identify systems that always preserve TRS, while graphs displaying two disjoint loops indicate systems that allow to break TRS for certain choices of phases.
	
	As coherent couplings, we first focus on systems with only BS or TMS terms
	\begin{align}
	   \mathcal{H} & \equiv \sum_{j,\ell} J_{j,\ell} e^{\i\theta_{j,\ell}} a_j^\dagger a_\ell + \sum_{m,n} \lambda_{m,n} e^{\i\psi_{m,n}} a_m a_n + \mathrm{H.c.}
	   \label{eq:HamiltonianPlaquette}
	\end{align}
	In a later step, we extend the argument to systems that include local squeezing.
	TRS implies that there exists a gauge transformation, $a_j\to a_j e^{\mathrm{i}\phi_j}$, that makes the coefficients of the Hamiltonian real, i.e.,~we need to check, if we can find a set of  $\{\phi_j\}$ that removes any possible phases $\theta_{j,\ell}$, $\psi_{j,\ell}$. By asking if TRS is preserved, we ask if the following system of equations has a solution for $\phi_j$
    \begin{align}
       \theta_{j,\ell}-\phi_j+\phi_\ell & = 0, \label{eq:phasesBS} \\
       \psi_{j,\ell}+\phi_j+\phi_\ell & = 0. \label{eq:phasesTMS}
    \end{align}
    For a simple ring with $N$ modes, the number of equations above equals $N$.
    This system of equations can be expressed in terms of the adjacency matrix $A$ of a signed, directed graph representing the graphs of Fig.~3 in the main text, in which $A_{j,\ell} = -1$ for BS and $j<\ell$, and $A_{j,\ell} = +1$ for TMS couplings and $j<\ell$. $A_{j,\ell} = 0$ otherwise. Eqs.~\eqref{eq:phasesBS} and \eqref{eq:phasesTMS} expressed in terms of the adjacency matrix become
    \begin{align}
       (\mathbb{1} + \bm A) \bm{\Phi} = \bm{\Theta}
    \end{align}
    with $\bm{\Phi} \equiv (\phi_1,\dots,\phi_N)^\mathrm{T}$ and $\bm{\Theta}$ a vector containing the phases $\theta_{j,\ell}$, $\psi_{j,\ell}$.
    If TRS is preserved, the system of Eqs.~\eqref{eq:phasesBS} and \eqref{eq:phasesTMS} has at least one solution, i.e.,~$(\mathbb{1} + \bm A)$ has full rank, if TRS can be broken via a non-vanishing flux (some non-vanishing $\theta_{j,\ell}$ or $\psi_{j,\ell}$), these equations do not have a solution.
    
    By recursively replying Gaussian elimination we prove next that full loops signal that TRS is preserved for any  $\theta_{j,\ell}$, $\psi_{j,\ell}$, while two disjoint loops allow to break TRS through non-vanishing phases $\theta_{j,\ell}$, $\psi_{j,\ell}$.
    
\subsection{Disjoint loops allow to break TRS}
First, we show here that if the graph breaks up into two disjoint loops, the system can break TRS.
Recursively applying Gaussian elimination to $(\mathbb{1}+\bm A)$, we show that the matrix rank is smaller than the number of modes $N$ when the graph decomposes into two loops. If all sites are either coupled via beamsplitter of parametric interactions, we can, without loss of generality, reorder the matrix $(\mathbb{1}+\bm A)$, so that it takes the following form
\begin{align}
   (\mathbb{1}+\bm A) & = \begin{pmatrix}
      1 & \pm 1 & 0 & 0 & \dots & 0 & 0 \\
      0 & 1 & \pm 1 & 0 & \dots & 0 & 0 \\
      0 & 0 & 1 & \pm 1 & \dots & 0 & 0 \\
      & & & & \ddots & & & \\
      0 & 0 & 0 & 0 & \dots & 1 & \pm 1 \\
      1 & 0 & 0 & 0 & \dots & 0 & \pm 1 \\
   \end{pmatrix}.
\end{align}
We start by examining systems whose graphs decompose into two disjoint graphs. We note that this requires an even number of parametric couplings (pairs of $+1$, $+1$ in $\bm\Phi$) and any number of beamsplitter couplings.

\subsubsection{Beamsplitter coupling followed by beamsplitter interaction:}
We consider the case that both mode pairs $j$ and $(j-1)$ as well as $j$ and $(j+1)$ are coupled via BS links.
In this case, we can eliminate the $j$th row in $(\mathbb{1}+\bm A)$ such that the resulting matrix looks like a direct connection between $(j-1)$ and $(j+1)$
\begin{align}
   (\mathbb{1}+\bm A) & = 
   \begin{pmatrix}
      \ddots & & & & \\
      \dots  & 1 & -1 & 0 & \dots \\
      \dots  & 0 & 1 & -1 & \dots \\
      & & & \ddots &
   \end{pmatrix}
   \to
   \begin{pmatrix}
      \ddots & & & & \\
      \dots  & 1 & -1 & 0 & \dots \\
      \dots  & 1 & 0 & -1 & \dots \\
      & & & \ddots &
   \end{pmatrix}.
\end{align}
If the loop \emph{only} consists of beamsplitter interactions, then we can repeat this elimination step throughout the matrix until the last row equals the second to last row. Subsequently, the last row can be eliminated, the system does not have full rank, therefore no solution and we can, through a non-vanishing phase $\theta_{j,\ell}$, $\psi_{j,\ell}$ break TRS. This is a well know result~\cite{Koch2010_SI}.

\subsubsection{Beamsplitter coupling followed by parametric coupling (or vice versa):}
Here we can show that beamsplitter coupling between $(j-1)$ and $j$ followed by parametric coupling between $j$ and $(j+1)$ can, in $(\mathbb{1}+\bm A)$, be reduced to a row that has the same form as parametric coupling between $(j-1)$ and $(j+1)$
\begin{align}
   (\mathbb{1}+\bm A) & = 
   \begin{pmatrix}
      \ddots & & & & \\
      \dots  & 1 & -1 & 0 & \dots \\
      \dots  & 0 & 1 & 1 & \dots \\
      & & & \ddots &
   \end{pmatrix}
   \to
   \begin{pmatrix}
      \ddots & & & & \\
      \dots  & 1 & -1 & 0 & \dots \\
      \dots  & 1 & 0 & 1 & \dots \\
      & & & \ddots &
   \end{pmatrix}.
\end{align}

\subsubsection{An even number of parametric couplings:}
An even number of parametric couplings looks like beamsplitter couplings. If they follow right after each other, the following argument holds:
\begin{align}
   (\mathbb{1}+\bm A) & = 
   \begin{pmatrix}
      \ddots & & & & \\
      \dots  & 1 & 1 & 0 & \dots \\
      \dots  & 0 & 1 & 1 & \dots \\
      & & & \ddots &
   \end{pmatrix}
   \to
   \begin{pmatrix}
      \ddots & & & & \\
      \dots  & 1 & 1 & 0 & \dots \\
      \dots  & 1 & 0 & -1 & \dots \\
      & & & \ddots &
   \end{pmatrix}.
\end{align}
If there are beamsplitter couplings in between, these can be reduced firstly, for instance,
\begin{align}
   (\mathbb{1}+\bm A) & =
   \begin{pmatrix}
      \ddots & & & & \\
      \dots  & 1 & 1 & 0 & \dots & 0 & 0 & 0 & \dots \\
      \dots  & 0 & 1 & -1 & \dots & 0 & 0 & 0 & \dots \\
             & & & & \ddots & & & & \\
      \dots  & 0 & 0 & 0 & \dots & 1 & -1 & 0 & \dots \\
      \dots  & 0 & 0 & 0 & \dots & 0 & 1 & 1 & \dots \\
       & & & & & & & \ddots
   \end{pmatrix}\notag
   \to
   \begin{pmatrix}
      \ddots & & & & \\
      \dots  & 1 & 1 & 0 & \dots & 0 & 0 & 0 & \dots \\
      \dots  & 1 & 0 & 1 & \dots & 0 & 0 & 0 & \dots \\
             & & & & \ddots & & & & \\
      \dots  & 0 & 0 & 0 & \dots & 1 & -1 & 0 & \dots \\
      \dots  & 0 & 0 & 0 & \dots & 0 & 1 & 1 & \dots \\
       & & & & & & & \ddots
   \end{pmatrix} \\
   & \to \dots \to
   \begin{pmatrix}
      \ddots & & & & \\
      \dots  & 1 & 1 & 0 & \dots & 0 & 0 & 0 & \dots \\
      \dots  & 1 & 0 & 1 & \dots & 0 & 0 & 0 & \dots \\
             & & & & \ddots & & & & \\
      \dots  & 1 & 0 & 0 & \dots & 0 & 1 & 0 & \dots \\
      \dots  & 0 & 0 & 0 & \dots & 0 & 1 & 1 & \dots \\
       & & & & & & & \ddots
   \end{pmatrix}
   \to
   \begin{pmatrix}
      \ddots & & & & \\
      \dots  & 1 & 1 & 0 & \dots & 0 & 0 & 0 & \dots \\
      \dots  & 1 & 0 & 1 & \dots & 0 & 0 & 0 & \dots \\
             & & & & \ddots & & & & \\
      \dots  & 1 & 0 & 0 & \dots & 0 & 1 & 0 & \dots \\
      \dots  & 1 & 0 & 0 & \dots & 0 & 0 & -1 & \dots \\
       & & & & & & & \ddots
   \end{pmatrix}.
\end{align}
Therefore, for an even number of parametric couplings, the matrix $(\mathbb{1}+\bm A)$ can be reduced such that the last and second to last row are the same (both look like beamsplitter couplings between mode $1$ and $N$) and one row can be eliminated. Thus, the matrix does not have full rank.

\subsubsection{Single-mode squeezing terms/ local parametric terms:}
Examining the graph of a system containing single-mode squeezing $\propto \eta_j e^{\psi_j} a_j^2 +\mathrm{H.c.}$, we note that as soon as the single-mode squeezing is part of a simple ring, it decomposes the graph into two disjoint loops. This is because single-mode squeezing closes the loop on the same site\footnote{In this case, the number of rows of $(\mathbb{1}+\bm A)$ exceeds the number of modes in the ring.}.
In this case, TRS can again be broken through non-vanishing phases, since single-mode squeezing acts locally and fixes one phase $\phi_j$ of one of the operators $a_j$ in Eq.~\eqref{eq:HamiltonianPlaquette} (e.g.~if the Hamiltonian contains a term $\eta_j e^{\psi_j} a_j^2+\mathrm{H.c.}$ then this fixes the phase $\phi_j=\psi_j/2$ of $a_j$).

\subsubsection{Multiple couplings per link in a single ring}
The only simple ring with multiple (distinct) couplings per link, is the qNR dimer (main text), combining BS and TMS couplings which does not break TRS, since
\begin{align}
   (\mathbb{1}+\bm A) = \begin{pmatrix}
      1 & -1 \\
      1 & 1
   \end{pmatrix}
   \to
   \begin{pmatrix}
      1 & 0 \\
      0 & 1
   \end{pmatrix},
\end{align}
has full rank.
This is the only possibility for multiple links in a simple ring.

\subsection{Full loops and their conglomeration preserve TRS}
We noted that disjoint loops require an even number of TMS, while a full loop occurs for an odd number of parametric couplings. In the previous calculations we showed that TRS can be broken when the number of TMS couplings is even, by bringing $(\mathbb{1}+\bm A)$ to a form in which the second-to-last and last row were equal so that we could eliminate one to show that the matrix does not have full rank. With an odd number of couplings the second-to-last and last row are not equal, so that we cannot eliminate the last row and the matrix has full rank. This implies that TRS is always preserved when the graph of the ring displays one full loop.

In a system consisting of multiple rings, we can show using the same elimination technique that as soon as one ring with a graph displaying two disjoint loops is contained, TRS can be broken. Otherwise, TRS is automatically preserved.
	
\section{Deriving the conditions for qNR}
We stated in the main text that TRS alone is not sufficient for qNR, we also require the ring to consist of an even number of modes to ensure that all modes can pair up. Here, we prove that a ring of an even number of modes is the \emph{only} possibility for qNR.
For that purpose, we employ a reduction technique similar to the one we used in the previous section for the phases, only now, we work directly with the equations of motions and eliminate intermediate sites. This is mathematically analogous to the process of adiabatic elimination but is no approximation. A similar technique has recently been developed in Ref.~\cite{Ranzani2015_SI,Naaman2022_SI}.
Mathematically, our reduction technique performs an LU-decomposition.

\begin{figure}[htbp]
   \centering
   \includegraphics[width=\textwidth]{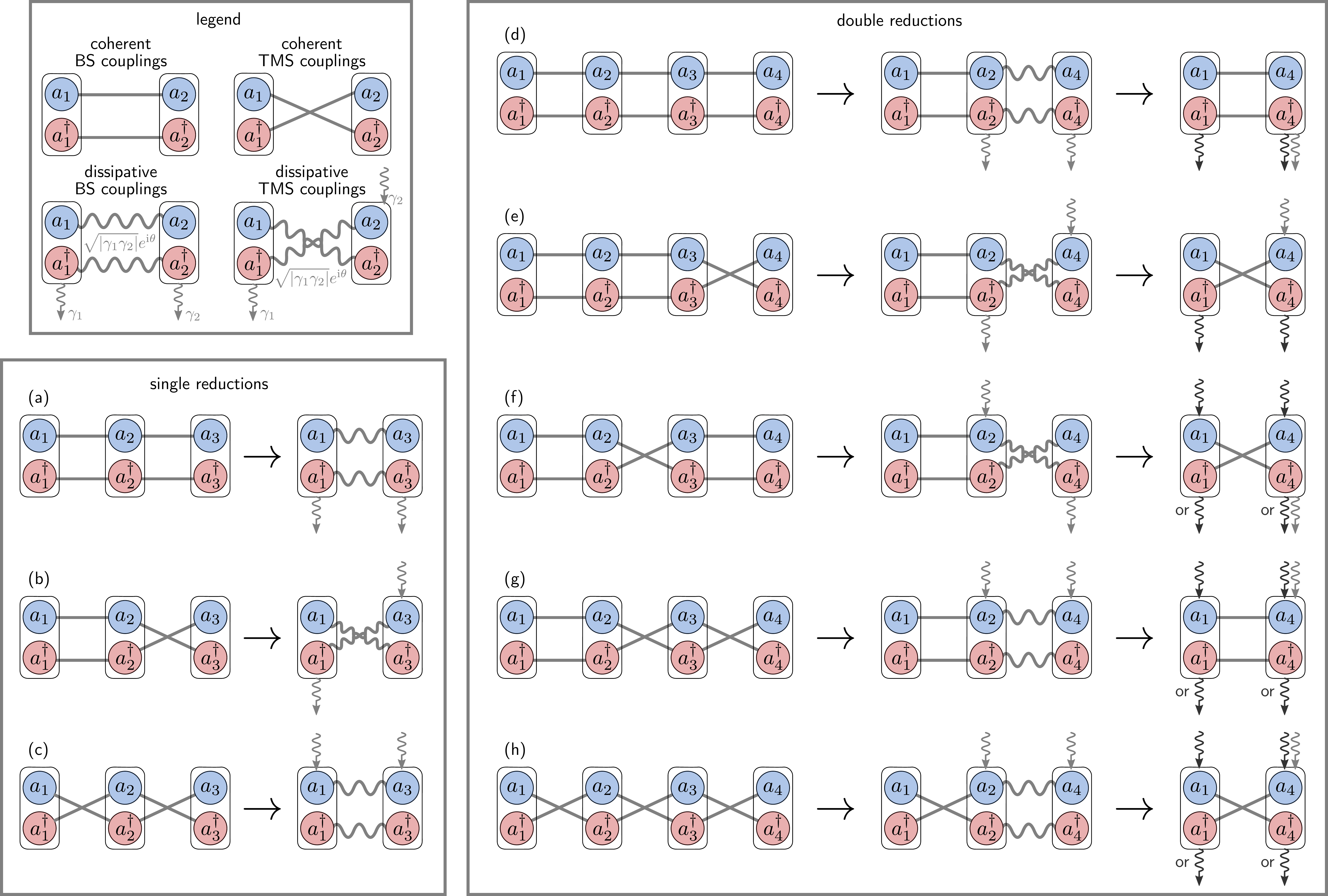}
   \caption{\textbf{Reduction rules.}
   Operators $a_i$ (or similarly amplitudes $\alpha_i=\braket{a_i}$) that are coupled in the equations of motion are connected via lines. Straight lines indicate coherent couplings while wiggled lines symbolise dissipative couplings, i.e.~couplings that appear with the same pre-factor as a decay rate. (a)-(c)~An odd number of reduction steps produces incoherent couplings, (d)-(h)~while an even number of reductions maps back to coherent couplings so that these rules can be applied recursively. Each reduction introduces local decay (gain) indicated by the wiggly lines pointing away (towards) the mode in question. The added decay/ gain is related to the effective coupling strength after elimination as indicated in the legend.
   }
   \label{fig:reductionRules}
\end{figure}%

First, we derive the reduction rules reducing any possible TRS preserving rings to four elementary systems. Examining the properties of each of these systems, we find only one displays qNR and hence any system that can be reduced to this system, displays qNR. Focusing on simple rings, there is only a finite number of possibilities how different couplings can be combined. We derive a set of rules how to contract these couplings by eliminating intermediate modes. Specifically, we show that there is a difference in the number of contractions we perform: a even number of contraction steps manifests in effectively dissipative coupling, i.e.~the coupling enters with the same prefactors as a decay rate would, while an even number of reduction steps yields effectively coherent coupling with added dissipation or gain. We therefore only need to check reductions in which we eliminate a single mode (single reductions) and such in which we eliminate, one after the other, two modes (double reductions), see Fig.~\ref{fig:reductionRules}. After this, the process repeats recursively.

\subsection{Single reductions}
There are three possible single reductions: BS-BS, BS-TMS and TMS-TMS. The results of these reductions are summarised step by step in Figs.~\ref{fig:reductionRules}~(a)-(c). In the following, the coupling strengths are numbered from left to right according to Fig.~\ref{fig:reductionRules} and $J_j$ stands for BS coupling between sites $j$ and $j+1$, while $\lambda_j$ stands for TMS coupling between sites $j$ and $j+1$. Each mode is assumed to have decay rate $\gamma_j$. In the interest of brevity, we consider first moments in the fields only and denote them by $\alpha_j\equiv\langle a_j\rangle$. Extensions beyond first moments are possible along similar lines.

\subsubsection{BS-BS}
The Langevin equations of motions for subsequent BS coupling according to Fig.~\eqref{fig:reductionRules}~(a) are given by
\begin{align}
   \dot \alpha_1 & = -\frac{\gamma_1}{2} \alpha_1 -\mathrm{i} J_1 \alpha_2 - \sqrt{\gamma_1} \alpha_{1,\mathrm{in}} \\
   \dot \alpha_2 & = -\frac{\gamma_2}{2} \alpha_2 - \mathrm{i} J_1^* \alpha_1 - \mathrm{i} J_2 \alpha_3 - \sqrt{\gamma_2} \alpha_{2,\mathrm{in}} \\
   \dot \alpha_3 & = -\frac{\gamma_3}{2} \alpha_3 - \mathrm{i} J_2^* \alpha_2 - \sqrt{\gamma_3} \alpha_{3,\mathrm{in}}.
\end{align}
Fourier transforming $\alpha_j(t)\equiv \int_{-\infty}^\infty \alpha_j(\omega) e^{-\mathrm{i}\omega t}$, and inserting the expression for $\alpha_2(\omega)$ into the equations for $\alpha_1(\omega)$ and $\alpha_3(\omega)$, we obtain,
\begin{align}
   \begin{pmatrix} -\mathrm{i}\omega\alpha_1(\omega) \\ -\mathrm{i}\omega\alpha_3(\omega) \end{pmatrix}
   & =
   \underbrace{
   \begin{pmatrix}
   -\left(\frac{\gamma_1}{2}
      + \frac{\lvert J_1\rvert^2}{\frac{\gamma_2}{2}-\mathrm{i}\omega}\right)
      &
   - \frac{J_1 J_2}{\frac{\gamma_2}{2}-\mathrm{i}\omega} \\[2ex]
   - \frac{J_1^* J_2^*}{\frac{\gamma_2}{2}-\mathrm{i}\omega}
   &
   -\left(\frac{\gamma_3}{2}
      + \frac{\lvert J_2\rvert^2}{\frac{\gamma_2}{2}-\mathrm{i}\omega}\right)
   \end{pmatrix}
   }_{\equiv H_\mathrm{red}^{(a)}(\omega)}
   \begin{pmatrix}\alpha_1(\omega) \\ \alpha_3(\omega) \end{pmatrix}
   +
   \begin{pmatrix}
   \mathrm{i} \frac{J_1\sqrt{\gamma_2}}{\frac{\gamma_2}{2}-\mathrm{i}\omega} \alpha_{2,\mathrm{in}}(\omega) 
   - \sqrt{\gamma_1} \alpha_{1,\mathrm{in}}(\omega) \\[2ex]
   \mathrm{i} \frac{J_2^*\sqrt{\gamma_2}}{\frac{\gamma_2}{2}-\mathrm{i}\omega} \alpha_{2,\mathrm{in}}(\omega)
   - \sqrt{\gamma_3} \alpha_{3,\mathrm{in}}(\omega)
   \end{pmatrix}
\end{align}
where we identify a new dynamical matrix $H_\mathrm{red}^{(a)}(\omega)$ in the frequency domain.

The susceptibility matrix of this reduced system is calculated from $H_\mathrm{red}^{(a)}(\omega)$ in the normal way via $\chi_\mathrm{red}^{(a)}(\omega)=(\mathrm{i}\omega\mathbb{1}+H_\mathrm{red}^{(a)}(\omega))^{-1}$ so $H_\mathrm{red}^{(a)}(\omega)$ indeed plays the role of a dynamical matrix.
The elimination of mode $2$ has introduced an additional local decay term
$\frac{\lvert J_1\rvert^2}{\frac{\gamma_2}{2}-\mathrm{i}\omega}$
(which is a decay rate on resonance, $\omega=0$)
and coupling between modes $1$ and $3$ which lacks the typical factor $\mathrm{i}$ but instead enters with the same sign as the decay rate and which we therefore refer to as dissipative coupling. The resulting dynamical matrix is graphically represented in Fig.~\ref{fig:reductionRules}~(a).

\subsubsection{BS-TMS}
We proceed in a similar spirit for the other possible cases.
For a system of modes coupled via BS followed by TMS couplings, the relevant equations of motions are given by
\begin{align}
   \dot \alpha_1 & = -\frac{\gamma_1}{2} \alpha_1 -\mathrm{i} J_1 \alpha_2 - \sqrt{\gamma_1} \alpha_{1,\mathrm{in}} \\
   \dot \alpha_2 & = -\frac{\gamma_2}{2} \alpha_2 - \mathrm{i} J_1^* \alpha_1 - \mathrm{i} \lambda_2 \alpha_3^\dagger - \sqrt{\gamma_2} \alpha_{2,\mathrm{in}} \\
   \dot \alpha_3^\dagger & = -\frac{\gamma_3}{2} \alpha_3 + \mathrm{i} \lambda_2^* \alpha_2 - \sqrt{\gamma_3} \alpha_{3,\mathrm{in}}^\dagger.
\end{align}
Eliminating again mode $2$, the only key difference compared to the previous case is the sign in front of the effective coupling and the new local term
\begin{align}
   \begin{pmatrix} -\mathrm{i}\omega\alpha_1(\omega) \\ -\mathrm{i}\omega\alpha_3^\dagger(\omega) \end{pmatrix}
   & =
   \underbrace{
   \begin{pmatrix}
   -\left(\frac{\gamma_1}{2}
      + \frac{\lvert J_1\rvert^2}{\frac{\gamma_2}{2}-\mathrm{i}\omega}\right)
      &
   - \frac{J_1 \lambda_2}{\frac{\gamma_2}{2}-\mathrm{i}\omega} \\[2ex]
   \frac{J_1^* \lambda_2^*}{\frac{\gamma_2}{2}-\mathrm{i}\omega}
   &
   -\left(\frac{\gamma_3}{2}
      - \frac{\lvert \lambda_2\rvert^2}{\frac{\gamma_2}{2}-\mathrm{i}\omega}\right)
   \end{pmatrix}
   }_{\equiv H_\mathrm{red}^{(a)}(\omega)}
   \begin{pmatrix}\alpha_1(\omega) \\ \alpha_3^\dagger(\omega) \end{pmatrix}
   +
   \begin{pmatrix}
   \mathrm{i} \frac{J_1\sqrt{\gamma_2}}{\frac{\gamma_2}{2}-\mathrm{i}\omega} \alpha_{2,\mathrm{in}}(\omega) 
   - \sqrt{\gamma_1} \alpha_{1,\mathrm{in}}(\omega) \\[2ex]
   -\mathrm{i} \frac{\lambda_2^*\sqrt{\gamma_2}}{\frac{\gamma_2}{2}-\mathrm{i}\omega} \alpha_{2,\mathrm{in}}(\omega)
   - \sqrt{\gamma_3} \alpha_{3,\mathrm{in}}^\dagger(\omega)
   \end{pmatrix}.
\end{align}
The effective dynamical matrix in this case is shown in Fig.~\ref{fig:reductionRules}~(b).

\subsubsection{TMS-TMS}
Combining a series of TMS couplings, we obtain the following equations of motion
\begin{align}
   \dot \alpha_1 & = -\frac{\gamma_1}{2} \alpha_1 -\mathrm{i} \lambda_1 \alpha_2^\dagger - \sqrt{\gamma_1} \alpha_{1,\mathrm{in}} \\
   \dot \alpha_2^\dagger & = -\frac{\gamma_2}{2} \alpha_2 + \mathrm{i} \lambda_1^* \alpha_1 + \mathrm{i} \lambda_2^* \alpha_3 - \sqrt{\gamma_2} \alpha_{2,\mathrm{in}}^\dagger \\
   \dot \alpha_3 & = -\frac{\gamma_3}{2} \alpha_3 - \mathrm{i} \lambda_2 \alpha_2 - \sqrt{\gamma_3} \alpha_{3,\mathrm{in}}.
\end{align}
Eliminating mode $2$
\begin{align}
   \begin{pmatrix} -\mathrm{i}\omega\alpha_1(\omega) \\ -\mathrm{i}\omega\alpha_3(\omega) \end{pmatrix}
   & =
   \underbrace{
   \begin{pmatrix}
   -\left(\frac{\gamma_1}{2}
      - \frac{\lvert \lambda_1\rvert^2}{\frac{\gamma_2}{2}-\mathrm{i}\omega}\right)
      &
   \frac{\lambda_1 \lambda_2^*}{\frac{\gamma_2}{2}-\mathrm{i}\omega} \\[2ex]
   \frac{\lambda_1^* \lambda_2}{\frac{\gamma_2}{2}-\mathrm{i}\omega}
   &
   -\left(\frac{\gamma_3}{2}
      - \frac{\lvert \lambda_2\rvert^2}{\frac{\gamma_2}{2}-\mathrm{i}\omega}\right)
   \end{pmatrix}
   }_{\equiv H_\mathrm{red}^{(a)}(\omega)}
   \begin{pmatrix}\alpha_1(\omega) \\ \alpha_3(\omega) \end{pmatrix}
   +
   \begin{pmatrix}
   \mathrm{i} \frac{\lambda_1\sqrt{\gamma_2}}{\frac{\gamma_2}{2}-\mathrm{i}\omega} \alpha_{2,\mathrm{in}}(\omega) 
   - \sqrt{\gamma_1} \alpha_{1,\mathrm{in}}(\omega) \\[2ex]
   \mathrm{i} \frac{\lambda_2\sqrt{\gamma_2}}{\frac{\gamma_2}{2}-\mathrm{i}\omega} \alpha_{2,\mathrm{in}}(\omega)
   - \sqrt{\gamma_3} \alpha_{3,\mathrm{in}}^\dagger(\omega)
   \end{pmatrix}.
\end{align}
This case is shown in Fig.~\ref{fig:reductionRules}~(c).

\subsection{Double reductions}
For the double reductions, we perform another subsequent reduction step in a longer chain of modes (at least four modes) as shown in Fig.~\ref{fig:reductionRules}.
There are five distinct cases to examine as shown in Figs.~\ref{fig:reductionRules}~(d)-(h). Any other case can be obtained from the cases shown here through reflection of the system.

\subsubsection{BS-BS-BS}
We utilize the results of the previous section for the case BS-BS and complement them by coupling to another mode
\begin{align}
   \begin{pmatrix} -\mathrm{i}\omega\alpha_1(\omega) \\ -\mathrm{i}\omega\alpha_2(\omega) \\ -\mathrm{i}\omega\alpha_4(\omega) \end{pmatrix}
   & =
   \begin{pmatrix}
   -\frac{\gamma_1}{2} & - \mathrm{i} J_1 & 0 \\
   -\mathrm{i} J_1^*
   & -\left(\frac{\gamma_2}{2}
      + \frac{\lvert J_2\rvert^2}{\frac{\gamma_3}{2}-\mathrm{i}\omega}\right)
      &
   - \frac{J_2 J_3}{\frac{\gamma_3}{2}-\mathrm{i}\omega} \\[2ex]
   0
   & - \frac{J_2^* J_3^*}{\frac{\gamma_3}{2}-\mathrm{i}\omega}
   &
   -\left(\frac{\gamma_4}{2}
      + \frac{\lvert J_3\rvert^2}{\frac{\gamma_2}{2}-\mathrm{i}\omega}\right)
   \end{pmatrix}
   \begin{pmatrix}\alpha_1(\omega) \\ \alpha_2(\omega) \\ \alpha_4(\omega) \end{pmatrix}
   +
   \begin{pmatrix}
   - \sqrt{\gamma_1} \alpha_{1,\mathrm{in}}(\omega) \\[2ex]
   \mathrm{i} \frac{J_2\sqrt{\gamma_2}}{\frac{\gamma_3}{2}-\mathrm{i}\omega} \alpha_{3,\mathrm{in}}(\omega) 
   - \sqrt{\gamma_2} \alpha_{2,\mathrm{in}}(\omega) \\[2ex]
   \mathrm{i} \frac{J_3^*\sqrt{\gamma_2}}{\frac{\gamma_3}{2}-\mathrm{i}\omega} \alpha_{3,\mathrm{in}}(\omega)
   - \sqrt{\gamma_4} \alpha_{4,\mathrm{in}}(\omega)
   \end{pmatrix}.
\end{align}
Eliminating $\alpha_2$ and defining
$\xi_1(\omega)\equiv\left(\frac{\gamma_3}{2}-\mathrm{i}\omega\right)^{-1}$
and
$\xi_2(\omega)\equiv \left(\frac{\gamma_2}{2} - \mathrm{i}\omega
+ \xi_1(\omega)\lvert J_2\rvert^2\right)^{-1}$, we obtain
\begin{align}
   \begin{pmatrix} -\mathrm{i}\omega\alpha_1(\omega) \\ -\mathrm{i}\omega\alpha_4(\omega) \end{pmatrix}
   & = \begin{pmatrix}
   -\frac{\gamma_1}{2}-\xi_2(\omega) \lvert J_1\rvert^2 & \mathrm{i} \xi_1(\omega) \xi_2(\omega) J_1 J_2 J_3 \\
   \mathrm{i} \xi_1(\omega) \xi_2(\omega) J_1^* J_2^* J_3^*
   & -\left(\frac{\gamma_4}{2}+\xi_1(\omega) \lvert J_3\rvert^2\right)
   + \xi_1^2(\omega)\xi_2(\omega) \lvert J_2 J_3\rvert^2
   \end{pmatrix}
   \begin{pmatrix} \alpha_1(\omega) \\ \alpha_4(\omega) \end{pmatrix}
   + \mathrm{input}.
\end{align}
The resulting effective coupling between $\alpha_1$ and $\alpha_4$, on resonance, has the same form as coherent BS coupling with a negative coupling constant and added local decay\footnote{%
The added local term in the equation for $\alpha_4$ is negative on resonance, as expected for a decay term, since $-\left(\frac{\gamma_4}{2}+\xi_1(\omega) \lvert J_3\rvert^2\right)
+ \xi_1^2(\omega)\xi_2(\omega) \lvert J_2 J_3\rvert^2=
-\frac{\gamma_4}{2}-\frac{1}{\frac{\gamma_3}{2}-\mathrm{i}\omega}
\frac{\lvert J_3\rvert^2}{1+\lvert J_2\rvert^2/[(\frac{\gamma_2}{2}-\mathrm{i}\omega)(\frac{\gamma_3}{2}-\mathrm{i}\omega)]}$ which is $<0$ on resonance, $\omega=0$.
}. This is depicted graphically in Fig.~\ref{fig:reductionRules}~(d).

\subsubsection{BS-BS-TMS}
We proceed analogously for a combination of BS and BS couplings followed by TMS couplings. Eliminating $\alpha_3$ yields
\begin{align}
   \begin{pmatrix} -\mathrm{i}\omega\alpha_1(\omega) \\ -\mathrm{i}\omega\alpha_2(\omega) \\ -\mathrm{i}\omega\alpha_4^\dagger(\omega) \end{pmatrix}
   & =
   \begin{pmatrix}
   -\frac{\gamma_1}{2} & - \mathrm{i} J_1 & 0 \\
   -\mathrm{i} J_1^*
   & -\left(\frac{\gamma_2}{2}
      + \frac{\lvert J_2\rvert^2}{\frac{\gamma_3}{2}-\mathrm{i}\omega}\right)
      &
   - \frac{J_2 \lambda_3}{\frac{\gamma_3}{2}-\mathrm{i}\omega} \\[2ex]
   0
   & \frac{J_2^* \lambda_3^*}{\frac{\gamma_3}{2}-\mathrm{i}\omega}
   &
   -\left(\frac{\gamma_4}{2}
      - \frac{\lvert \lambda_3\rvert^2}{\frac{\gamma_2}{2}-\mathrm{i}\omega}\right)
   \end{pmatrix}
   \begin{pmatrix}\alpha_1(\omega) \\ \alpha_2(\omega) \\ \alpha_4^\dagger(\omega) \end{pmatrix}
   +
   \mathrm{input}.
\end{align}
With $\xi_1(\omega)$ and $\xi_2(\omega)$ defined above,
\begin{align}
   \begin{pmatrix} -\mathrm{i}\omega\alpha_1(\omega) \\ -\mathrm{i}\omega\alpha_4^\dagger(\omega) \end{pmatrix}
   & = \begin{pmatrix}
   -\frac{\gamma_1}{2}-\xi_2(\omega) \lvert J_1\rvert^2 & \mathrm{i} \xi_1(\omega) \xi_2(\omega) J_1 J_2 \lambda_3 \\
   -\mathrm{i} \xi_1(\omega) \xi_2(\omega) J_1^* J_2^* \lambda_3^*
   & -\left(\frac{\gamma_4}{2}-\xi_1(\omega) \lvert \lambda_3\rvert^2\right)
   - \xi_1^2(\omega)\xi_2(\omega) \lvert J_2 \lambda_3\rvert^2
   \end{pmatrix}
   \begin{pmatrix} \alpha_1(\omega) \\ \alpha_4^\dagger(\omega) \end{pmatrix}
   + \mathrm{input}'.
\end{align}
This takes the form of TMS couplings with added local decay on the first and decay/gain on the fourth site (depending on how the terms balance out) as shown in Fig.~\ref{fig:reductionRules}~(e).

\subsubsection{BS-TMS-BS}
BS followed by TMS, followed by BS coupling yields a very similar result. After the first reduction step of $\alpha_3$, we find
\begin{align}
   \begin{pmatrix} -\mathrm{i}\omega\alpha_1(\omega) \\ -\mathrm{i}\omega\alpha_2(\omega) \\ -\mathrm{i}\omega\alpha_4^\dagger(\omega) \end{pmatrix}
   & =
   \begin{pmatrix}
   -\frac{\gamma_1}{2} & - \mathrm{i} J_1 & 0 \\
   -\mathrm{i} J_1^*
   & -\left(\frac{\gamma_2}{2}
      - \frac{\lvert \lambda_2\rvert^2}{\frac{\gamma_3}{2}-\mathrm{i}\omega}\right)
      &
   \frac{\lambda_2 J_3}{\frac{\gamma_3}{2}-\mathrm{i}\omega} \\[2ex]
   0
   & - \frac{\lambda_2^* J_3^*}{\frac{\gamma_3}{2}-\mathrm{i}\omega}
   &
   -\left(\frac{\gamma_4}{2}
      + \frac{\lvert J_3\rvert^2}{\frac{\gamma_2}{2}-\mathrm{i}\omega}\right)
   \end{pmatrix}
   \begin{pmatrix}\alpha_1(\omega) \\ \alpha_2(\omega) \\ \alpha_4^\dagger(\omega) \end{pmatrix}
   +
   \mathrm{input}.
\end{align}
After eliminating $\alpha_2$ as well, we obtain with $\xi_1(\omega)\equiv (\frac{\gamma_3}{2}-\mathrm{i}\omega)^{-1}$ and $\xi_2(\omega)\equiv(\frac{\gamma_2}{2}-\mathrm{i}\omega-\xi_1(\omega)\lvert \lambda_2\rvert^2)^{-1}$,
\begin{align}
   \begin{pmatrix} -\mathrm{i}\omega\alpha_1(\omega) \\ -\mathrm{i}\omega\alpha_4^\dagger(\omega) \end{pmatrix}
   & = \begin{pmatrix}
   -\frac{\gamma_1}{2}-\xi_2(\omega) \lvert J_1\rvert^2 & -\mathrm{i} \xi_1(\omega) \xi_2(\omega) J_1 \lambda_2 J_3 \\
   \mathrm{i} \xi_1(\omega) \xi_2(\omega) J_1^* \lambda_2^* J_3^*
   & -\left(\frac{\gamma_4}{2}-\xi_1(\omega) \lvert \lambda_3\rvert^2\right)
   - \xi_1^2(\omega)\xi_2(\omega) \lvert \lambda_2 J_3\rvert^2
   \end{pmatrix}
   \begin{pmatrix} \alpha_1(\omega) \\ \alpha_4^\dagger(\omega) \end{pmatrix}
   + \mathrm{input}'.
\end{align}
This resembles TMS coupling with local decay added on the first and gain added on the last site, see Fig.~\ref{fig:reductionRules}~(f).

\subsubsection{BS-TMS-TMS}
We repeat the same steps for a combination of BS with two TMS couplings. First we use the result from the TMS-TMS reduction step to eliminate $\alpha_3^\dagger(\omega)$
\begin{align}
   \begin{pmatrix} -\mathrm{i}\omega\alpha_1(\omega) \\ -\mathrm{i}\omega\alpha_2(\omega) \\ -\mathrm{i}\omega\alpha_4(\omega) \end{pmatrix}
   & =
   \begin{pmatrix}
   -\frac{\gamma_1}{2} & - \mathrm{i} J_1 & 0 \\
   -\mathrm{i} J_1^*
   & -\left(\frac{\gamma_2}{2}
      - \frac{\lvert \lambda_2\rvert^2}{\frac{\gamma_3}{2}-\mathrm{i}\omega}\right)
      &
   \frac{\lambda_2 \lambda_3^*}{\frac{\gamma_3}{2}-\mathrm{i}\omega} \\[2ex]
   0
   & \frac{\lambda_2^* \lambda_3}{\frac{\gamma_3}{2}-\mathrm{i}\omega}
   &
   -\left(\frac{\gamma_4}{2}
      - \frac{\lvert \lambda_3\rvert^2}{\frac{\gamma_2}{2}-\mathrm{i}\omega}\right)
   \end{pmatrix}
   \begin{pmatrix}\alpha_1(\omega) \\ \alpha_2(\omega) \\ \alpha_4(\omega) \end{pmatrix}
   +
   \mathrm{input}'.
\end{align}
Eliminating $\alpha_2$, we obtain with $\xi_1(\omega)\equiv (\frac{\gamma_3}{2}-\mathrm{i}\omega)^{-1}$ and $\xi_2(\omega)\equiv(\frac{\gamma_2}{2}-\mathrm{i}\omega-\xi_1(\omega)\lvert \lambda_2\rvert^2)^{-1}$,
\begin{align}
   \begin{pmatrix} -\mathrm{i}\omega\alpha_1(\omega) \\ -\mathrm{i}\omega\alpha_4(\omega) \end{pmatrix}
   & = \begin{pmatrix}
   -\frac{\gamma_1}{2}-\xi_2(\omega) \lvert J_1\rvert^2 & -\mathrm{i} \xi_1(\omega) \xi_2(\omega) J_1 \lambda_2 \lambda_3^* \\
   -\mathrm{i} \xi_1(\omega) \xi_2(\omega) J_1^* \lambda_2^* \lambda_3
   & -\left(\frac{\gamma_4}{2}-\xi_1(\omega) \lvert \lambda_3\rvert^2\right)
   + \xi_1^2(\omega)\xi_2(\omega) \lvert \lambda_2 \lambda_3\rvert^2
   \end{pmatrix}
   \begin{pmatrix} \alpha_1(\omega) \\ \alpha_4(\omega) \end{pmatrix}
   + \mathrm{input}.
\end{align}
We effectively obtain BS couplings with added local decay on the first and local gain on the last site, see Fig.~\ref{fig:reductionRules}~(g).

\subsubsection{TMS-TMS-TMS}
Finally, we consider four modes all connected via TMS couplings. Eliminating in the first step $\alpha_3$, we obtain, recalling the result of a TMS-TMS reduction,
\begin{align}
   \begin{pmatrix} -\mathrm{i}\omega\alpha_1(\omega) \\ -\mathrm{i}\omega\alpha_2^\dagger(\omega) \\ -\mathrm{i}\omega\alpha_4^\dagger(\omega) \end{pmatrix}
   & =
   \begin{pmatrix}
   -\frac{\gamma_1}{2} & - \mathrm{i} \lambda_1 & 0 \\
   \mathrm{i} \lambda_1^*
   & -\left(\frac{\gamma_2}{2}
      - \frac{\lvert \lambda_2\rvert^2}{\frac{\gamma_3}{2}-\mathrm{i}\omega}\right)
      &
   \frac{\lambda_2^* \lambda_3}{\frac{\gamma_3}{2}-\mathrm{i}\omega} \\[2ex]
   0
   & \frac{\lambda_2 \lambda_3^*}{\frac{\gamma_3}{2}-\mathrm{i}\omega}
   &
   -\left(\frac{\gamma_4}{2}
      - \frac{\lvert \lambda_3\rvert^2}{\frac{\gamma_2}{2}-\mathrm{i}\omega}\right)
   \end{pmatrix}
   \begin{pmatrix}\alpha_1(\omega) \\ \alpha_2^\dagger(\omega) \\ \alpha_4^\dagger(\omega) \end{pmatrix}
   +
   \mathrm{input}.
\end{align}
Eliminating $\alpha_2$, we obtain with $\xi_1(\omega)\equiv (\frac{\gamma_3}{2}-\mathrm{i}\omega)^{-1}$ and $\xi_2(\omega)\equiv(\frac{\gamma_2}{2}-\mathrm{i}\omega-\xi_1(\omega)\lvert \lambda_2\rvert^2)^{-1}$,
\begin{align}
   \begin{pmatrix} -\mathrm{i}\omega\alpha_1(\omega) \\ -\mathrm{i}\omega\alpha_4^\dagger(\omega) \end{pmatrix}
   & = \begin{pmatrix}
   -\frac{\gamma_1}{2}+\xi_2(\omega) \lvert \lambda_1\rvert^2 & -\mathrm{i} \xi_1(\omega) \xi_2(\omega) \lambda_1 \lambda_2^* \lambda_3 \\
   \mathrm{i} \xi_1(\omega) \xi_2(\omega) \lambda_1^* \lambda_2 \lambda_3^*
   & -\left(\frac{\gamma_4}{2}-\xi_1(\omega) \lvert \lambda_3\rvert^2\right)
   + \xi_1^2(\omega)\xi_2(\omega) \lvert \lambda_2 \lambda_3\rvert^2
   \end{pmatrix}
   \begin{pmatrix} \alpha_1(\omega) \\ \alpha_4^\dagger(\omega) \end{pmatrix}
   + \mathrm{input}.
\end{align}
This effectively takes the form of TMS coupling with locally added gain on both sites, see Fig.~\ref{fig:reductionRules}~(h).

\subsection{Many-fold reductions: the qNR condition}
Further reduction steps proceed analogously to the single and double reductions shown here. Since an even number of reduction steps reduces the dynamical matrix effectively to one with coherent couplings and added local decay or gain, we can recursively apply single and double reductions until we reduced a ring of arbitrary size to coupling between just a pair of modes.
In principle, even complicated networks can be analysed in this way, eliminating intermediate modes proceeding loop by loop.

\begin{figure}[htbp]
   \centering
   \includegraphics[width=.9\textwidth]{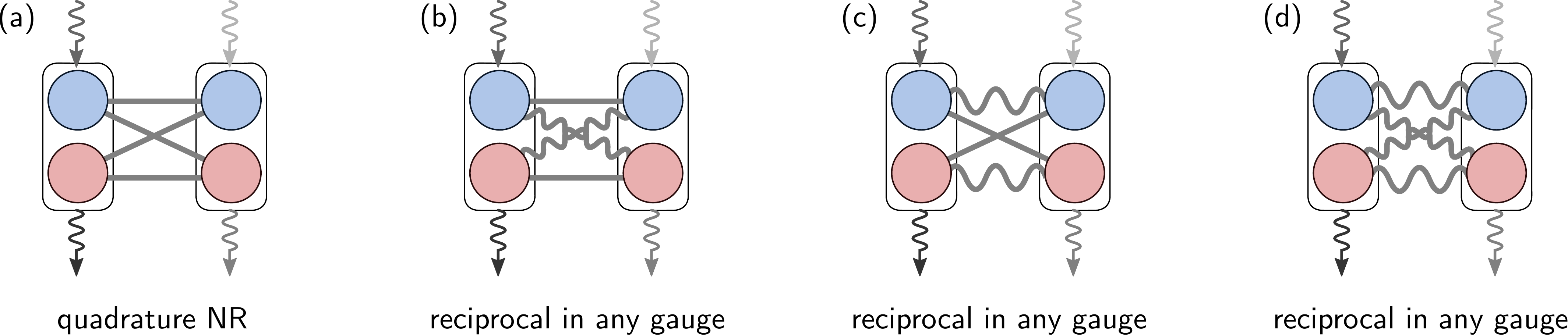}
   \caption{\textbf{Possible reduced systems of TRS preserving rings.}
   (a)~The only possible system that displays non-reciprocity in the site basis is a version of the dimer with inhomogeneously added local decay or gain. Diagram node colors are identical to~Fig.~\ref{fig:reductionRules}. Systems with an even number of modes and an odd number of TMS couplings (and all other BS) reduce to this system.
   Any other ring reduces to one of the systems (b)-(d) which are reciprocal in all possible gauges.
   }
   \label{fig:reducedMinimalSystems}
\end{figure}%

The reduction of larger systems to smaller systems let us derive general conditions for qNR systems. We only need to check systems with an odd number of TMS couplings since such allow us to preserve TRS. Focussing on rings and performing the reduction until we are left with only two modes, we obtain four different possible reduced systems, see Fig.~\ref{fig:reducedMinimalSystems}. One of them is a more general form of the qNR dimer discussed in the main text with inhomogeneously added local decay and gain. Directly examining these systems for the possibility of qNR, we find that indeed \emph{only} the modified qNR dimer exhibits qNR, so any system that can be reduced to a version of the qNR dimer---a system combining BS and TMS couplings and possibly local gain and loss---is a qNR system. This leads us to the sufficient condition for qNR rings. Since only an even number of reductions steps will map the couplings to a form that resemble coherent couplings (with added local decay/ gain), only rings consisting of an even number of modes can lead to qNR.
Together with the TRS condition, which required an odd number of TMS couplings, we have now found both necessary and sufficient conditions for qNR rings: the rings defined by these conditions are the complete set of qNR rings.

To make this last inference, we note that gauge transformations commute with the reductions performed above (this is because the reductions are performed for $\alpha_j$ and $\alpha_j^\dagger$ in the same way) such that transport between modes of the reduced system will transform in the same way as for the `full' system we started with.

%

\end{document}